\documentclass[twocolumn]{aastex63}
\UseRawInputEncoding
\usepackage{hyperref}
\usepackage{breakurl}
\usepackage{amsmath}
\shorttitle{A new sample of warm extreme debris disks}
\shortauthors{Mo\'or et al.}
\begin{document}

\title{A new sample of warm extreme debris disks from the ALLWISE catalog}

\correspondingauthor{Attila Mo\'or}
\email{moor@konkoly.hu}

\author{Attila Mo\'or}
\affil{Konkoly Observatory, Research Centre for Astronomy and
Earth Sciences, E\"otv\"os Lor\'and Research Network (ELKH) \\ Konkoly-Thege Mikl\'os \'ut 15-17, 1121 Budapest, Hungary}
\affil{ELTE E\"otv\"os Lor\'and University, Institute of Physics, P\'azm\'any 
P\'eter s\'et\'any 1/A, 1117 Budapest, Hungary}

\author{P\'eter \'Abrah\'am}
\affiliation{Konkoly Observatory, Research Centre for Astronomy and Earth Sciences, E\"otv\"os Lor\'and Research Network (ELKH) \\ 
Konkoly-Thege Mikl\'os \'ut 15-17, 1121 Budapest, Hungary}
\affil{ELTE E\"otv\"os Lor\'and University, Institute of Physics, P\'azm\'any 
P\'eter s\'et\'any 1/A, 1117 Budapest, Hungary}

\author{Gyula Szab\'o}
\affiliation{ELTE E\"otv\"os Lor\'and University, Gothard Astrophysical Observatory, Szombathely, Hungary}
\affiliation{MTA-ELTE Exoplanet Research Group, 9700 Szombathely, Szent Imre h. u. 112, Hungary}

\author{Kriszti\'an Vida}
\affiliation{Konkoly Observatory, Research Centre for Astronomy and Earth Sciences, E\"otv\"os Lor\'and Research Network (ELKH) \\ 
Konkoly-Thege Mikl\'os \'ut 15-17, 1121 Budapest, Hungary}
\affil{ELTE E\"otv\"os Lor\'and University, Institute of Physics, P\'azm\'any 
P\'eter s\'et\'any 1/A, 1117 Budapest, Hungary}

\author{Gianni Cataldi}
\affiliation{National Astronomical Observatory of Japan, Osawa 2-21-1, Mitaka, Tokyo 181-8588, Japan}
\affiliation{Department of Astronomy, Graduate School of Science, The University of Tokyo, 7-3-1 
Hongo, Bunkyo-ku, Tokyo 113-0033, Japan}

\author{Al\'\i z Derekas}
\affiliation{ELTE E\"otv\"os Lor\'and University, Gothard Astrophysical Observatory, Szombathely, Hungary}
\affiliation{MTA-ELTE Exoplanet Research Group, 9700 Szombathely, Szent Imre h. u. 112, Hungary}

\author{Thomas Henning}
\affiliation{Max-Planck-Institut f\"ur Astronomie, K\"onigstuhl 17, D-69117 Heidelberg, Germany}

\author{Karen Kinemuchi}
\affiliation{Apache Point Observatory and New Mexico State University, Sunspot NM 88349, USA}

\author{\'Agnes K\'osp\'al}
\affiliation{Konkoly Observatory, Research Centre for Astronomy and Earth Sciences, E\"otv\"os Lor\'and Research Network (ELKH) \\ 
Konkoly-Thege Mikl\'os \'ut 15-17, 1121 Budapest, Hungary}
\affiliation{Max-Planck-Institut f\"ur Astronomie, K\"onigstuhl 17, D-69117 Heidelberg, Germany}
\affil{ELTE E\"otv\"os Lor\'and University, Institute of Physics, P\'azm\'any 
P\'eter s\'et\'any 1/A, 1117 Budapest, Hungary}

\author{J\'ozsef Kov\'acs}
\affiliation{ELTE E\"otv\"os Lor\'and University, Gothard Astrophysical Observatory, Szombathely, Hungary}
\affiliation{MTA-ELTE Exoplanet Research Group, 9700 Szombathely, Szent Imre h. u. 112, Hungary}

\author{Andr\'as P\'al}
\affiliation{Konkoly Observatory, Research Centre for Astronomy and Earth Sciences, E\"otv\"os Lor\'and Research Network (ELKH) \\ 
Konkoly-Thege Mikl\'os \'ut 15-17, 1121 Budapest, Hungary}
\affil{ELTE E\"otv\"os Lor\'and University, Institute of Physics, P\'azm\'any 
P\'eter s\'et\'any 1/A, 1117 Budapest, Hungary}

\author{Paula Sarkis}
\affiliation{Max-Planck-Institut f\"ur Astronomie, K\"onigstuhl 17, D-69117 Heidelberg, Germany}

\author{B\'alint Seli}
\affiliation{Konkoly Observatory, Research Centre for Astronomy and Earth Sciences, E\"otv\"os Lor\'and Research Network (ELKH) \\ 
Konkoly-Thege Mikl\'os \'ut 15-17, 1121 Budapest, Hungary}
\affiliation{E\"otv\"os Lor\'and University, Department of Astronomy, P\'azm\'any P\'eter 
s\'et\'any 1/A, 1117 Budapest, Hungary}

\author{Zs\'ofia M. Szab\'o}
\affiliation{E\"otv\"os Lor\'and University, Department of Astronomy, P\'azm\'any P\'eter 
s\'et\'any 1/A, 1117 Budapest, Hungary}
\affil{Konkoly Observatory, Research Centre for Astronomy and Earth Sciences, E\"otv\"os Lor\'and Research Network (ELKH) \\ 
Konkoly-Thege Mikl\'os \'ut 15-17, 1121 Budapest, Hungary}

\author{Katalin Tak\'ats}
\affiliation{Departamento de Ciencias F\'\i sicas, Universidad Andr\'es Bello, 
Avda. Rep\'ublica 252, 32349 Santiago, Chile }
%%



\begin{abstract}
Extreme debris disks (EDDs) are rare systems with peculiarly large amounts 
of warm dust that may stem from recent giant impacts between 
planetary embryos during the final phases of terrestrial planet growth.
Here we report on the identification and characterization of six new EDDs.
These disks surround F5-G9 type main-sequence stars with ages $>$100\,Myr, 
have dust temperatures higher than 300\,K and fractional luminosities
between 0.01 and 0.07. Using time-domain photometric data at 3.4 and 
4.6\,{\micron} from the WISE all sky surveys, we conclude that four of these
disks exhibited variable mid-infrared emission between 2010 and 2019. Analyzing 
the sample of all known EDDs, now expanded to 17 objects, we find that 
14 of them showed changes at 3--5\,{\micron} over the past decade suggesting that mid-infrared 
variability is an inherent characteristic of EDDs. We also report that 
wide-orbit pairs are significantly more common in EDD systems than in 
the normal stellar population. While current models of rocky planet 
formation predict that the majority of giant collisions occur in the 
first 100\,Myr, we find that the sample of EDDs is dominated by 
systems older than this age. This raises the possibility that the era of 
giant impacts may be longer than we think, or that some other mechanism(s) can 
also produce EDDs. We examine a scenario where the observed warm dust 
stems from the disruption and/or collisions of comets delivered from an 
outer reservoir into the inner regions, and explore what role the 
wide companions could play in this process.  
\end{abstract}




\section{Introduction} \label{sec:intro}

Debris disks are comprised of planetesimals and of their erosional
products down to micrometer-sized dust particles \citep{wyatt2008,krivov2010}. 
From this mixture of solids of different sizes, only the smallest particles are observable. 
The majority of debris disks have been detected based on the infrared (IR) excess caused 
by thermal emission from their optically thin dust material. Large far-IR surveys with 
the {\sl Spitzer Space Telescope} \citep[{\sl Spitzer,}][]{werner2004} and the 
{\sl Herschel Space Observatory} \citep{pilbratt2010} indicated that 
at least 20\% of main-sequence A--K type stars are encircled by cold dust ($\lesssim$100\,K) 
implying the presence of debris belts at tens of astronomical units from the central stars 
\citep{su2006,thureau2014,montesinos2016,sibthorpe2018,hughes2018}. 
These structures are analogous to the Kuiper-belt of 
our solar system. Their dust particles, which are continuously removed by interactions with 
the radiation forces of the star and/or with the stellar wind, are thought to be replenished by
a collisional cascade involving larger bodies. 

Analysis of the spectral energy distribution 
(SED) implies that a fraction of systems with cold debris dust also host warmer ($>$150\,K) 
dust particles, suggesting that an inner debris ring resembling our asteroid belt 
also exists around the star \citep{kennedy2014,chen2014,ertel2020}. We also 
know of some systems that possess only warm dust \citep[e.g.,][]{beichman2011}.
The majority of these warm dust belts can be explained 
as a product of long term collisional evolution of a planetesimal belt co-located with the 
dust \citep{geiler2017}. Alternatively, or in addition to the in situ dust generation, 
disrupting comets delivered from an outer reservoir \citep{bonsor2012,nesvorny2010} 
or grains transported inward from an outer debris cloud under the influence of 
Poynting-Robertson and stellar wind drag \citep{reidemeister2011,kennedy2015,rigley2020} 
can also produce warm dust particles in the inner regions. 

The last fifteen years have seen the discovery of several FGK-type 
main-sequence stars that host unusually dust-rich warm debris disks with 
typical dust temperature of $>$300\,K and with fractional 
luminosity ($f_d = L_{\rm disk}/L_*$) of $>$0.01. These fractional 
luminosities are three orders of magnitude higher than those of typical 
warm debris systems, and suggest that these {\sl extreme debris disks} 
\citep[EDDs,][]{balog2009} are at least partly optically thick for the 
stellar illumination. EDDs 
share some additional common properties. Mid-IR spectroscopic 
observations revealed that they tend to show strong solid state emission 
features, implying the presence of small, submicron-sized crystalline dust 
particles \citep{rhee2008,olofsson2012,lisse2020}. Even more uniquely, their 
photometric monitoring with {\sl Spitzer} demonstrated that most EDDs 
display significant variability on monthly to yearly timescales 
at 3--5\,{\micron} \citep{meng2014,meng2015,su2019}. The ages of known EDDs 
fall predominantly in the range between 10 and 200\,Myr, the only exception 
is BD+20~307 with an age of $>$1\,Gyr \citep{weinberger2011}.

The peculiarly high dust content of EDDs is orders of magnitude above what 
can be explained by a steady state grinding down process of an in situ 
planetesimal population, that started when the system was formed 
\citep[e.g.,][]{wyatt2007}. Instead, the observed properties point to 
a recent, episodic increase in dust production, which may be related to a 
giant collision of planetary embryos in the inner 1--2\,au region \citep{melis2010,su2019}. 
Similar collisions -- including the Moon-forming event between the proto-Earth 
and a planetary embryo -- are thought to be common during the
final accumulation of terrestrial planets in the early ($\lesssim$100\,Myr) 
Solar system \citep{chambers2013}. Such impacts produce a large quantity 
of debris material that escapes the planets entirely, forming a dust disk that
causes significant mid-IR excess for a period \citep{jackson2012,genda2015}. In 
this scenario, the observed rapid time variations can be linked to the orbital and collisional 
evolution of the dust and vapor cloud that emerged from the impact event 
\citep{su2019}. Submicron-sized grains produced in optically thick clumps 
of vapor condensates can explain the observed prominent 10\,{\micron} 
solid-state features \citep{su2020}.

By studying EDDs, we can learn more about the immediate
aftermath of major transient dust production events possibly associated to 
giant impacts. Such investigations thus also have the potential to 
improve our knowledge on the formation of terrestrial planets. However, EDDs are rare 
objects, currently only 11\footnote{This number refers only to EDDs hosted by FGK-type main-sequence stars.
We note that recent years have seen the discovery of several young M dwarf stars 
with warm circumstellar disks having extremely high fractional luminosity
 \citep{flaherty2019,silverberg2020,zuckerman2019}.  
These systems may have a mixed nature while some of them are probably 
long lived accretion disks \citep[e.g.][]{silverberg2020} some may harbor
debris disks \citep{zuckerman2019}.} such disks are known (Appendix~\ref{appendix:c}). To understand this
phenomenon in greater detail, we need a larger sample. In the present work 
we aim to identify additional EDDs around Sun-like stars and study their properties 
with special regard to infrared variability and age distribution. The discovery 
of currently known EDDs in the literature was mainly based on infrared data from the {\sl Infrared Astronomical 
Satellite} \citep[{\sl IRAS},][]{neugebauer1984} survey and 
from dedicated {\sl Spitzer} programs targeting young open clusters.
In our study we use photometry obtained with the {\sl Wide-Field Infrared Survey 
Explorer} \citep[{\sl WISE},][]{wright2010}, that scanned the whole sky in four 
infrared bands, at 3.4\,{\micron} (W1), 4.6\,{\micron} (W2), 12\,{\micron} (W3), and 22\,{\micron} (W4). 
Thanks to its substantially 
better sensitivity in the latter two bands than {\sl IRAS} had at 12 and 25\,{\micron}, 
{\sl WISE} offers an excellent opportunity to search for warm debris 
disks over the entire sky, as it has already been demonstrated by several studies 
\citep[e.g.,][]{kennedy2013,de-miera2014,patel2014,patel2017,cotten2016}. 
The two shorter wavelength bands enable the detection and characterization 
of hot dust at temperatures above 400\,K.
The {\sl WISE} satellite is still in operation and has been continuing the 
all sky surveys at 3.4 and 4.6\,{\micron} providing photometry 
spaced by 6\,months. Our study on infrared variability of EDDs is principally 
based on this data set.

This paper is structured as follows. Section~\ref{sec:sampleselection} reviews
our investigation to search for new EDD systems utilizing the AllWISE infrared 
\citep{cutri2013} and the Tycho-Gaia Astrometric Solution \citep[TGAS,][]{michalik2015,lindegren2016} 
astrometric catalogs. Our procedure -- in which we took special care to filter out 
possible false disk identifications -- resulted in six disks. 
In Section~\ref{sec:hoststars} we discuss the characterization of the host stars 
including the age estimates of the systems. The determination 
of the disk properties is described in Section~\ref{sec:characteizationofedds}, while 
the mid-infrared variability of the selected objects is studied in Section~\ref{sec:diskvariability}.
We then discuss our results (Section~\ref{sec:discussion}) 
and provide a summary of the work (Section~\ref{sec:summary}).

%%%%%%%%%%%%%%%%%%%% TABLE 1 %%%%%%%%%%%%%%%%%%%%%%%%%%%%%%%%%%

\begin{deluxetable*}{lccccccc}
\tabletypesize{\tiny}
\tablecaption{Stellar and disk properties \label{tab:props}} 
\tablecolumns{8}
\tablehead{
\colhead{Parameters} &
\colhead{TYC\,4515}  &
\colhead{TYC\,5940}  &
\colhead{TYC\,8105}  &
\colhead{TYC\,4946}  &
\colhead{TYC\,4209}  &
\colhead{TYC\,4479}  &
\colhead{References}     
}
\startdata
\cutinhead{Identifiers}
TYCHO     & TYC~4515-485-1 &  TYC~5940-1510-1 &  TYC~8105-370-1 &  TYC~4946-1106-1 &  TYC~4209-1322-1 &  TYC~4479-3-1 &  \citet{hog2000} \\
Gaia DR2  & 552973538962350080 & 2942418533272205312 & 5554553591848668672 & 3596395748683517440 & 2161325369818713984 & 2210856513228282496 & Gaia DR2 \\
Gaia EDR3 & 552973538964681088 & 2942418533272205312 & 5554553591848668672 & 3596395748683517440 & 2161325369818713984 & 2210856513228282496 & Gaia EDR3 \\
\cutinhead{Astrometric Information}
$\alpha$ R.A. (2000) &  \phantom{$+$}05:04:07.2  & \phantom{$+$}06:05:13.6 & \phantom{$+$}06:11:03.5 & \phantom{$+$}12:13:34.2 & \phantom{$+$}18:17:03.9 & \phantom{$+$}23:52:50.6 & SIMBAD \\
$\delta$ Decl. (2000) & $+$77:58:57.5 & $-$19:13:08.3 & $-$47:11:29.4 & $-$05:35:43.4 & $+$64:33:55.0 & $+$67:30:37.6 & SIMBAD \\
$\mu_{\alpha^*}$ (mas~yr$^{-1}$) & $+$6.710$\pm$0.012 & $+$0.011$\pm$0.010 & $+$6.277$\pm$0.012 & $-$27.754$\pm$0.023 & $+$12.051$\pm$0.014 & $+$38.619$\pm$0.016 & Gaia EDR3 \\
$\mu_{\delta}$ (mas~yr$^{-1}$) & $-$23.625$\pm$0.015 & $-$11.719$\pm$0.012 & $+$12.577$\pm$0.014 & {$-$4.047$\pm$0.018} & $-$3.106$\pm$0.013 & $-$39.644$\pm$0.014 & Gaia EDR3 \\
$\pi$ (mas) & $+$3.564$\pm$0.015 & $+$3.784$\pm$0.013 & $+$5.411$\pm$0.010 & $+$4.093$\pm$0.021 & $+$3.647$\pm$0.010 & $+$6.142$\pm$0.014 & Gaia EDR3 \\
$RUWE$\tablenotemark{a} & 0.872 & 0.964 & 0.796 & 1.214 & 0.744 & 0.888 & Gaia EDR3 \\
\cutinhead{Photometric Data}
$B_{\rm T}$    &  11.315$\pm$0.054 &  12.949$\pm$0.258 &  12.486$\pm$0.191 &  11.199$\pm$0.064 &  12.281$\pm$0.160 &  11.853$\pm$0.078 & \citet{hog2000} \\
$B$            &  11.154$\pm$0.100 &  12.694$\pm$0.021 &  12.448$\pm$0.031 &  11.143$\pm$0.011 &  11.980$\pm$0.133 &  11.769$\pm$0.075 & \citet{henden2016} \\
$BP_{\rm Gaia}$           & 10.8965$\pm$0.0028 &  12.1920$\pm$0.0035 &  11.9305$\pm$0.0038 &  10.8298$\pm$0.0028 &  11.7268$\pm$0.0031 &  11.3735$\pm$0.0030 & Gaia EDR3 \\
$V_{\rm T}$    &  10.807$\pm$0.051 &  11.923$\pm$0.156 &  11.746$\pm$0.136 &  10.921$\pm$0.073 &  11.684$\pm$0.131 &  11.153$\pm$0.069 & \citet{hog2000} \\
$V$            &  10.705$\pm$0.087 &  12.007$\pm$0.070 &  11.696$\pm$0.026 &  10.669$\pm$0.013 &  11.560$\pm$0.049 &  11.156$\pm$0.048 & \citet{henden2016} \\
$G_{\rm Gaia}$  &  10.6430$\pm$0.0028 & 11.8630$\pm$0.0028 & 11.5505$\pm$0.0028 &  10.5926$\pm$0.0028 & 11.4509$\pm$0.0028 &  11.0459$\pm$0.0028 & Gaia EDR3 \\
$RP_{\rm Gaia}$ & 10.2366$\pm$0.0038 & 11.3693$\pm$0.0040 &  11.0042$\pm$0.0042 &  10.2026$\pm$0.0038 &  11.0122$\pm$0.0039 &  10.5491$\pm$0.0038 & Gaia EDR3 \\
$J$            &  9.815$\pm$0.022 &  10.787$\pm$0.024 &  10.426$\pm$0.024 &  9.793$\pm$0.023 &  10.556$\pm$0.023 &  9.972$\pm$0.023 & \citet{cutri2003} \\
$H$            &  9.620$\pm$0.021 &  10.518$\pm$0.022 &  10.028$\pm$0.022 &  9.586$\pm$0.022 &  10.267$\pm$0.023 &  9.663$\pm$0.031 & \citet{cutri2003} \\
$K_{\rm s}$    &  9.538$\pm$0.019 &  10.402$\pm$0.019 &  9.934$\pm$0.021 &  9.520$\pm$0.023 &  10.139$\pm$0.016 &  9.624$\pm$0.018  & \citet{cutri2003} \\
\cutinhead{Kinematics}
$v_{\rm r,our}$ (km~s$^{-1}$)    & $+$12.50$\pm$0.50 & $+$28.50$\pm$0.20 & $+$29.00$\pm$0.20 & $-$11.30$\pm$0.50 & $-$13.20$\pm$0.60 & $+$0.22$\pm$0.30 &  This  work \\  
$v_{\rm r,Gaia}$ (km~s$^{-1}$)   &  $-$2.65$\pm$0.89 & $+$30.52$\pm$1.38 & $+$29.06$\pm$0.65 & $-$12.29$\pm$0.36 & $-$13.46$\pm$1.14 & \nodata & Gaia DR2 \\
$U$\tablenotemark{b} (km~s$^{-1}$) & $-$32.7$\pm$0.3 & $-$8.1$\pm$0.1 & $-$16.9$\pm$0.1 & $-$27.2$\pm$0.2 & $+$3.0$\pm$0.1 & $-$21.2$\pm$0.1 & This  work \\
$V$\tablenotemark{b} (km~s$^{-1}$) & $-$11.3$\pm$0.3 & $-$27.4$\pm$0.1 & $-$25.9$\pm$0.2 & $-$12.3$\pm$0.3& $-$4.1$\pm$0.5 & $-$7.0$\pm$0. & This  work \\
$W$\tablenotemark{b} (km~s$^{-1}$) & $-$3.7$\pm$0.2  & $-$14.5$\pm$0.1 & $-$5.8$\pm$0.1 & $-$16.5$\pm$0.4 & $-$20.1$\pm$0.3 & $-$36.2$\pm$0.1 & This  work \\
Distance (pc)\tablenotemark{c}  &  278.8$\pm$1.1 & 262.1$^{+1.2}_{-1.0}$ & 184.1$^{+0.4}_{-0.3}$ & 242.1$^{+1.4}_{-1.1}$ & 270.8$^{+0.7}_{-0.6}$ & 162.0$^{+0.4}_{-0.3}$ & \citet{cbj2020} \\
\cutinhead{Stellar properties}
Spectral Type          &  F5V & G5V & G9V & F6V & G1V & G6V & This work \\
$T_{\rm eff}$ (K)      &  6540$\pm$100 & 5660$\pm$100 & 5350$\pm$100 &  6350$\pm$60 & 5890$\pm$120 & 5620$\pm$100 & This  work \\
$\log{g}$ (cgs)        & 4.28$\pm$0.10 & 4.49$\pm$0.10 & 4.48$\pm$0.10  & 4.13$\pm$0.06 & 4.40$\pm$0.10 & 4.30$\pm$0.10 &  This  work \\  
$[{\rm Fe/H}]$ (dex) & $+$0.07$\pm$0.10 & $+$0.06$\pm$0.10  & $+$0.08$\pm$0.10 & $-$0.02$\pm$0.05 & $+$0.01$\pm$0.12 & $-$0.29$\pm$0.10 &  This  work \\
$L_*$ (L$_\odot$)      & 3.72$\pm$0.12 & 0.94$\pm$0.03 & 0.65$\pm$0.02 & 2.55$\pm$0.08 & 1.43$\pm$0.04 & 0.76$\pm$0.02 &  This  work \\
$M_*$ (M$_\odot$)      & 1.32$\pm$0.04 & 0.99$^{+0.04}_{-0.03}$ & 0.94$\pm$0.03 & 1.17$^{+0.04}_{-0.03}$ & 1.05$\pm$0.03 & 0.84$\pm$0.03 &  This  work \\
$EW_{\rm Li}$ ({m\AA})\tablenotemark{d} &  $<$15.0 & 157$\pm$8 & 173$\pm$6  & 38$\pm$8 & 101$\pm$15 & 13$\pm$4 &  This  work \\
$P_{\rm rot}$ (day)    &  \nodata & 3.76 & 5.04 & \nodata  & 5.07 & 32.1 &  This  work \\
Age (Myr)              & $>$150 & 120$\pm$20 & 130$\pm$30 & $>$150 & 275$\pm$50 & 5000$\pm$2000 &  This  work \\
Multiplicity           & Y & N & Y & Y & Y & Y & This work \\
\cutinhead{Disk properties}
T$_{\rm disk,bb}$ (K)     &  500$\pm$25 & 420$\pm$16 & 300$\pm$20 & 680$\pm$45 & 530$\pm$15 & 400$\pm$25 & This work \\
f$_{\rm d}$               &  0.015 & 0.037 & 0.052 & 0.010 & 0.070 & 0.014 & This work \\
R$_{\rm disk,bb}$ (au)    &  0.60 & 0.43 & 0.69 & 0.27 & 0.33 & 0.42 & This work \\
f$_{\rm d}$/f$_{\rm max}$ (10$^5$) &  1.1 & 1.9 & 0.8 & 3.4 & 20.0 & 25.4 & This work \\
mid-IR variations & N & Y & Y & N & Y & Y & This work \\
\enddata 
\tablenotetext{a}{Renormalised Unit Weight Error (RUWE) is a quality indicator for Gaia astrometric data 
(for more details see the technical note GAIA-C3-TN-LU-LL-124-01, \url{https://www.cosmos.esa.int/web/gaia/public-dpac-documents}). RUWE values of $<$1.4  
indicate reliable astrometric solutions.}
\tablenotetext{b}{In computation of the $U, V, W$ velocity components, we used the radial velocity data obtained by us ($v_{\rm r,our}$).}
\tablenotetext{c}{ Geometric distances from \citet{cbj2020}.}
\tablenotetext{d}{Lithium equivalent widths after correction for blending with the iron line (Sect.~\ref{sec:ages}).}
\end{deluxetable*}

%%%%%%%%%%%%%%%%%%%% TABLE 1 %%%%%%%%%%%%%%%%%%%%%%%%%%%%%%%%%%

\section{Identification of new EDDs} \label{sec:sampleselection}
Aimed at revealing new EDDs hosted by Sun-like (F5-K7-type) stars, we took mid-infrared photometric 
data from the AllWISE catalog and combined them with astrometric information from 
the TGAS database.
The AllWISE data release is 
based on observations from the {\sl WISE} cryogenic and NEOWISE post-cryogenic survey phases of the {\sl WISE} mission 
obtained in 2010--2011. 
The catalog contains positions and four band (W1--W4) mid-IR photometric information for $\sim$750\,million 
sources.  To reduce the chance of source confusion and contamination from strong 
diffuse background emission, we discarded all sources located within 5{\degr} of the 
galactic plane. Correspondingly to our targeted spectral type range, we selected objects 
with $J-K_s$ color index between 0.22 and 0.8. The conversion between spectral types and 
colors was adopted from \citet{pecaut2013}. To discard suspicious {\sl WISE} photometric data 
from the compiled sample, we applied a number of additional selection criteria. 
We only used sources where (1) the W1, W2, and W3 band flux measurements have
signal-to-noise ratios (S/Ns)$>$10; (2) the W4 band measurement has an 
S/N of $>$3; and (3) the reduced $\chi^2$ (\texttt{w1rchi2-w4rchi2}) of the profile-fit 
photometry measurement is $<$3 in each band.
%We only used sources where the W1, W2, and W3 band flux measurements have signal-to-noise 
%ratios (S/N) $>$10, i.e. having photometric quality (\texttt{ph\_qual}) of 'A', and the  
%W4 band measurement has an S/N of $>$3 (\texttt{ph\_qual} = 'A' or 'B'); 
%and where the reduced $\chi^2$ of the profile-fit photometry measurement 
%(\texttt{w1rchi2-w4rchi2}) is $<$3 in each band. 
Moreover, we removed all objects with contamination 
and confusion flags (\texttt{cc\_flags}) of either 'D', 'P', 'H' or 'O' in any of the four 
bands, i.e., where the photometry is probably affected by diffraction, persistence, 
halo, or ghost artifacts. In the finally selected sample, we corrected the potential overestimate of the real flux 
in saturated W2 photometry at 4.6\,{\micron} by applying the formula proposed by 
\citet[][eq.~5]{cotten2016}.

The TGAS catalog provides a 5D astrometric solution for over 2\,million sources. From this data set 
we retained only those objects that are located within 300\,pc of the Sun, and have a good quality parallax 
measurement with $\pi/\sigma_\pi > 5$. We also eliminated giant stars from the list by removing 
TGAS sources  with $J$ band absolute magnitude $M_{J} < 5\times(J-K_s)$. We took near-IR photometry 
from the Two Micron All-Sky Survey \citep[2MASS,][]{skrutskie2006}.
The applied criteria were developed based on \citet{rhee2007}, who 
used $B$ and $V$ band data, but -- due to possible reddening -- we based our formula  
on near-infrared photometry, which is less sensitive to extinction.

Accounting for the epoch difference between 
the TGAS (J2015.0) and {AllWISE} (J2010.0) catalogs, we corrected
the positions of the selected TGAS objects for proper motion.
After applying all of the above selection criteria, we cross-matched the two samples
adopting a match radius of 0\farcs3. This resulted in a combined data set including 78650 objects.

To identify new EDDs, we looked for objects exhibiting significant excess 
emission in the W2 and W3 bands. The W2 band excess was requested because we aim at
examining the mid-IR variability of the revealed disks. To assess possible excesses 
in both bands, we compared the measured $K_{\rm s}-W2$ and $K_{\rm s}-W3$ color indices 
with the predicted ones. The predicted $K_{\rm s}-W2$ and $K_{\rm s}-W3$ ($P_{\rm K_s-W_i}$, where 
$i$ is 2 or 3) values were inferred from the $J-K_{\rm s}$ color of the stars using color--color 
sequences quoted in table~5 of \citet{pecaut2013}. The uncertainty of this prediction 
($\sigma_{\rm P_{\rm K_s-W_i}}$) is estimated 
to be 2\%. The significance of the excesses was then computed as 
\begin{equation}
 \chi_{\rm E(K_s-W_i)} = \frac{E(K_s-W_i)}{\sigma_{\rm E(K_s-W_i)}} = 
 \frac{K_s - W_i - P_{\rm K_s-W_i}}{\sigma_{\rm E(K_s-W_i)}},
\end{equation}
where $\sigma_{\rm E(K_s-W_i)}$ is the quadratic sum of the measured uncertainties in the 
$K_{\rm s}$ and $W_{\rm i}$  bands, and 
$\sigma_{\rm P_{\rm K_s-W_i}}$, i.e.: $\sigma_{\rm E(K_s-W_i)} = 
\sqrt{\sigma_{\rm K_s}^2 + \sigma_{\rm W_i}^2 + \sigma_{\rm P_{\rm K_s-W_i}}^2}$.
We selected those objects as EDD candidates where both $\chi_{\rm E(K_s-W_2)}$ 
and $\chi_{\rm E(K_s-W_3)}$ were higher than 5. This left us with 27 
objects. 

Looking at the candidates, we found that two of them were previously 
known EDDs (HD\,23514 and BD+20~307). For a further ten stars, the spectral type 
found in the literature does not correspond to the interval we selected. 
Four among them have A0$-$F2 spectral type, while the other six objects 
are nearby M-type stars. By comparing with the proper atmosphere models, 
the latter six no longer show significant excess.
For ten objects the spectral energy distribution based on  W1--W4 data is parallel 
to the stellar photosphere defined by the 2MASS measurements, but at a higher 
absolute level. Based on the literature all of these objects are eclipsing binaries.
According to the AllWISE catalog, six of them were 
probably variable in the W2 band as indicated by 
the value of their variability flag (\texttt{var\_flag}) that has been set 
to the highest possible value of 9. Four of these objects probably displayed 
substantial changes in the W3 band, too. 
For these ten binaries, our predicted photospheric colors ($P_{\rm K_s-W_2}$, 
$P_{\rm K_s-W_3}$)  are probably inaccurate, 
and our excess detection is false. We note that two of these objects were 
already removed from our list because they have a spectral type that is too early.     
The {\sl WISE} photometry of CD$-$48~8486, a G-type member of the Upper Centaurus Lupus 
group \citep{mamajek2002}, also indicates an SED shifted upward with respect to the 
2MASS near-infrared data, suggesting that the excess identification is dubious. 
Earlier mid-infrared photometric and spectroscopic observations of this source with the 
{\sl Spitzer}
confirm this suspicion: the object shows no significant
excess at wavelengths $<$25\,{\micron} \citep{silverstone2006,carpenter2009}.

The literature lists 11 EDDs with F--K-type host stars (Appendix~\ref{appendix:c}). 
Out of this sample we recovered two objects, HD\,23514 and BD+20~307. The other nine objects 
were excluded by our searching algorithm because of various reasons. 
HD\,15407 and HD\,166191 lie too close to the Galactic plane ($|b|<5{\degr}$), 
while RZ\,Psc, V488\,Per, HD\,113766, ID\,8 and P\,1121 are not included in the TGAS catalog.
HD\,145263 has an earlier spectral type than the other targets, its $J-K_s$ color index 
is outside our specified range, while the disk of TYC\,8241\,2652\,1 had been substantially depleted 
before the start of WISE mission (Appendix~\ref{appendix:c}) and no longer exhibited strong excess 
in the W2 and W3 bands.

The detailed analysis, described above, left us with 6 EDD candidates out of the 27 initially 
selected sources. They are: TYC~4515-485-1 (hereafter TYC~4515), 
TYC~5940-1510-1 (TYC~5940), TYC~8105-370-1 (TYC~8105), TYC~4946-1106-1 (TYC~4946), 
TYC~4209-1322-1 (TYC~4209), and TYC~4479-3-1 (TYC~4479). 
However, due to the large {\sl WISE} beam source, confusion or contamination by 
extended emission cannot be ruled out even in these cases. In order to test this, we performed 
several additional checks, but based on the currently available infrared data 
we found no sign of confusion or significant contamination at any of the selected systems 
(Appendix~\ref{appendix:a}). 
Therefore, throughout the following analysis, we will assume that the excesses in 
all six cases come from warm circumstellar dust disks. Previously, one of these 
objects, TYC\,4479, was identified as a debris disk system \citep{cotten2016}, 
while the other five disks are new discoveries.

\begin{figure*}
\begin{center}
\includegraphics[angle=0,scale=.22,bb=4 70 478 480]{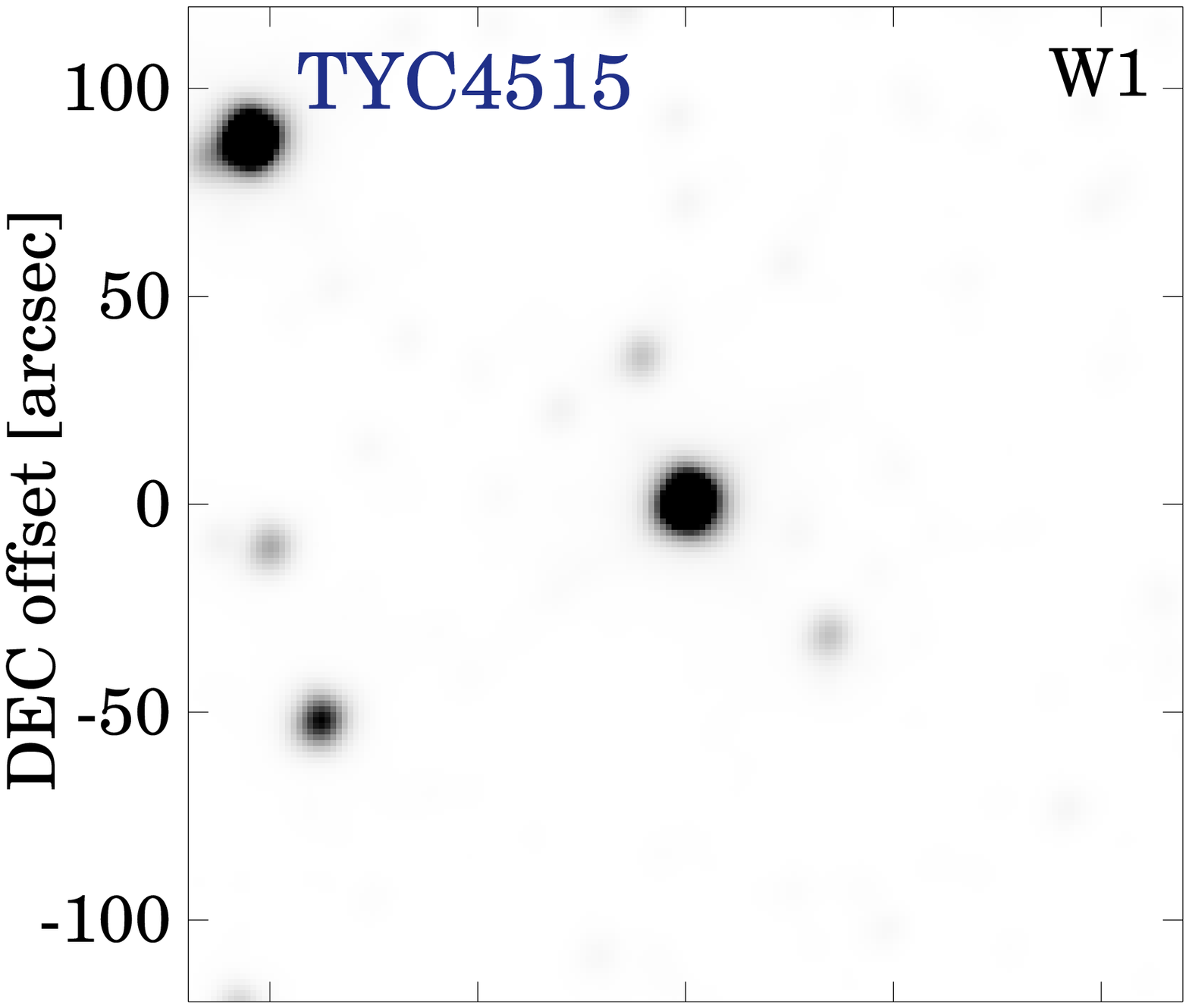}
\includegraphics[angle=0,scale=.22,bb=73 70 478 480]{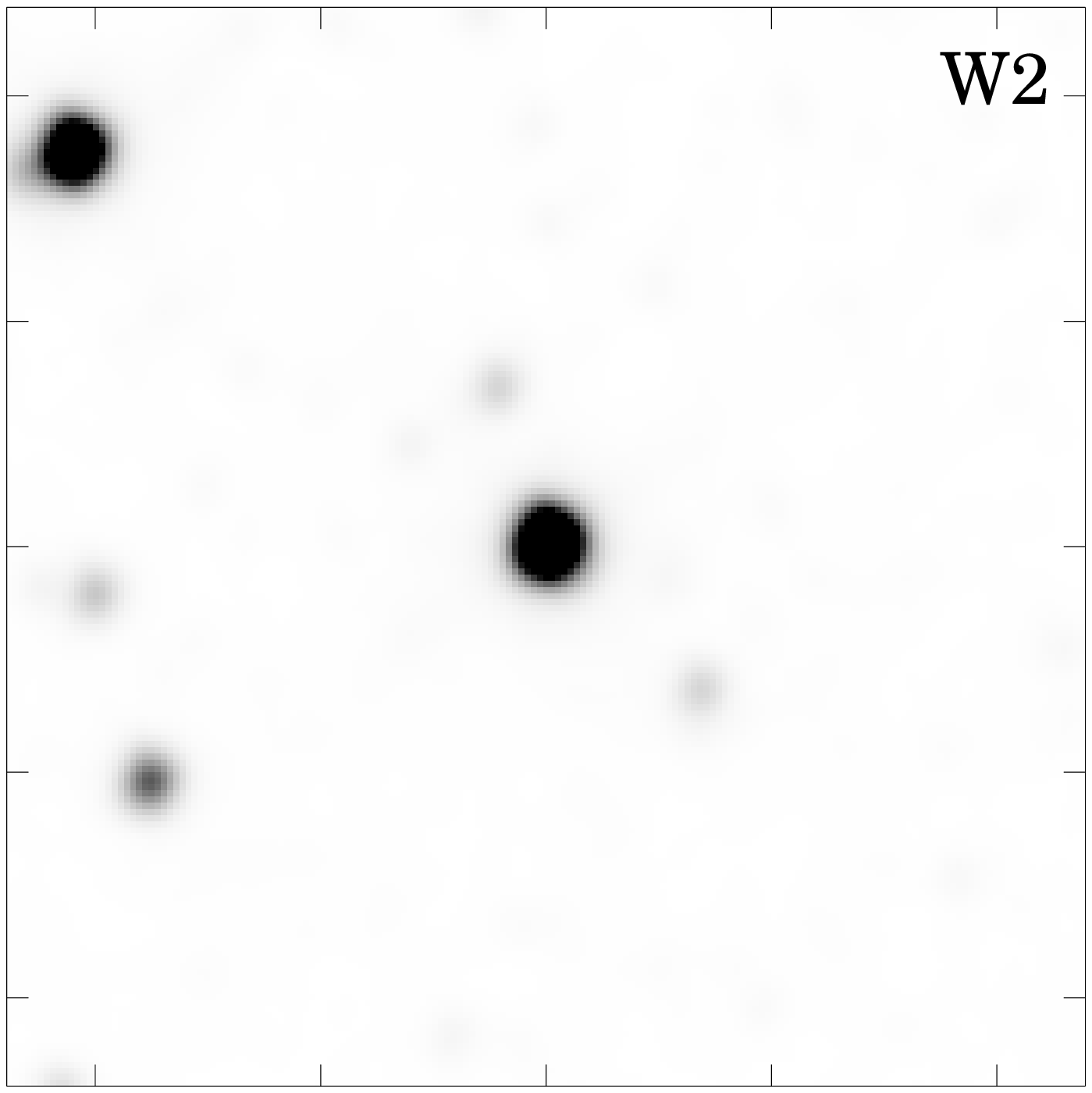}
\includegraphics[angle=0,scale=.22,bb=73 70 478 480]{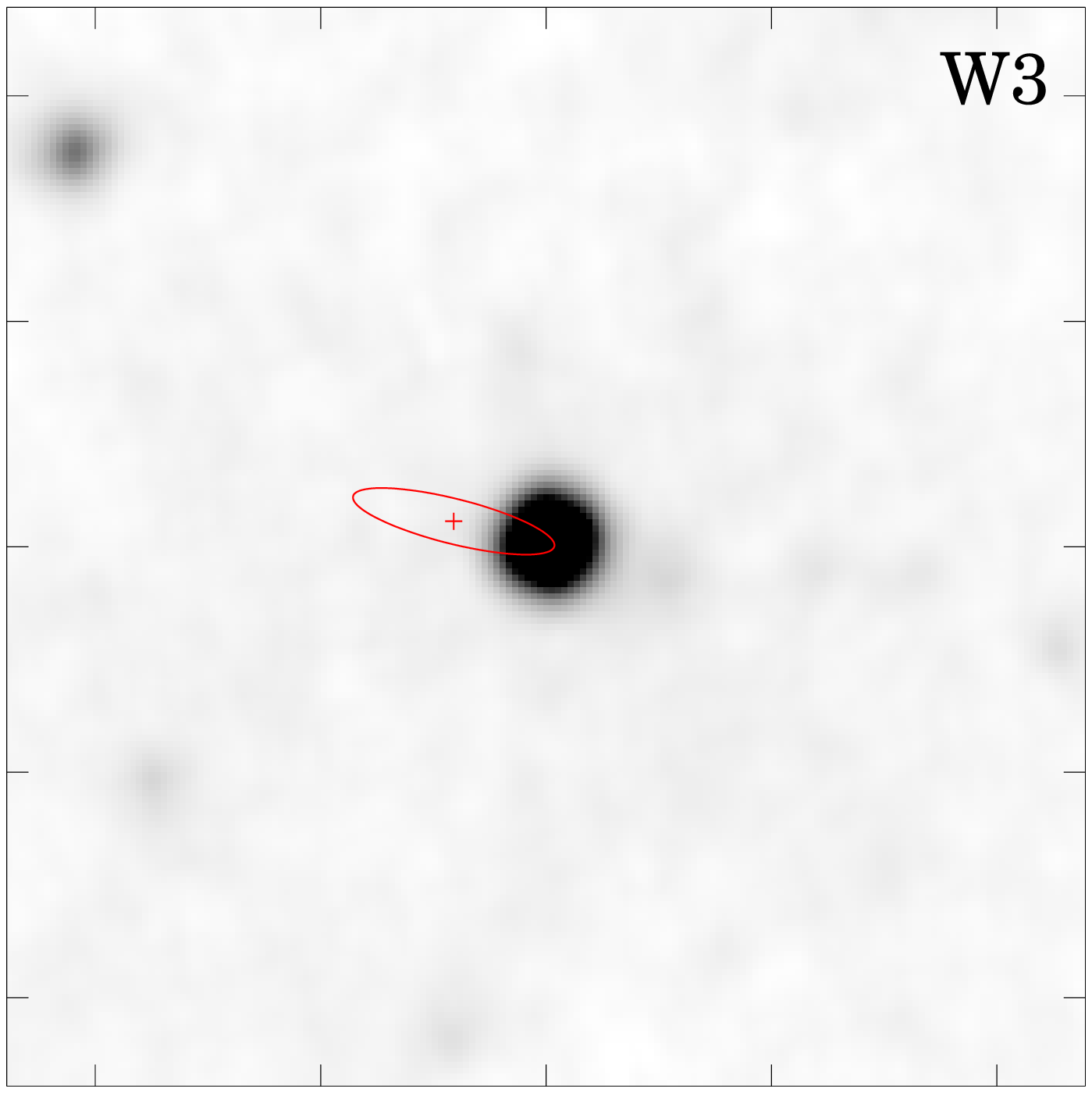}
\includegraphics[angle=0,scale=.22,bb=73 70 478 480]{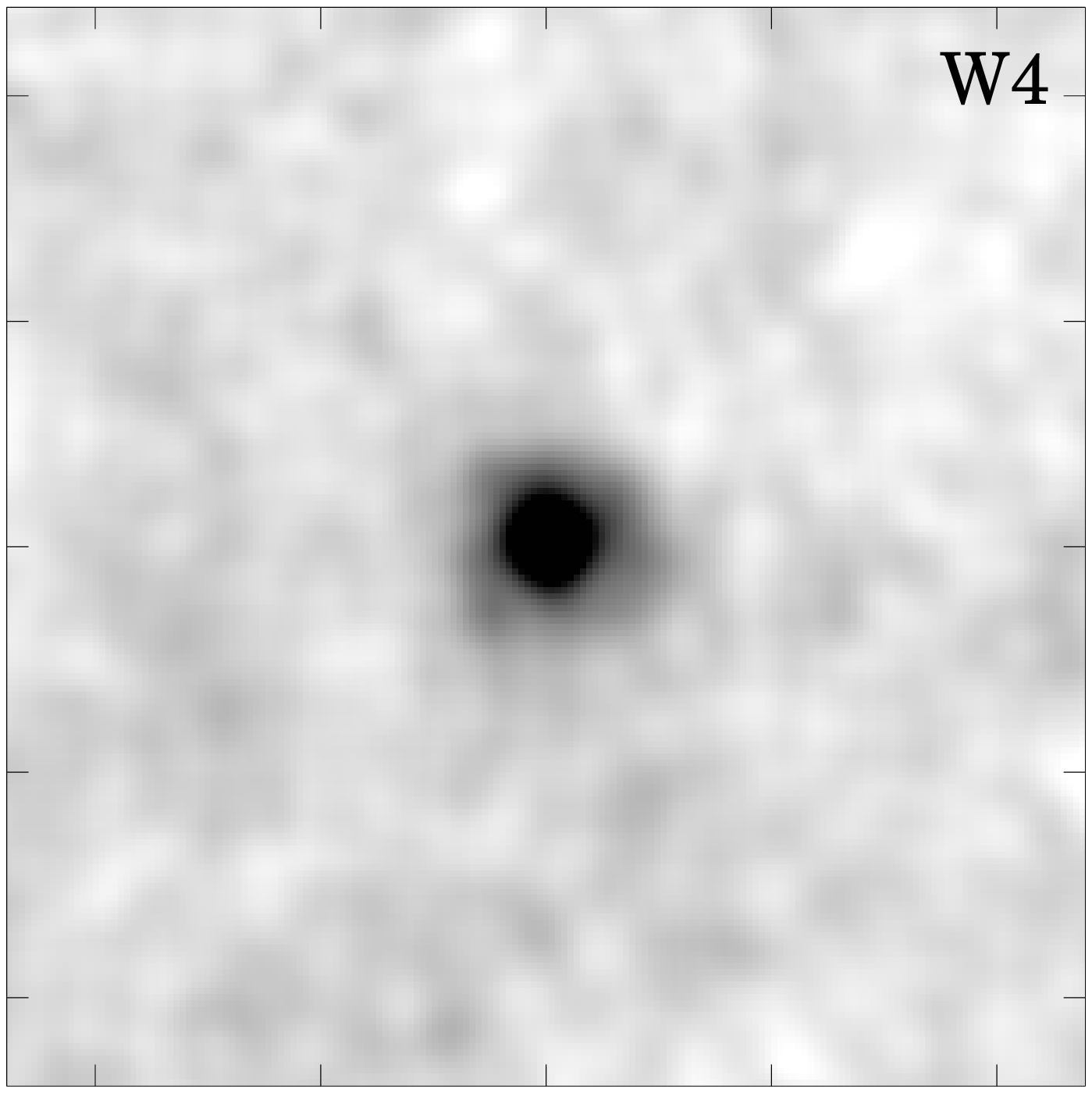}

\includegraphics[angle=0,scale=.22,bb=4 70 478 480]{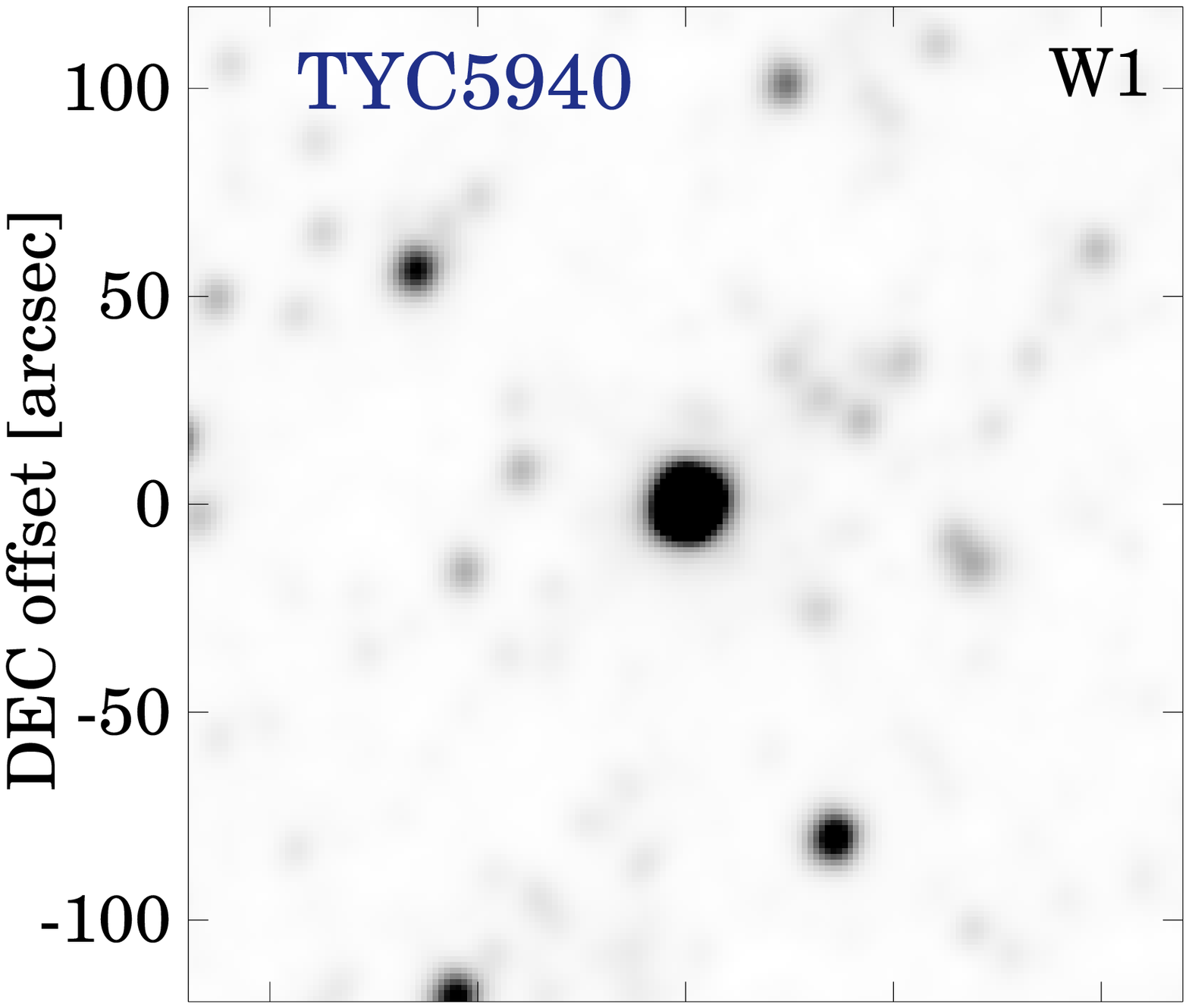}
\includegraphics[angle=0,scale=.22,bb=73 70 478 480]{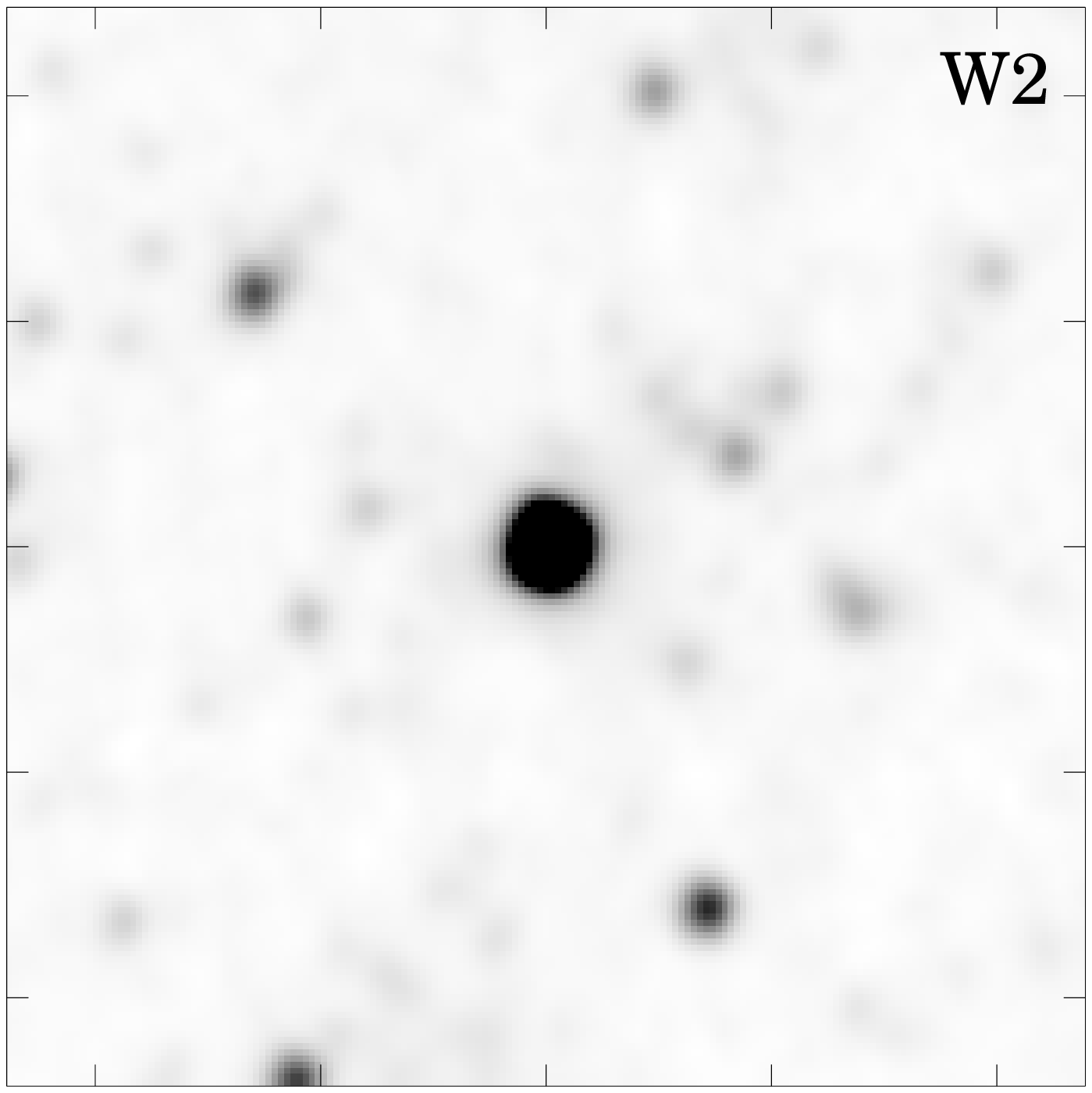}
\includegraphics[angle=0,scale=.22,bb=73 70 478 480]{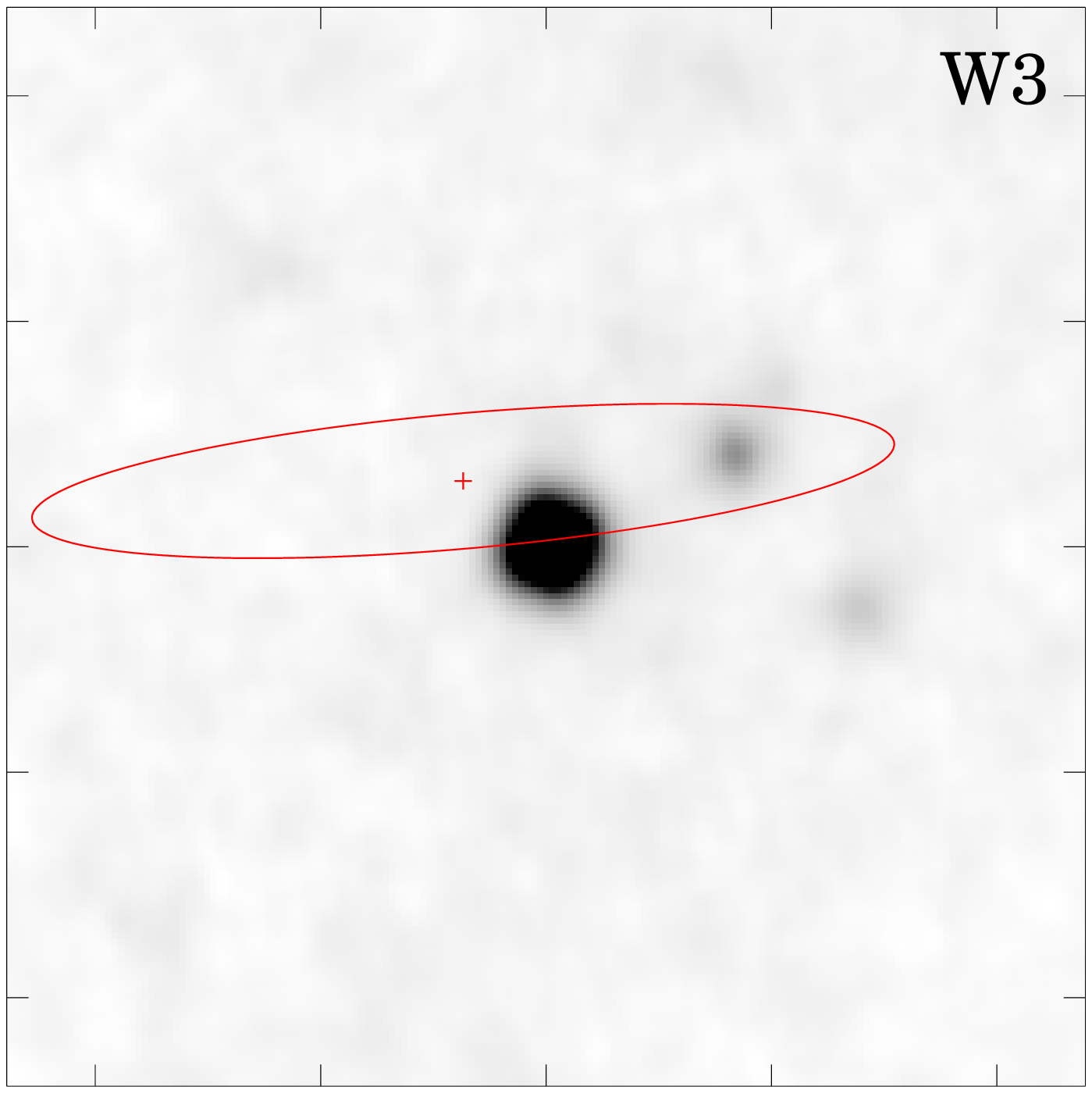}
\includegraphics[angle=0,scale=.22,bb=73 70 478 480]{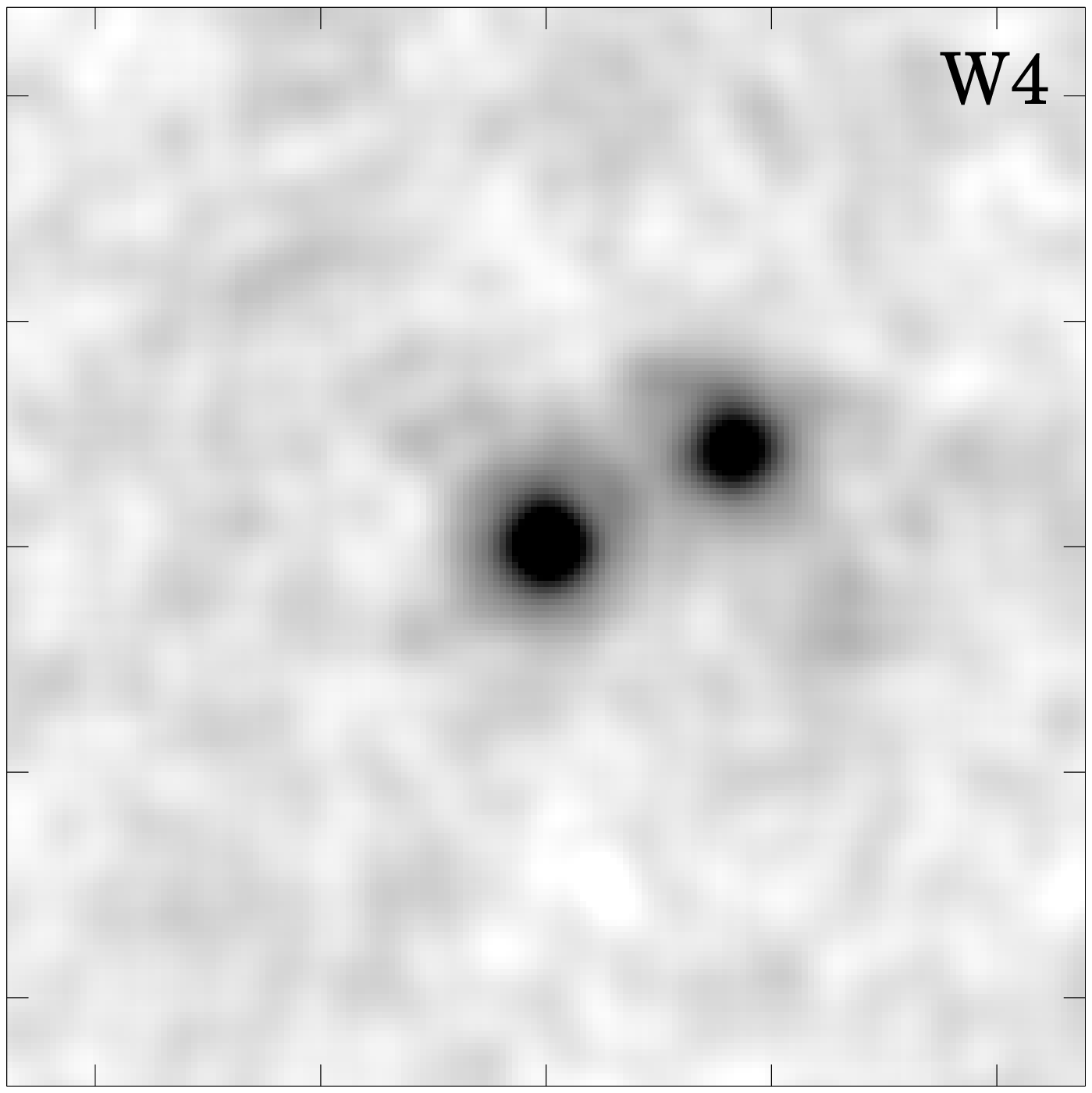}

\includegraphics[angle=0,scale=.22,bb=4 70 478 480]{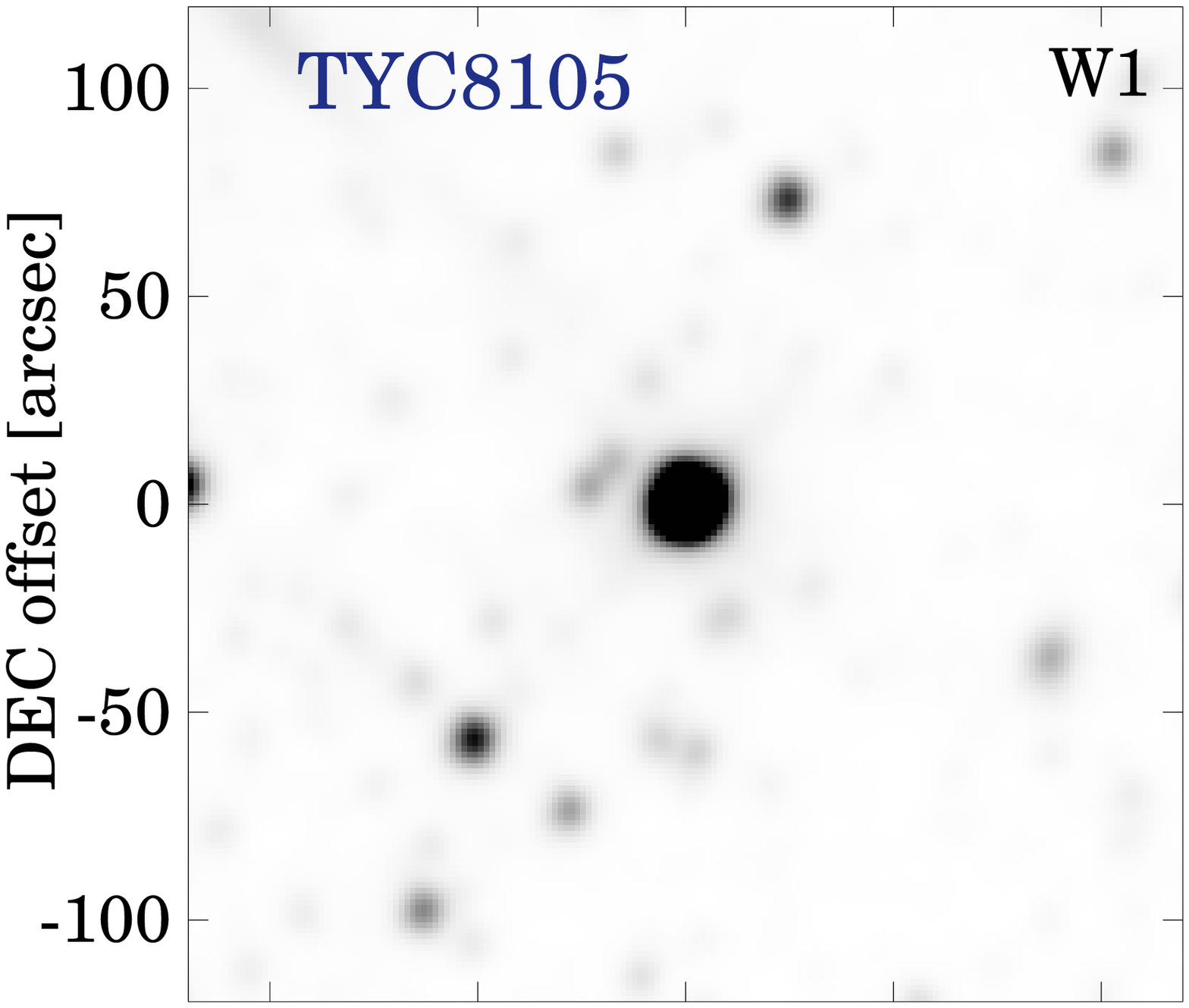}
\includegraphics[angle=0,scale=.22,bb=73 70 478 480]{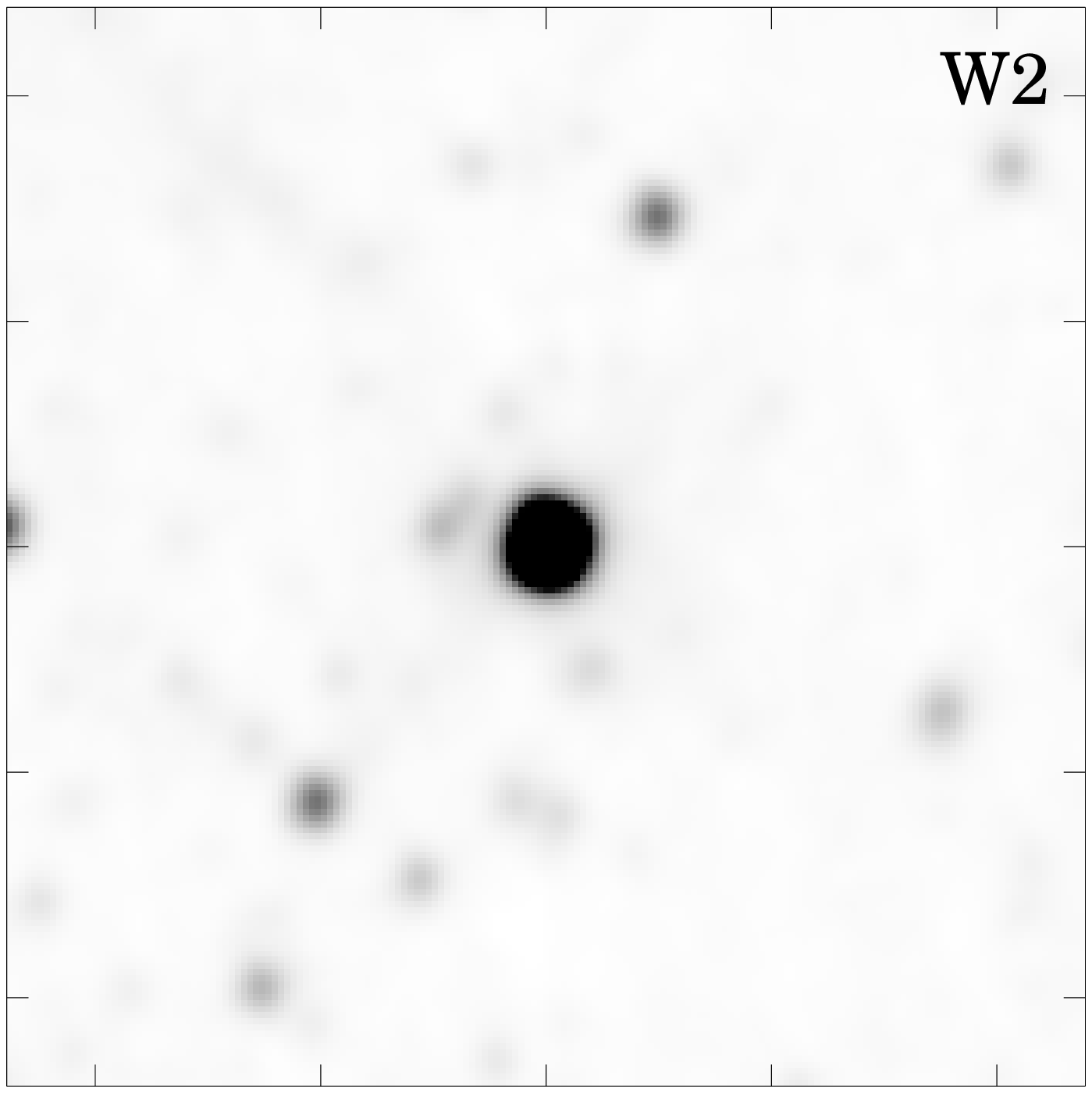}
\includegraphics[angle=0,scale=.22,bb=73 70 478 480]{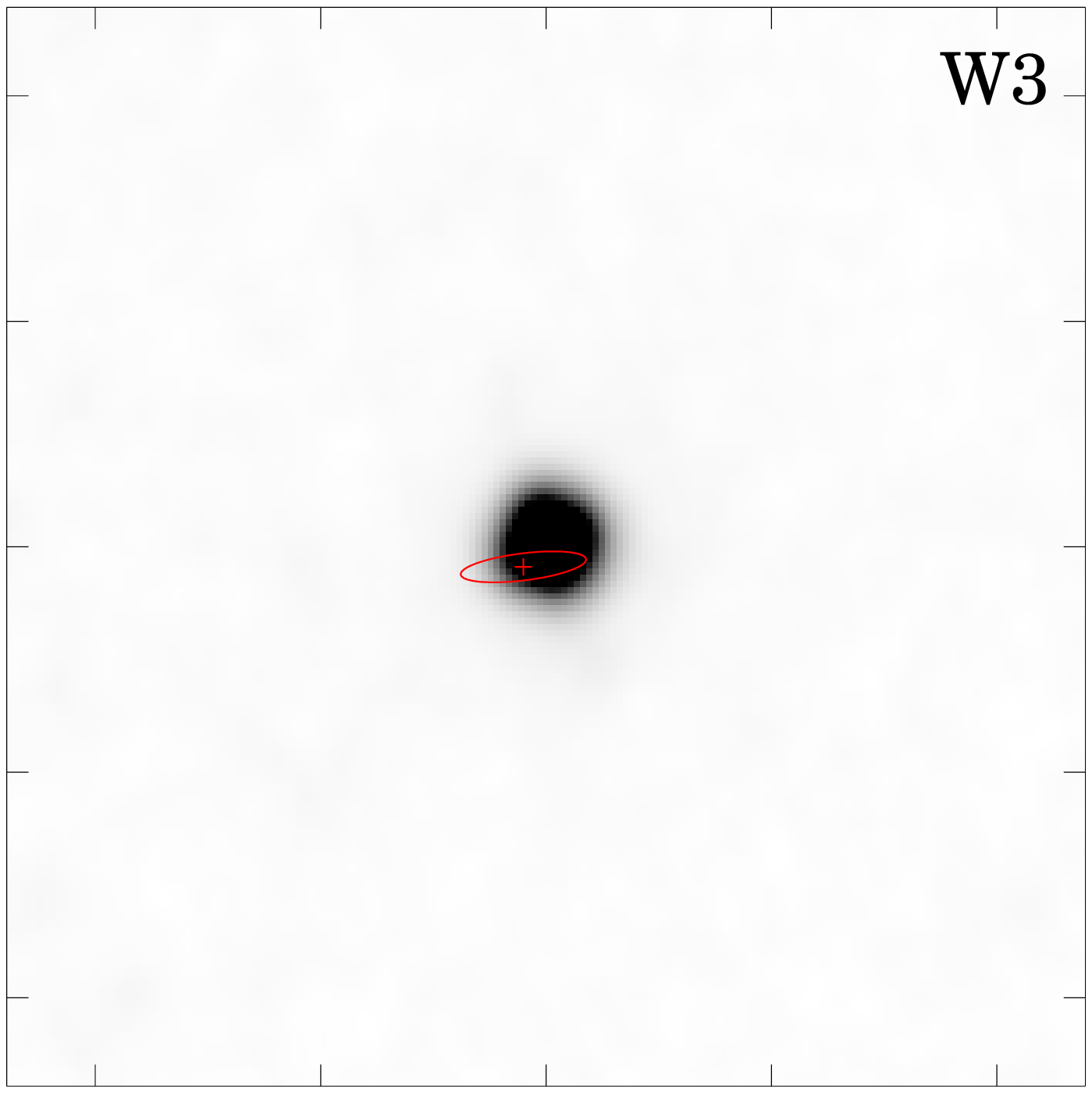}
\includegraphics[angle=0,scale=.22,bb=73 70 478 480]{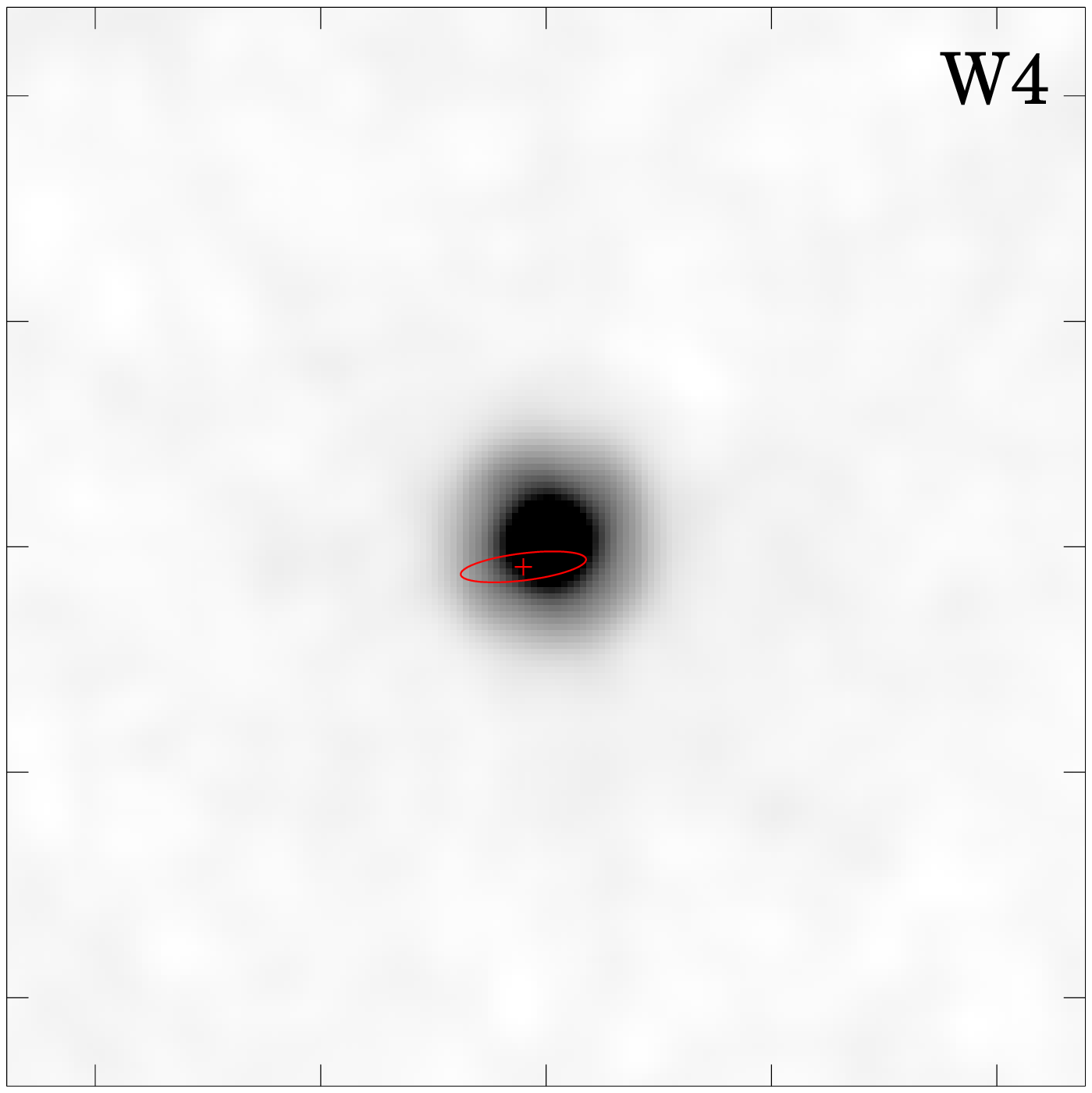}

\includegraphics[angle=0,scale=.22,bb=4 70 478 480]{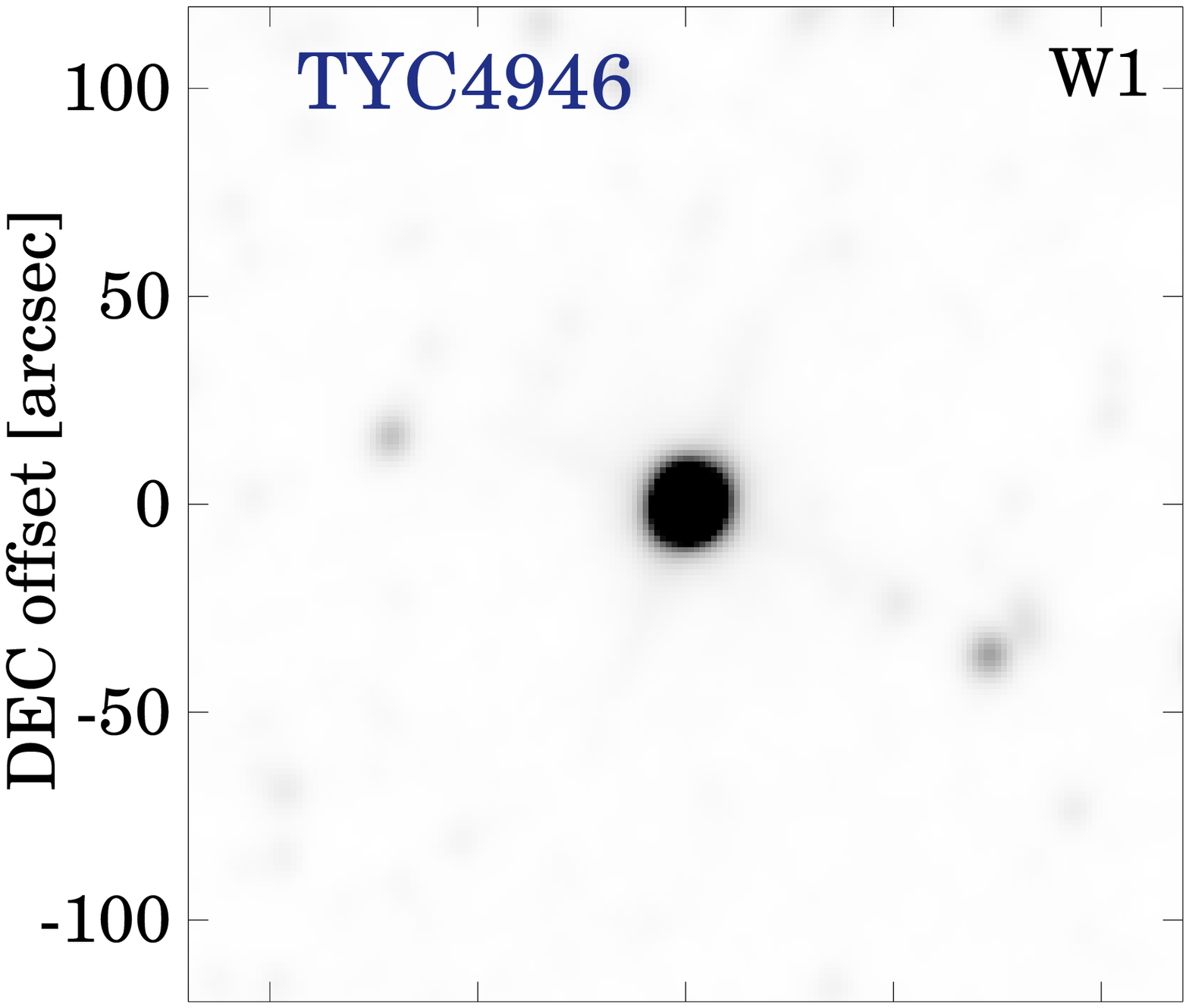}
\includegraphics[angle=0,scale=.22,bb=73 70 478 480]{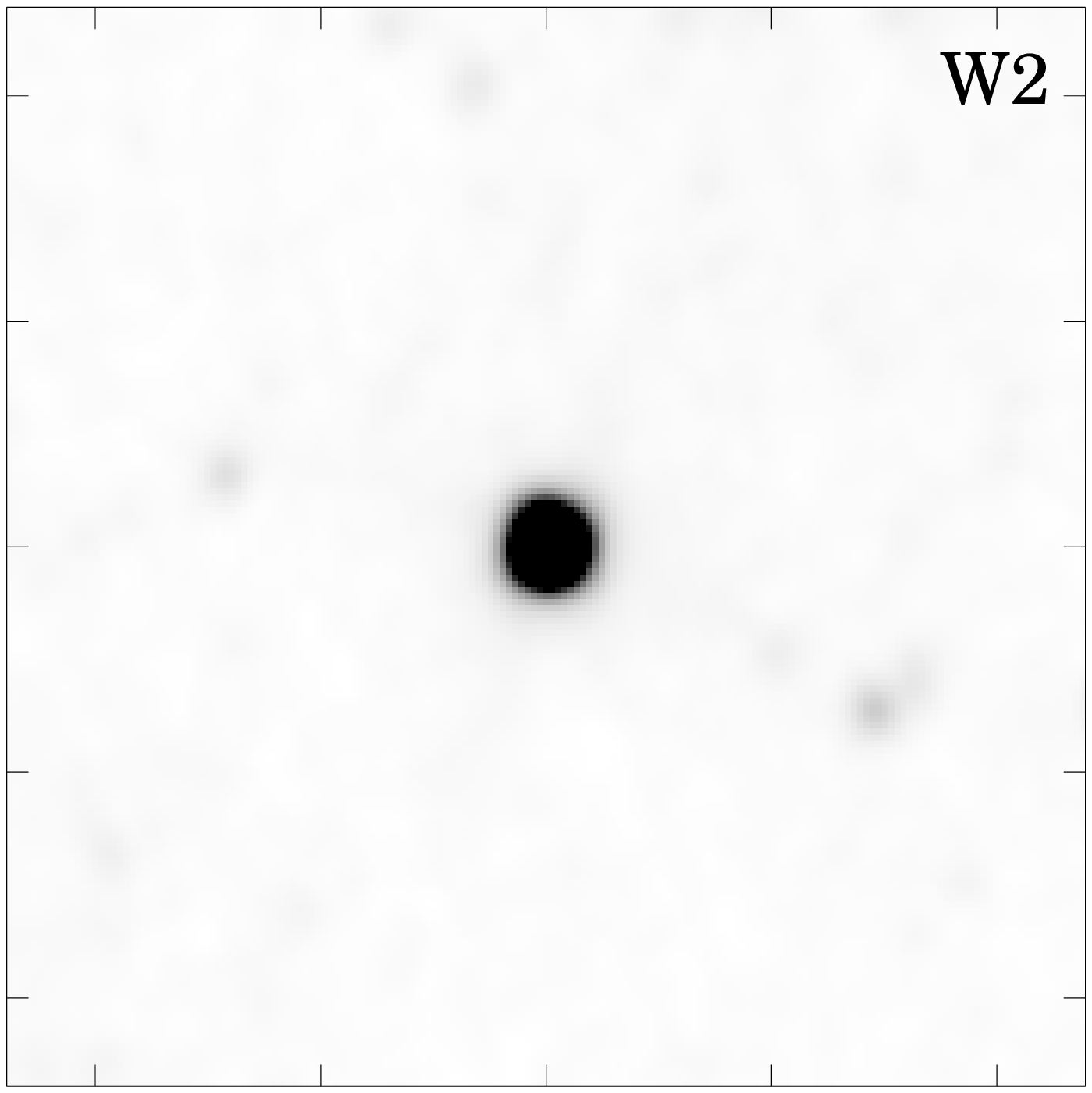}
\includegraphics[angle=0,scale=.22,bb=73 70 478 480]{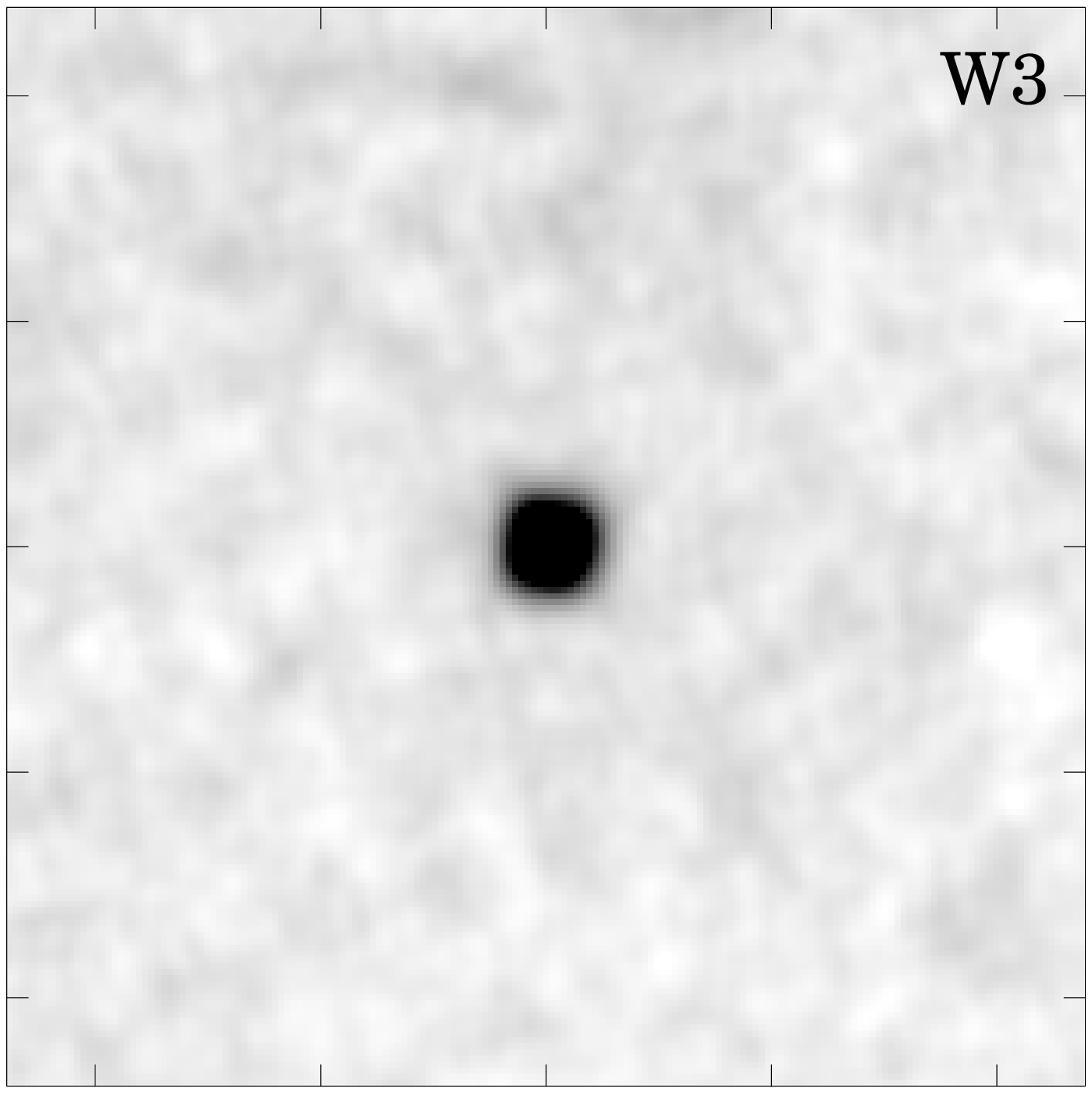}
\includegraphics[angle=0,scale=.22,bb=73 70 478 480]{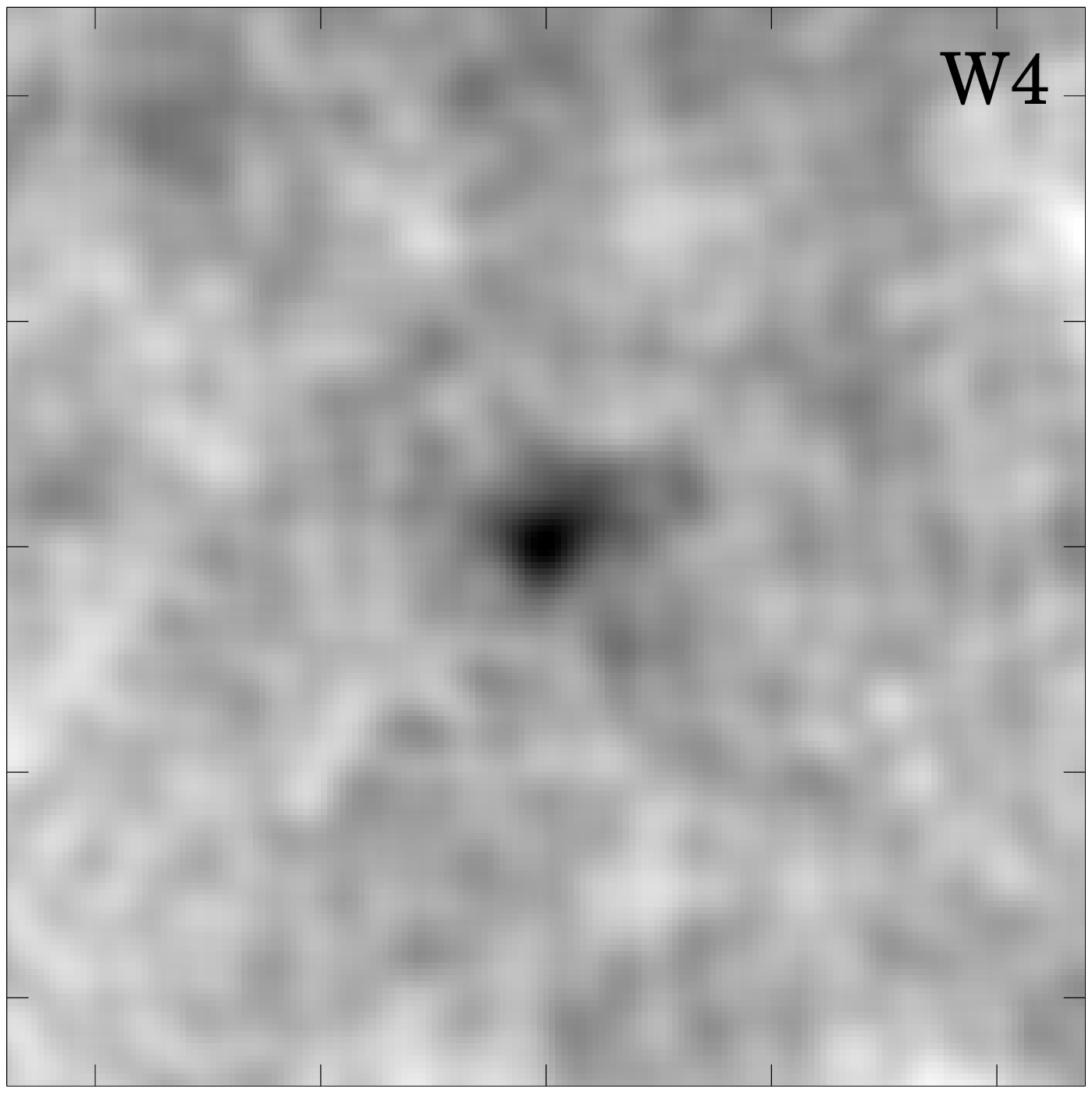}

\includegraphics[angle=0,scale=.22,bb=4 70 478 480]{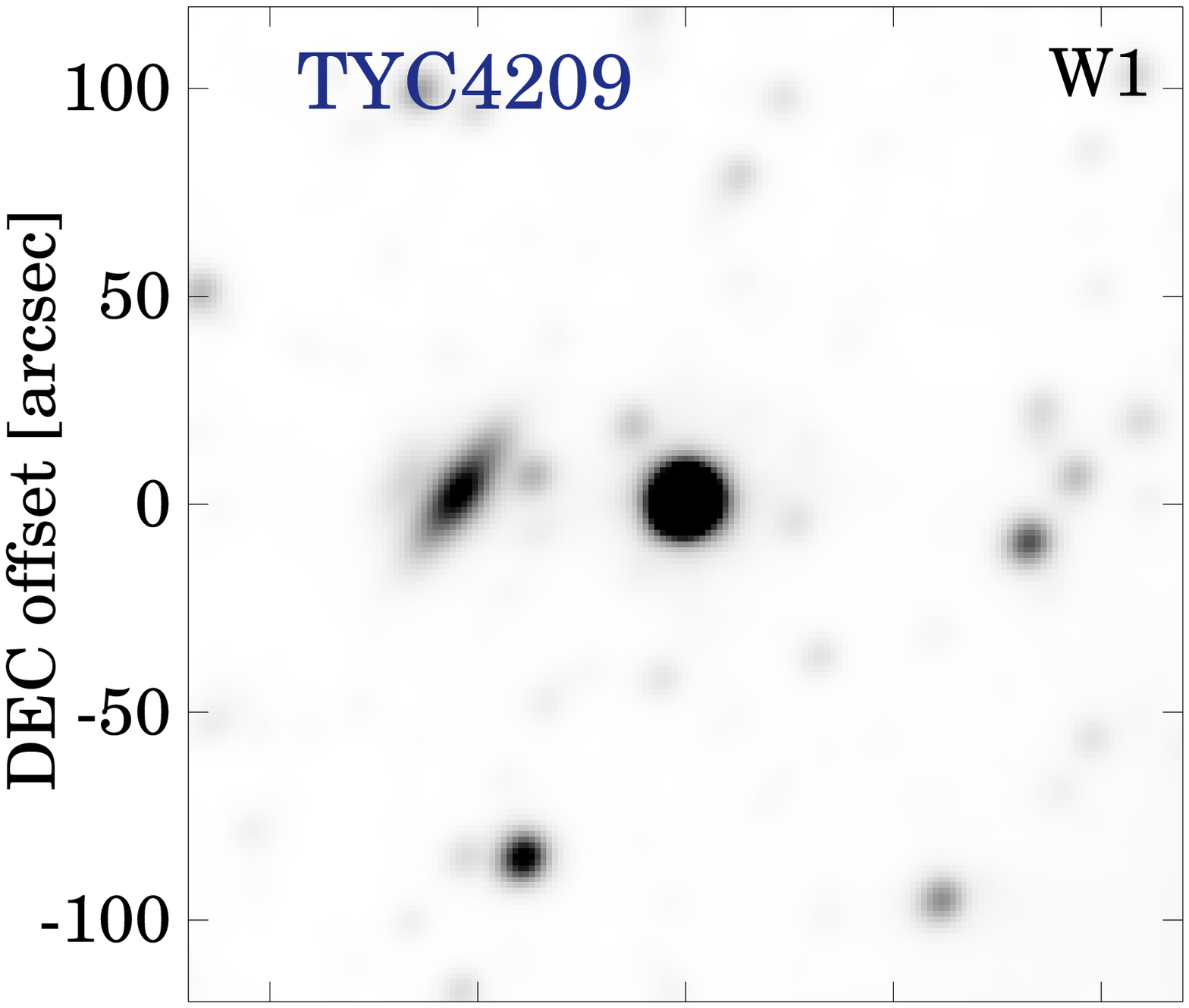}
\includegraphics[angle=0,scale=.22,bb=73 70 478 480]{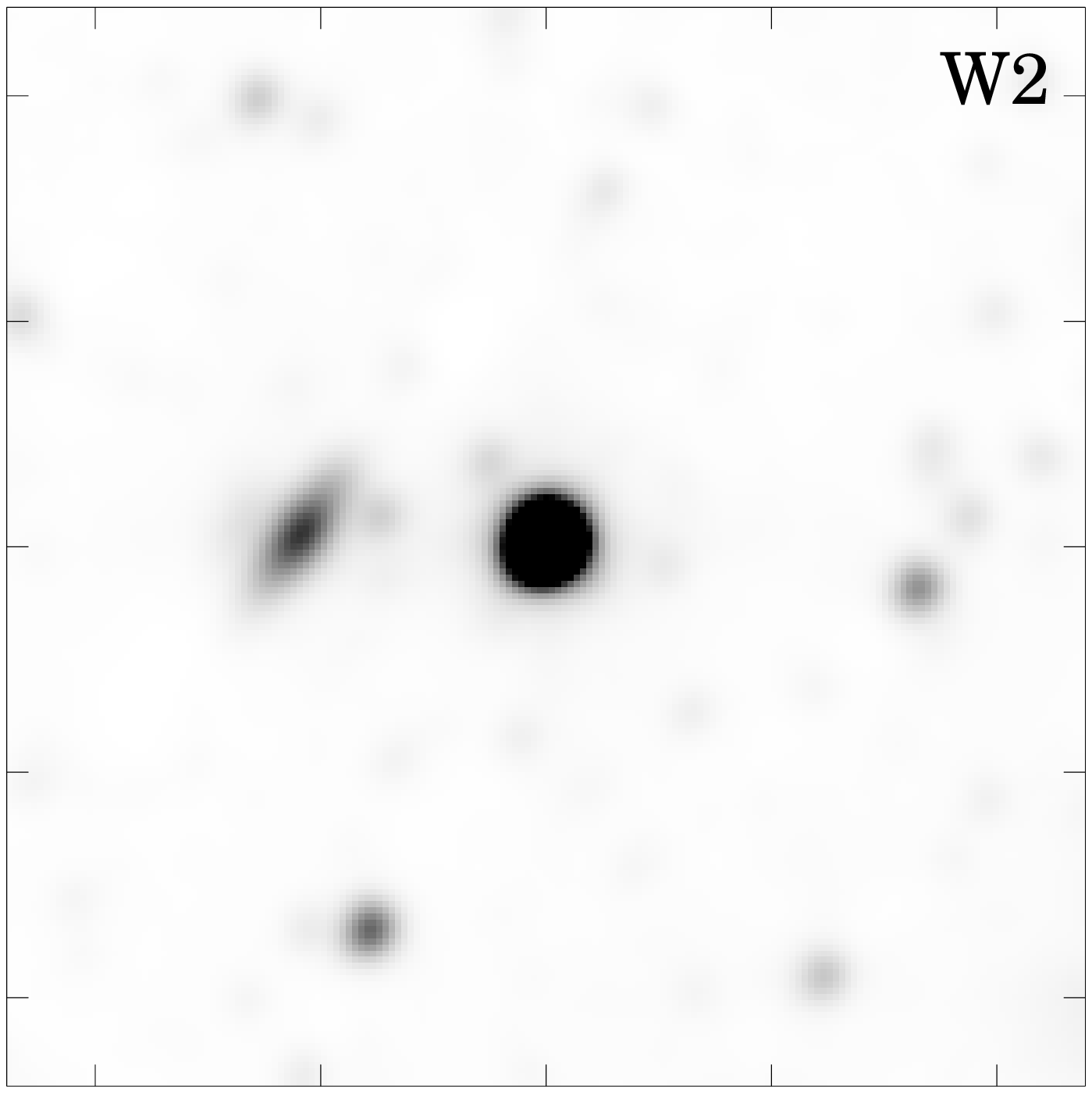}
\includegraphics[angle=0,scale=.22,bb=73 70 478 480]{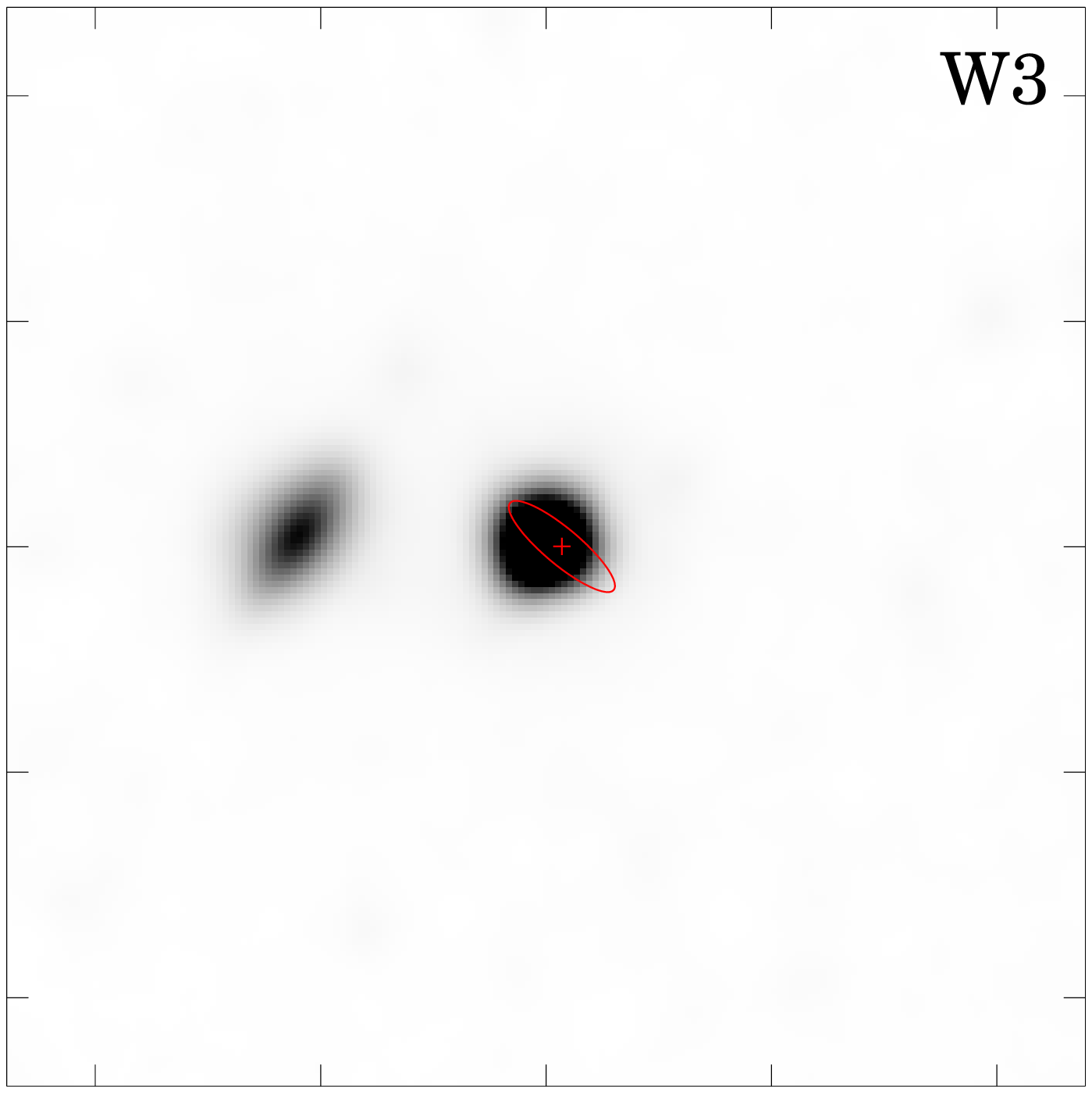}
\includegraphics[angle=0,scale=.22,bb=73 70 478 480]{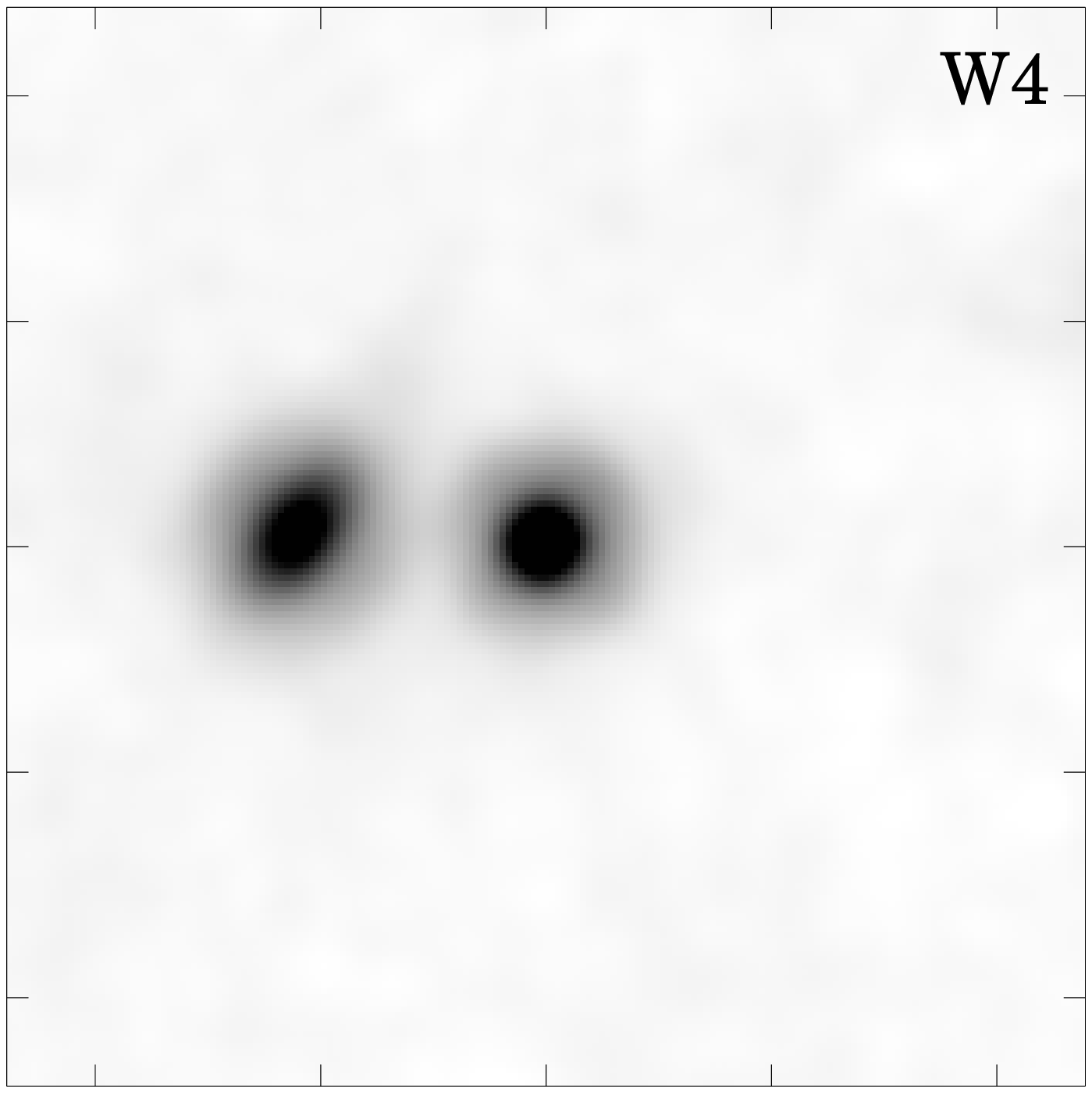}

\includegraphics[angle=0,scale=.22,bb=4 70 478 480]{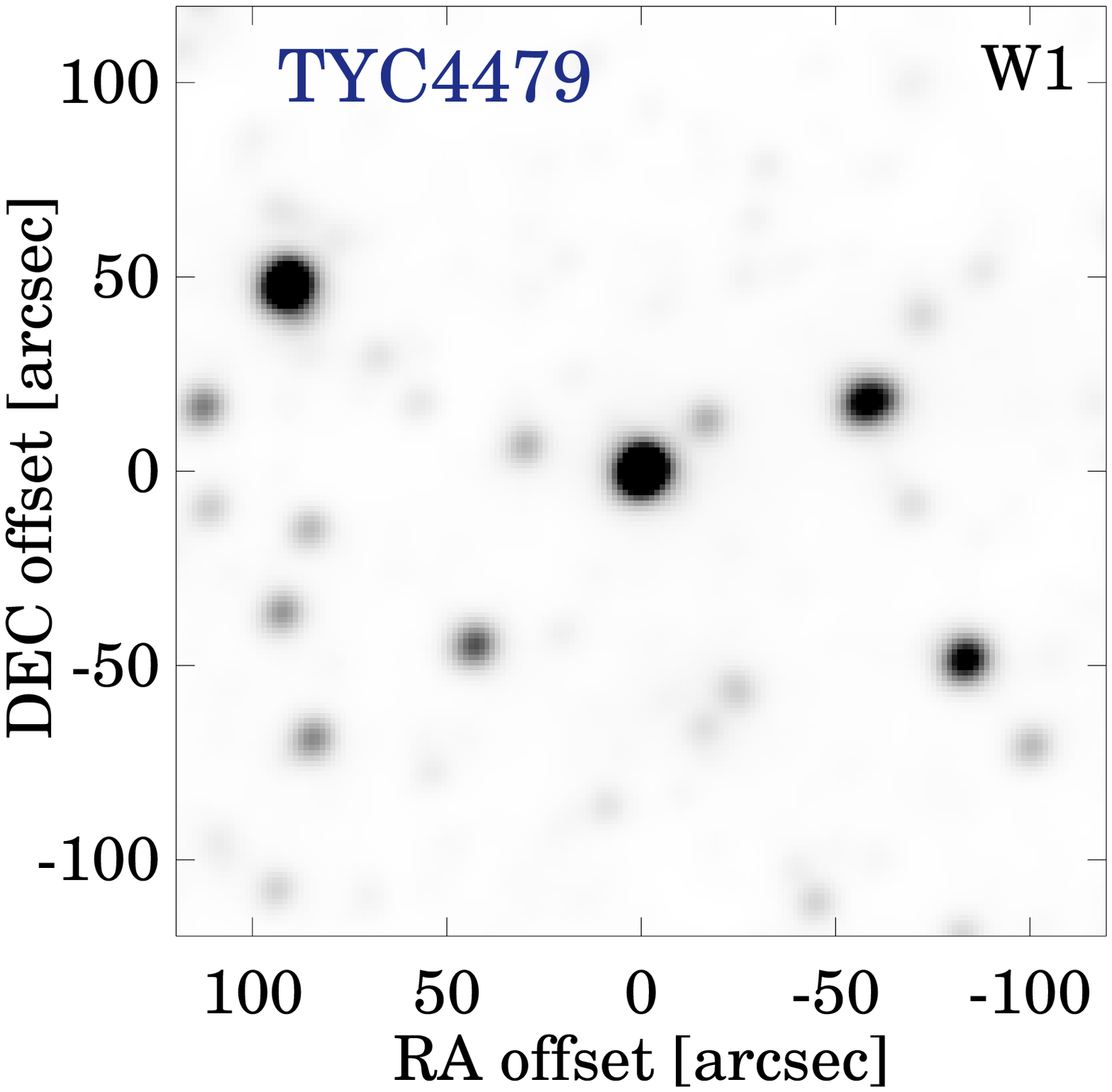}
\includegraphics[angle=0,scale=.22,bb=73 70 478 480]{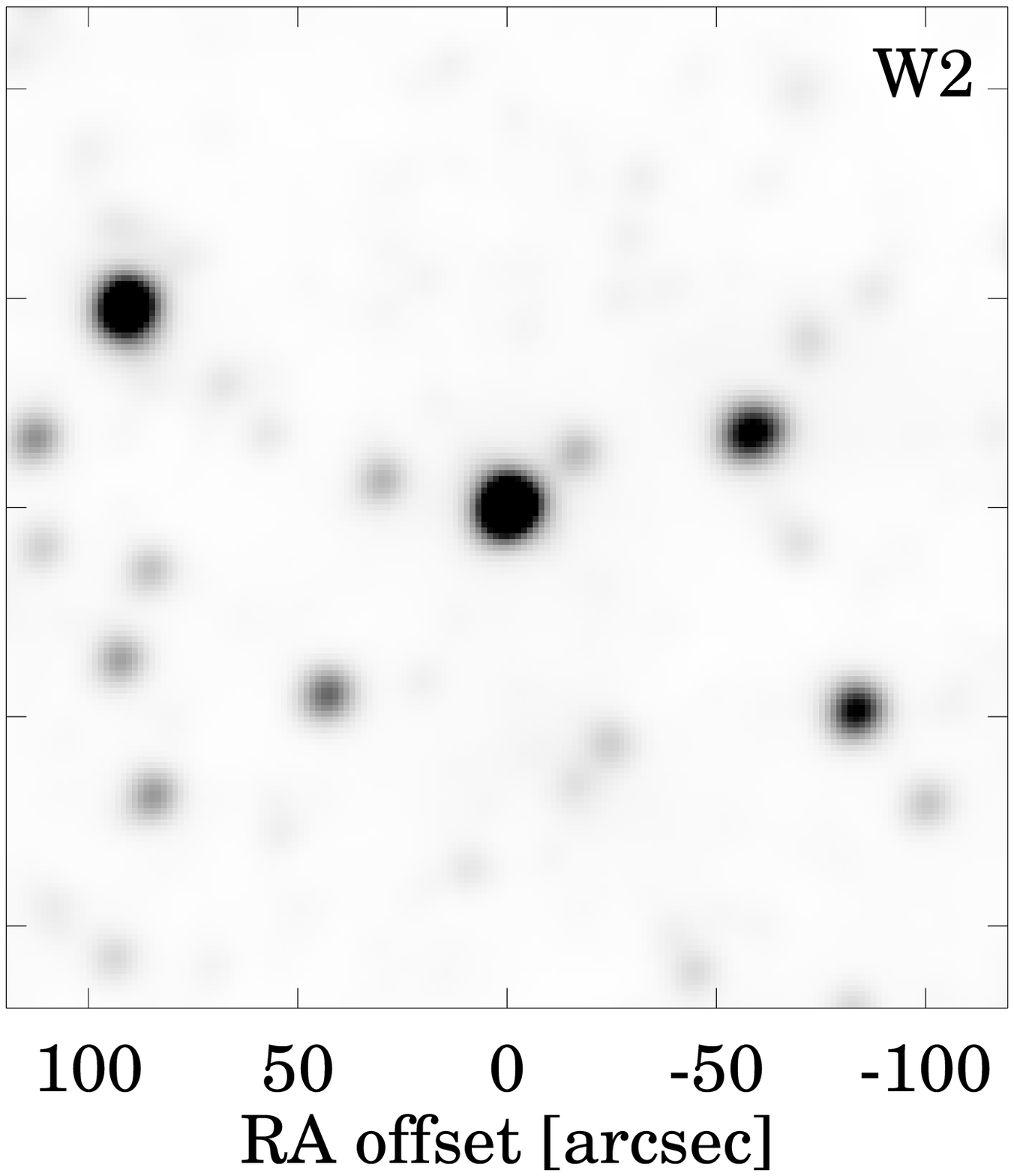}
\includegraphics[angle=0,scale=.22,bb=73 70 478 480]{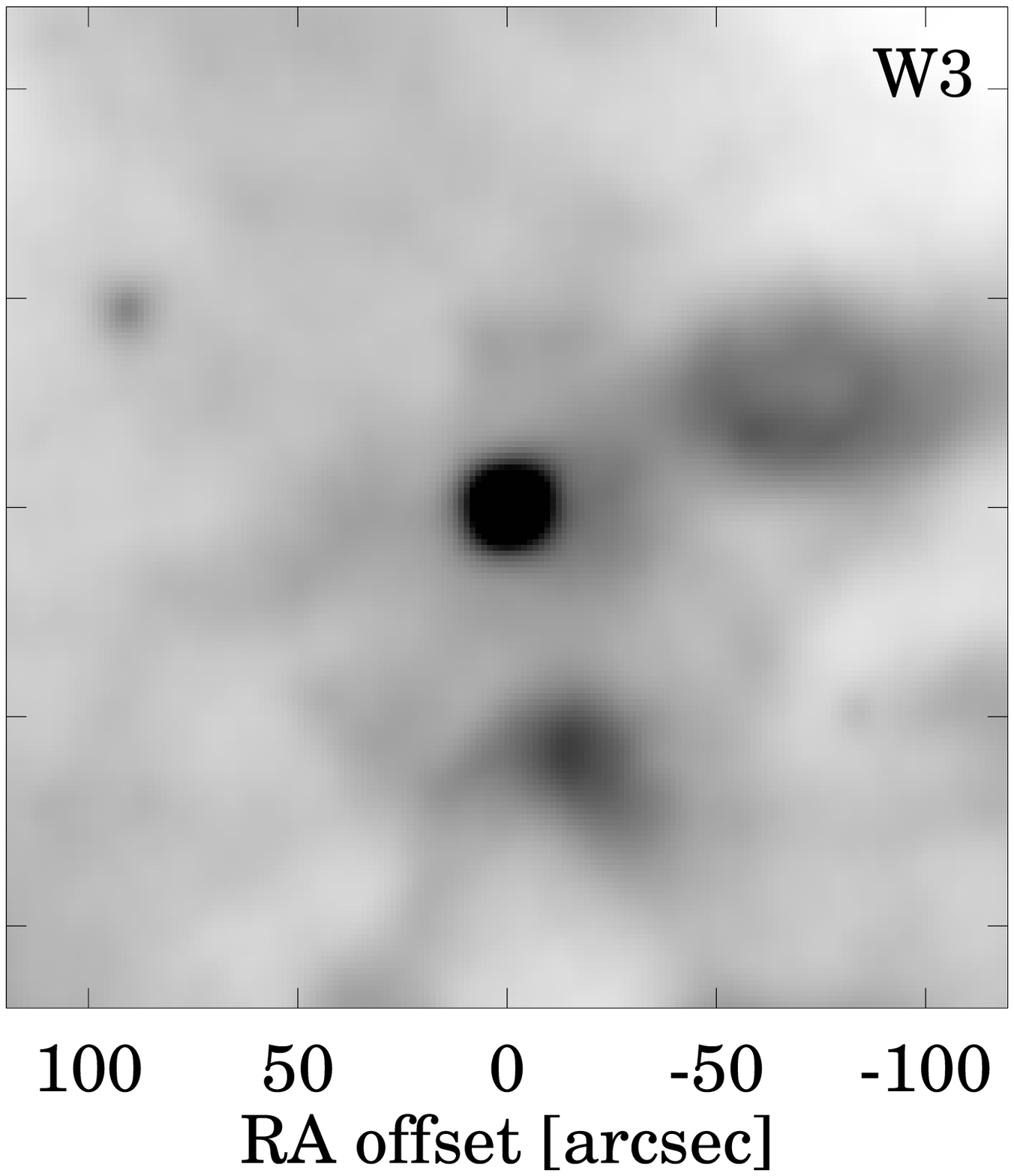}
\includegraphics[angle=0,scale=.22,bb=73 70 478 480]{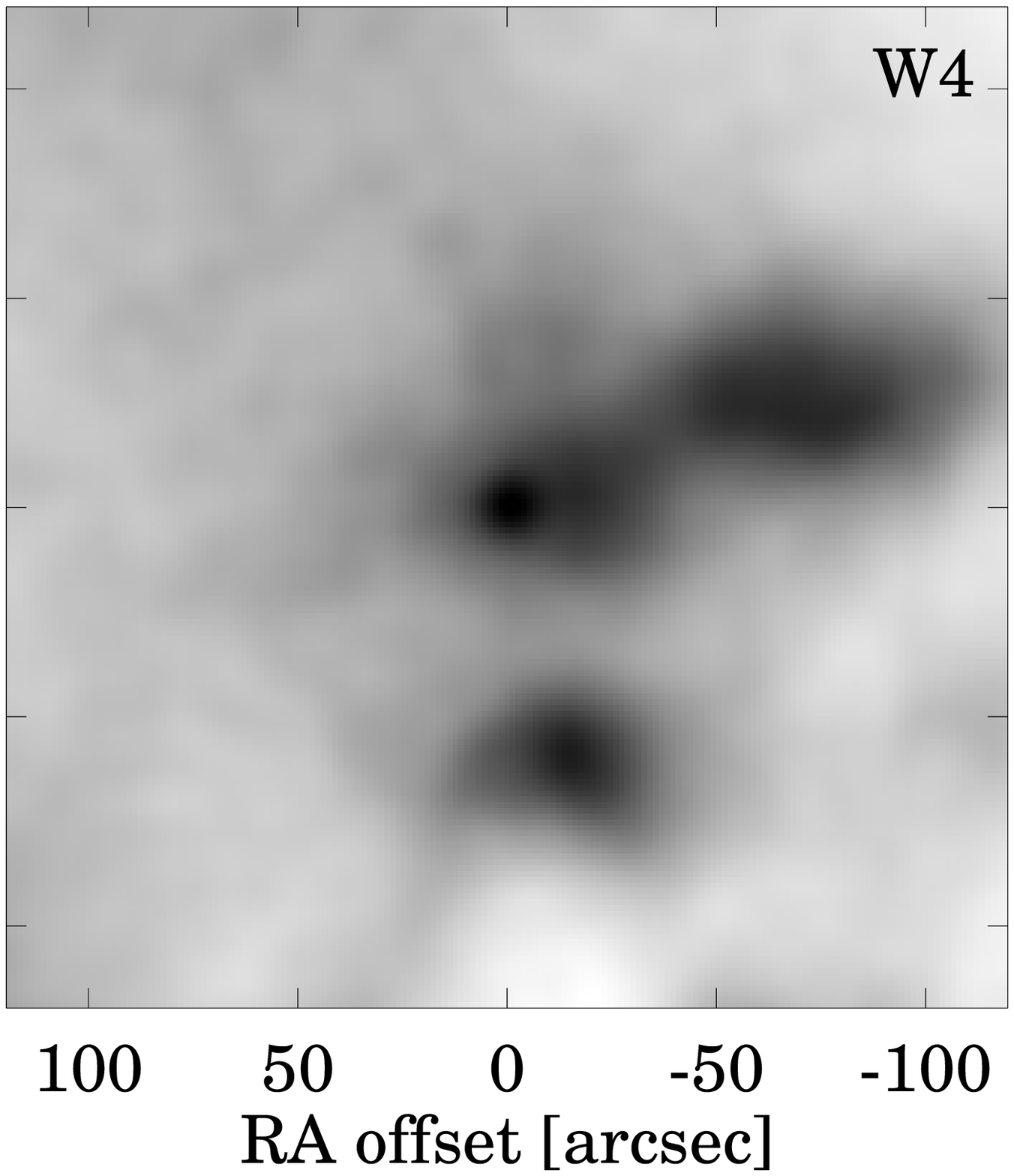}
\end{center}
\caption{W1--W4 band images (unWISE coadds of the {\sl WISE} all sky survey, 
see Appendix~\ref{appendix:a}) of the six selected objects. Positions 
of nearby IRAS FSC point sources (marked by red plus signs) with their 
1\,$\sigma$ (3\,$\sigma$ for TYC\,5940) error ellipses are overplotted in the 
W3 band panels. 
}
\label{fig:wiseimages}
\end{figure*}

Figure~\ref{fig:wiseimages} shows the vicinity of the six systems in all four 
{\sl WISE} bands (using the ``unWISE'' coadds, see Appendix~\ref{appendix:a}), 
while Table~\ref{tab:props} summarizes astrometric and photometric 
data of the host stars. Although, during the selection process, we used TGAS astrometry, in 
this table we list more precise astrometric parameters taken from the Gaia 
EDR3 catalog \citep{brown2020,lindegren2020}.

\section{Characterization of the host stars} \label{sec:hoststars}

\subsection{Spectroscopic observations} \label{sec:spectroscopy}
In order to estimate the fundamental parameters of the host stars of the   
identified EDDs, we performed high-resolution optical spectroscopy for all of them. 
As the observation log (Table~\ref{tab:obslog}) shows we used four different 
high-resolution \'echelle spectrographs to obtain the spectra. One of our 
targets, TYC\,4946, was measured twice.

TYC\,5940, TYC\,8105, and TYC\,4946
were targeted with the Fibre-fed Extended Range Optical Spectrograph 
\citep[FEROS,][]{kaufer1999} on the MPG/ESO 2.2\,m 
(La Silla, Chile) telescope. This instrument has a mean spectral resolution 
of 48\,000, covering the spectral region between 3500 and 9200{\,\AA}. In our 
observations, the object-sky operation mode was used, i.e., one of the two fibres 
was positioned at the star while the other at the sky. Reduction of FEROS 
spectra -- including bias subtraction, flat-field correction, background
subtraction, the definition and extraction of orders, and wavelength
calibration -- were performed using the FEROS Data Reduction System 
(DRS\footnote{\url{http://www.eso.org/sci/facilities/lasilla/instruments/feros/tools/DRS.html}}) 
pipeline implemented in the ESO-MIDAS environment. 

We obtained spectra of TYC\,4515 and TYC\,4946 at the Apache
Point Observatory (New Mexico, US), using the ARC Echelle Spectrograph (ARCES) on
the 3.5~m telescope with a resolution of $R = 31\,500$ in the spectral 
range $3200 - 10\,000$\,\AA. 
The spectra were reduced using a dedicated 
pipeline\footnote{\url{http://astronomy.nmsu.edu:8000/apo-wiki/attachment/wiki/ARCES/Thorburn\_ARCES\_manual.pdf} and 
\url{http://astronomy.nmsu.edu:8000/apo-wiki/attachment/wiki/ARCES/Kinemuchi\_ARCES\_cookbook.pdf}} 
based on IRAF
\footnote{IRAF is distributed by the National Optical Astronomy
Observatories, which are operated by the Association of Universities
for Research in Astronomy, Inc., under cooperative agreement with
the National Science Foundation.}, then normalized to the continuum level.

TYC\,4479 was observed using The FIbre-fed \'Echelle Spectrograph 
\citep[FIES,][]{frandsen1999,telting2014} on the 2.56\,m Nordic Optical 
Telescope (Observatorio del Roque de los Muchachos, La Palma, Spain) 
in its {\sl low-res} mode ($R = 25\,000$, with wavelength coverage from 
3700 to 7300{\,\AA}). Data processing was made with 
FIEStool\footnote{\url{http://www.not.iac.es/instruments/fies/fiestool/FIEStool.html}}, 
the dedicated pipeline of the FIES instrument.

%%%%%%%%%%%%%%%%%%%% TABLE 2 %%%%%%%%%%%%%%%%%%%%%%%%%%%%%%%%%%

\begin{deluxetable}{c|ccC}[h!]
\tablecaption{Log of spectroscopic observations \label{tab:obslog}}
\tablecolumns{4}
\tablewidth{0pt}
\tabletypesize{\scriptsize}
\tablehead{\colhead{Name} &
\colhead{Instrument} &
\colhead{Obs. date} &
\colhead{Exposure time (s)}
}
\startdata
TYC 4515 & ARCES  &  2013-03-25 & 780	\\ 
TYC 5940 & FEROS  &  2016-12-17 & 900	\\		 
TYC 8105 & FEROS  &  2016-12-17 & 900	\\
TYC 4946 & FEROS  &  2012-12-25 & 1022  \\
         & ARCES  &  2013-04-01 & 840   \\
TYC 4209 & ACE    &  2016-09-29 & 3$\times$5400 + 1$\times$3600 \\
TYC 4479 & FIES   &  2020-07-03 & 2760  \\
\enddata
\end{deluxetable}

%%%%%%%%%%%%%%%%%%%% TABLE 2 %%%%%%%%%%%%%%%%%%%%%%%%%%%%%%%%%%

Finally, the spectrum of TYC\,4209 was taken with the ACE spectrograph 
mounted on the 1\,m RCC telescope at Piszk\'estet\H o Observatory (Hungary). 
The fibre-fed spectrograph covers a range of 4150--9150{\AA}
with a spectral resolution of $\approx 20\,000$. For the wavelength
calibration, a thorium--argon (ThAr) lamp was used. 
Data reduction was performed using standard IRAF 
procedures in the \texttt{imred} and \texttt{echelle} packages.
Cosmic ray correction was done by the \texttt{crutil.cosmicrays}
utility. For the analysis, we combined 4 frames (one with a 3600\,s exposure time and three 
with a 5400\,s exposure time).

\subsection{Basic stellar properties} \label{sec:stellarprops}

The preparation of the reduced spectra consisted of identification and removal of
non-spectral features remaining from the reduction processes, and a correction of the
continuum normalization. These steps were carried out utilizing IRAF tasks and
the iSpec tool of \citet{blanco-cuaresma2014}. Before fitting the stellar parameters, the
radial velocity had to be determined and the necessary correction had to be
applied. These tasks were also performed with the iSpec tool.

For the determination of stellar parameters ($\log g$, $T_{\rm eff}$, and [Fe/H]) of
the target stars we used the SP\_Ace program (v1.1) written and developed by 
\citet{boeche2016}. 
SP\_Ace was designed for analyzing spectra with resolution between $\sim$1000 and 
$\sim$20\,000, but higher resolutions may be doable as well. 
In this range, the line profiles are plausible, 
since the effects of the applied approximations have been tested to not emerge.
The SP\_Ace methodology offers a fast and accurate solution of spectra in this range.

SP\_Ace is an equivalent width (EW) based tool \citep[the EW library was computed by MOOG,][]{sneden1973}, 
which performs the
spectral analysis under LTE assumption and based on 1D atmosphere models. It uses the
General Curve-Of-Growth library (GCOG) to compute the EWs of the lines, construct
models of spectra as a function of the stellar parameters and abundances, and
search for a model that minimizes the $\chi^2$ when compared to the
observed spectrum. In SP\_Ace, we covered the stellar parameter ranges of $3600 \,
\text{K} < T_{\rm eff} < 7400 \, \text{K}$, $0.2 < \log g < 5.4$ and  $-2.4 \,
\text{dex} < {\rm [Fe/H]} < 0.4 \, \text{dex}$.

To use this code, the spectra must be wavelength calibrated, continuum normalized, and radial velocity
corrected at rest frame. They must be in the wavelength ranges $5212-6860${\AA} and
$8400-8924${\AA} and must have less than 32\,000 pixels. Because of the last
requirement, when necessary we downgraded the spectral resolution. 
To test how the precision behaves after such downgrading, we analyzed publicly 
available spectra of stars with well known spectral parameters and found that 
there are no significant differences between the parameters derived from the spectra 
with different spectral resolution in the $\sim$20\,000$-$40\,000 resolution range.

The estimated stellar parameters and their uncertainties are summarized in Table~\ref{tab:props}.
As a final step, the error of the radial velocity is revised with the resulting
stellar parameters. The obtained radial velocities are also listed in Table~\ref{tab:props}.

In order to predict the stellar photospheric contribution to the total 
flux at mid-IR wavelengths, we fitted the optical and near-infrared photometry of the sources 
(taken from Table~\ref{tab:props})
by the appropriate ATLAS9 stellar atmosphere model \citep{castelli2004} 
selected based on the previously derived stellar parameters. 
In the fitting process, two parameters, the amplitude of the photospheric model and 
the interstellar reddening \citep[adopting the extinction curve from][]{ccm89} were 
estimated using a grid-based approach. Initially we used all 2MASS \citep{skrutskie2006} 
photometric data 
but we found that the longest wavelength $K_{\rm s}$ data points were systematically
higher than the fitted photosphere model, suggesting that the measured fluxes are not purely 
photospheric. Therefore, we discarded the $K_{\rm s}$ band photometry from this analysis.  
Although our targets are located farther away than 160\,pc, i.e., likely out of the Local Bubble, 
apart from the case of TYC\,4515 -- where an E(B-V) value of 0.04$\pm$0.02\,mag was obtained -- 
their reddening values were found to be negligible based on this fitting.
Combining these models with the Gaia\,EDR3 distances, we derived the stellar luminosities, 
also listed in Table~\ref{tab:props}.

\subsection{Stellar companions} \label{sec:companions}
To explore possible stellar companions, we used the Gaia EDR3 catalog to search for 
common proper motion and distance pairs of our targets.
Following the considerations from \citet{andrews2017}, we set 
a limit of 1\,pc for the maximum separation of the possible pairs. 
To ensure reliable quality astrometric solutions, we considered
 only stars whose parallax and proper motion measurements fulfilled 
 the following criteria: $\pi/\sigma_\pi > 5$ and 
 $ \frac{\sqrt{\mu_{\alpha*}^2  + \mu_\delta^2}}{\sqrt{\sigma_{\mu_{\alpha*}}^2  + \sigma_{\mu_\delta}^2}} > 5$, 
 respectively.
We identified companions by applying the method described in \citet{deacon2020}, which takes into account 
the possible orbital motion of the components.
We found probable wide separation companions to TYC\,4515, TYC\,8105, TYC\,4946, TYC\,4209, and TYC\,4479, 
i.e., to all stars with the exception of TYC\,5940.
All of the companion stars have $RUWE$\footnote{Renormalised Unit Weight Error (RUWE) 
is an indicator of the goodness of fit for Gaia astrometric data: values $<$1.4  
are considered reliable (for more details, see the technical note GAIA-C3-TN-LU-LL-124-01, 
\url{https://www.cosmos.esa.int/web/gaia/public-dpac-documents}).}$<$1.4, indicating a well-fitting astrometric solution.
Despite the large search radius, the projected separations of the identified pairs range between 1000 ($\sim$0.005\,pc) and 
6000\,au ($\sim$0.03\,pc). 
This means that both components can be formed by the fragmentation of a single protostellar core, that has a 
typical size of 0.1\,pc \citep{lada2008,duchene2013}.
Table~\ref{tab:pairs} shows the astrometric parameters  
for the five pairs.

%%%%%%%%%%%%%%%%%%%% TABLE 3 %%%%%%%%%%%%%%%%%%%%%%%%%%%%%%%%%%

\begin{deluxetable*}{l|cCCCCCcCCC}
\tabletypesize{\scriptsize}                                           
\tablecaption{Astrometric parameters and stellar properties of the probable common motion 
and distance pairs \label{tab:pairs}}                      
\tablecolumns{11}
\tablewidth{0pt} 
\tablehead{
\colhead{Name} &
\colhead{Gaia EDR3 id} &
\colhead{Separation} &
\colhead{$\mu_{\alpha^*}$} &
\colhead{$\mu_{\delta}$} &
\colhead{$\pi$} &
\colhead{$G$} &
\colhead{SpT} &
\colhead{$T_{\rm eff}$} &
\colhead{$L_*$} &
\colhead{$M_*$} \\
%%%%%%%%%%%%%%%%%%%%%%%%%%%
\colhead{} & 
\colhead{} &                                                                
\colhead{(au)} &
\colhead{({mas~yr$^{-1}$})} & 
\colhead{({mas~yr$^{-1}$})} & 
\colhead{(mas)} & 
\colhead{(mag)} &
\colhead{} &
\colhead{(K)} &
\colhead{(L$_\odot$)} &
\colhead{(M$_\odot$)}
}
\startdata  
TYC\,4515B  & 552973534667883904   & 1825 & $+$6.61$\pm$0.05  & $-$24.18$\pm$0.08 & 3.49$\pm$0.06 & 16.578 & M1   & 3670$\pm$60 & 0.0363$\pm$0.0014 & 0.50$\pm$0.02 \\
TYC\,8105B  & 5554553527426091136  & 4540 & $+$6.80$\pm$0.33  & $+$13.54$\pm$0.39 & 5.57$\pm$0.30 & 19.510\tablenotemark{a} & M5.5 & 2960$\pm$60 & 0.0024$\pm$0.0001 & 0.096$\pm$0.01 \\
TYC\,4946B  & 3596395748685297792  & 1010 & $-$27.66$\pm$0.34 & $-$5.07$\pm$0.26  & 4.08$\pm$0.28 & 18.812\tablenotemark{a} & M4.5 & $\sim$3120 & $\sim$0.005 &  $\sim$0.19 \\
TYC\,4209B  & 2161326842991191936  & 6010 & $+$12.13$\pm$0.17 & $-$3.38$\pm$0.17  & 3.85$\pm$0.11 & 18.337 & M3.5 & 3230$\pm$55 & 0.0077$\pm$0.0003 & 0.27$\pm$0.01 \\
TYC\,4479B  & 2210856513228282880  & 5040 & $+$38.90$\pm$0.04 & $-$39.53$\pm$0.04 & 6.12$\pm$0.03 & 15.867 & M2   & 3540$\pm$60 & 0.0182$\pm$0.0006 & 0.38$\pm$0.02 \\
\enddata
\tablenotetext{a}{Corrected G-band magnitude for sources with 6-parameter astrometric solutions based on \citet{riello2020}.} 
\end{deluxetable*}

%%%%%%%%%%%%%%%%%%%% TABLE 3 %%%%%%%%%%%%%%%%%%%%%%%%%%%%%%%%%%

For the companions of TYC\,4515, TYC\,8105, TYC\,4209, and TYC\,4479, we 
 estimated the effective temperatures based on 
their $r-z$, $r-J$, and $G_{BP}-G_{RP}$ color indices, using the calibration 
derived by \citet{amann2015}. The $G_{BP}$ and $G_{RP}$ band data were collected 
from the Gaia\,EDR3 catalog. The $J$ band measurements were extracted from the 2MASS 
survey \citep{skrutskie2006}. For TYC\,8105B, the $r$ and $z$ band photometry were taken from 
The Dark Energy Survey \citep[DR1,][]{abbott2018}, while in the cases of 
TYC\,4515B and TYC\,4209B, magnitudes measured in The Pan-STARRS Survey \citep[DR1,][]{chambers2016} 
were transformed into the SDSS system using the equations presented in \citet{tonry2012}.
For TYC\,4209B, that was observed both in The SDSS Photometric Catalog \citep[Release~12,][]{alam2015} and 
The Pan-STARRS Survey, we derived the final $r$ and $z$ photometry as the weighted average of 
these data (after the transformation of the Pan-STARRS observations to the SDSS photometric system).
Photometry for TYC\,4515B were dereddened by adopting the $E(B-V)$ values 
obtained for their primary components (Sect.~\ref{sec:stellarprops}).
The final $T_{\rm eff}$ values, which are the weighted averages of the three $T_{\rm eff}$ estimates 
corresponding to the three colors, are listed in Table~\ref{tab:pairs}. 

The spectral types of the companions were estimated based on the obtained effective temperatures using an updated 
version\footnote{\url{https://www.pas.rochester.edu/~emamajek/EEM\_dwarf\_UBVIJHK\_colors\_Teff.txt}}
of the conversion from \citet{pecaut2013}.
We found that the spectral types of these secondary components 
range between M1 and M5.5, thus, these are newly discovered M dwarfs.
We also derived the stellar luminosities. As a first step, we performed bolometric correction 
to the $G$, $r$, $z$, $J$, $H$, and $K_{\rm s}$ passbands using the proper equations 
of \citet{amann2015}. The final bolometric magnitudes were obtained as the weighted average 
of the six individual bolometric magnitudes. Then, considering the reddening and adopting the distances 
of the primary components (which have substantially lower uncertainties than those of 
the companions) we calculated the luminosities of the four companions (Table~\ref{tab:pairs}). 

For TYC\,4946B, where reliable photometry was available only in the $G$ band,  
the $T_{\rm eff}, L_*, M_*$ parameters and the spectral type were estimated based on its 
$G$ band absolute magnitude by interpolating in the updated 
table of \citet{pecaut2013}. The obtained 
parameters of this $\sim$M4.5-type star are also listed in Table~\ref{tab:pairs}.  

We also investigated the possible existence of closer binary pairs in our sample.
\citet{qian2019} listed TYC\,4946 as a possible spectroscopic binary, because its two observations in 
the LAMOST (The Large Sky AreaMulti-Object Fiber Spectroscopic Telescope) survey brought very different
radial velocity measurements ($v_{\rm r}$=$-37.99\pm0.72$\,km~s$^{-1}$ and $-16.30\pm3.43$\,km~s$^{-1}$). 
For this star, we obtained radial velocities of $-$10.5$\pm$0.7\,km~s$^{-1}$ and $-$12.1$\pm$0.7\,km~s$^{-1}$ 
from the observations carried out by the ARCES and FEROS spectrographs at two different epochs (Table~\ref{tab:obslog}).
The Gaia DR2 catalog \citep{gaia} -- in good agreement with our results -- quotes a $v_{\rm r}$ of $-$12.29$\pm$0.36\,km~s$^{-1}$.  
In the case of TYC\,4515, we extracted $v_{\rm r}$ of +12.5$\pm$0.5\,km~s$^{-1}$ from our high resolution 
spectrum (Table~\ref{tab:props}) which deviates significantly from its $v_{\rm r}$ of $-2.65\pm0.89$\,km~s$^{-1}$
 measured by Gaia. These two stars are candidate spectroscopic binaries, but further measurements are 
 needed to confirm this claim. 
If TYC\,4515 or TYC\,4946 has a spectroscopic binary pair, then they would be 
 triple systems. 
For TYC\,5940, TYC\,8105, and TYC\,4209, our $v_{\rm r}$ measurements are consistent with 
 the Gaia based values, thus, there is no indication for any close companion.

\subsection{Age of the stars} \label{sec:ages}
The ages of the host stars provide important information in assessing the 
evolutionary status of their disks. To infer this parameter, we combined 
a variety of age diagnostics based on lithium content, stellar rotation, 
and kinematic properties. In the case of TYC\,8105, isochrone fitting 
for its M-type companion was also performed.

\begin{figure} 
\begin{center}
\includegraphics[scale=.46,angle=0,bb=23 9 589 403]{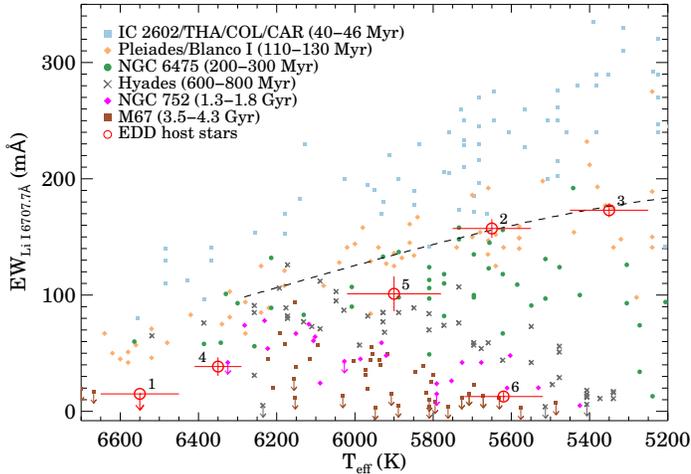}
\caption{ {Equivalent width of the lithium line (at 6707.7\,\AA) as a function of
effective temperature for our targets (1: TYC\,4515, 2: TYC\,5940, 3: TYC\,8105, 
4: TYC\,4946, 5: TYC\,4209, 6: TYC\,4479) and for members of some nearby open clusters 
and moving groups with well known ages. Data for open clusters were taken from 
\citet{sestito2005} via the WEBDA database\footnote{\url{https://webda.physics.muni.cz/}}, 
while for the Tucana-Horologium (THA), Columba (COL), and Carina (CAR) moving groups 
the data were from \citet{dasilva2009}. 
The dashed curve stands for a polynomial fit to the lithium data of Pleiades 
\citep{pecaut2016}. Ages of moving groups and open clusters were taken from the literature: 
IC\,2602 \citep{dobbie2010}; THA, COL, and CAR \citep{bell2015}; Pleiades \citep{stauffer1998,dahm2015}; 
Blanco\,I \citep{cargile2010,juarez2014}; NGC\,6475 \citep{villanova2009,cummings2018,bossini2019}; 
Hyades \citep[][and references therein]{douglas2019};
NGC\,752 \citep{daniel1994,twarog2015,agueros2018}; M\,67 \citep{giampapa2006,vandenberg2004,sarajedini2009,bossini2019}.  
}
\label{fig:ewli}
}
\end{center}
\end{figure}

In stars having a convective layer, as in our objects, photospheric lithium 
is thought to be depleted with time when mixed into deeper layers due to 
convection, and destroyed at a temperature of about 2.5$\times$10$^6$\,K. We 
measured the lithium content using the resonance doublet of \ion{Li}{1} 
at 6707.7{\AA}, whose equivalent widths (EW$_{\rm Li}$) were measured 
by fitting a Gaussian profile to the absorption feature in our high resolution 
spectra (see Sect.~\ref{sec:spectroscopy}). The Li absorption line is blended 
with a weak \ion{Fe}{1} line at 6707.4\,{\AA}. Taking into account this effect, 
the obtained equivalent widths were corrected using a method developed by 
\citet{soderblom1993}. For TYC\,4946, where more than one spectrum were available, 
the final EW$_{\rm Li}$ was computed as the average of the individual measurements. 
For TYC\,4515, where the lithium line was not detected, a 3$\sigma$ upper 
limit was calculated. The corrected EW$_{\rm Li}$ values are given in 
Tab.~\ref{tab:props} and displayed as a function of effective temperatures
in Fig.~\ref{fig:ewli}.
For comparison, members of some well-dated nearby open clusters and moving groups, 
representing an age range of $\sim$40--4300\,Myr, are also plotted. 

Late-type dwarf stars lose mass and angular momentum via magnetized stellar 
winds, leading to rotational spin down as they age \citep{schatzman1962,kraft1967,kawaler1988}. 
This enables using stellar rotation to estimate the ages of stars via a technique 
called gyrochronology \citep{barnes2003}. Starspots may cause periodic light changes 
from which the rotation period can be derived. 
In our sample, five stars (TYC\,4515, TYC\,5940, TYC\,8105, TYC\,4209, and TYC\,4479) 
were observed with the {\sl TESS} satellite, providing high 
accuracy photometry with 30 minutes cadence.
All five stars exhibit periodic variations likely due to rotational modulation by star-spots. 
By performing a frequency analysis, we obtained rotational periods as listed in 
Table~\ref{tab:props}. A detailed description of the {\sl TESS} data reduction 
and the frequency analysis is given in Appendix~\ref{appendix:b}. In Fig.~\ref{fig:prot} 
we plot the rotational periods as a function of the effective temperatures for those 
four G-type stars for which gyrochronology is applicable. Due to rotational 
evolution, stellar groups of different ages populate distinct, 
well-defined areas in this plot, as demonstrated for four open clusters: Pleiades 
($\sim$120\,Myr), Hyades and Praesepe (both 600--800\,Myr), and M67 (3.8--4.3\,Gyr). 
Having similar ages, the members of Hyades and Praesepe overlap in this diagram.

\begin{figure}[h!!!] 
\begin{center}
\includegraphics[scale=.46,angle=0,bb=23 9 589 403]{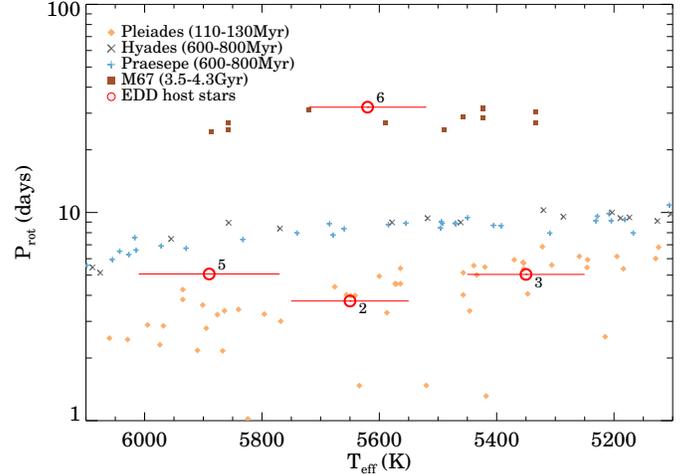}
\caption{Rotation periods as a function of effective temperatures for our targets and 
for single star members of three open clusters. Our targets (TYC\,5940, TYC\,8105, 
TYC\,4209, and TYC\,4479) have the same labels as in Fig.~\ref{fig:ewli}. 
Data for open cluster members were from the literature \citep{rebull2016,stauffer2016,barnes2016,douglas2019}.
\label{fig:prot}
}
\end{center}
\end{figure}

By comparing the effective temperatures and luminosities of the six host stars (Table~\ref{tab:props})
and the four companions (Table~\ref{tab:pairs}) with the corresponding mean properties of dwarf 
stars\footnote{\url{http://www.pas.rochester.edu/~emamajek/EEM\_dwarf\_UBVIJHK\_colors\_Teff.txt} 
(compiled by E. Mamajek)}, we found that all, but TYC\,8105\,B, are consistent with 
the main-sequence. TYC\,8105\,B is brighter than main-sequence dwarfs with comparable 
$T_{\rm eff}$, suggesting that it is in pre-main sequence evolution. To verify the age estimate 
 of the TYC\,8105 system, we performed isochrone-fitting (see below) for the companion 
 using isochrone tracks constructed by \citet{baraffe2015}.

\begin{figure*} 
\begin{center}
\includegraphics[scale=.50,angle=0]{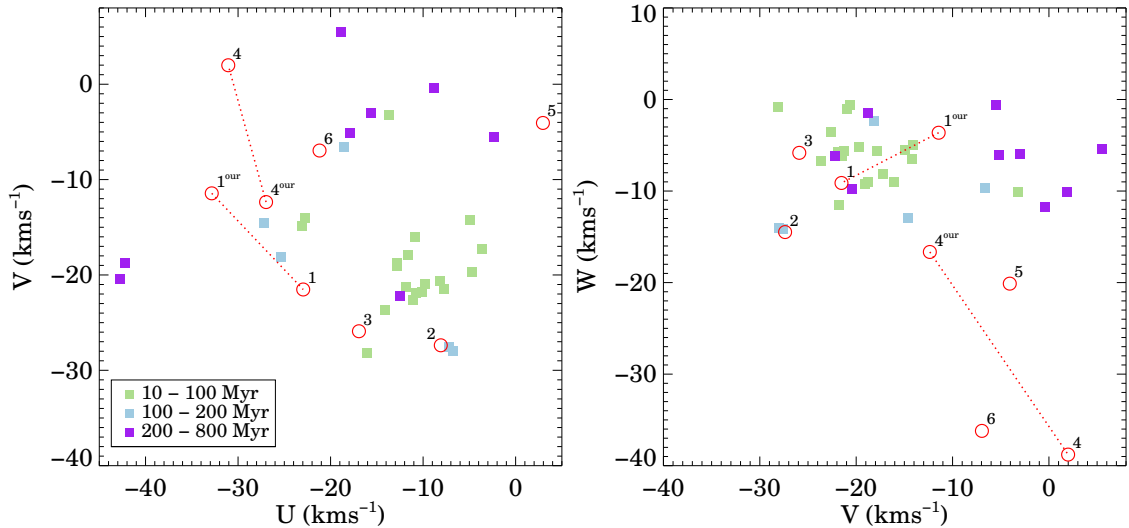}
\caption{ { (U, V)- and (V, W)-planes for the host stars of extreme debris disks 
(red circles). {For TYC\,4515 and TYC\,4946, correspondingly to the measured 
$v_{\rm r}$ range, two U,V,W values are shown (Sect.~\ref{sec:ages}).}   
1: TYC\,4515, 2: TYC\,5940, 3: TYC\,8105, 4: TYC\,4946, 5: TYC\,4209, 6: TYC\,4479. 
Filled squares display space velocities of known moving groups, associations, 
and open clusters located within 300\,pc taken from the 
literature \citep{gagne2018,soubiran2018}. Different colors stand for different age 
ranges.   
}
\label{fig:uvw}
}
\end{center}
\end{figure*}

Using astrometric data (coordinates, proper motions, and trigonometric parallax) 
from the Gaia EDR3 catalog, supplemented by our radial velocity measurements from 
Table~\ref{tab:props}, we computed the $U, V,$ and $W$ Galactic space velocity 
components of the targets. In this calculation, we used a right-handed coordinate system 
thus $U$ is positive toward the Galactic center, $V$ is positive in the direction of 
galactic rotation, and $W$ is positive toward the galactic North pole. The derived 
Galactic space velocities are listed in Tab.~\ref{tab:props} and displayed in 
Fig.~\ref{fig:uvw}. 
{For TYC\,4515 and TYC\,4946
where some radial velocity data in the literature differ substantially from our 
measurements (Sect.~\ref{sec:companions}) the UVW components were computed 
using the most different $v_{\rm r}$ data, too and displayed 
in Fig.~\ref{fig:uvw}.}
For reference, known moving groups, associations, and open clusters located within 
300\,pc are also plotted.

\paragraph{TYC\,5940 and TYC\,8105} are both G-type stars (Tab.~\ref{tab:props}).
As Fig.~\ref{fig:ewli} demonstrates, their lithium equivalent widths 
are consistent with the Pleiades stars ($\sim$120\,Myr), and lower than the values measured 
in the 40-50\,Myr old members of  nearby moving groups and of the IC\,2602 open cluster. 

In line with this, the position of both stars in the period-effective temperature diagram 
(Fig.~\ref{fig:prot}) also matches the sequence of the Pleiades.
To estimate the gyrochronology age, we applied four different calibrations for the 
rotation--age relation \citep{barnes2007,mamajek2008,meibom2009,angus2015}. 
For TYC\,5940, these calibrations yielded age estimates between 92 and 134\,Myr with an average of 118\,Myr,
while the gyroages for TYC\,8105 range from 125 to 197\,Myr with an average of 163\,Myr.

As Fig.~\ref{fig:uvw} outlines, the UVW velocities of both stars are consistent 
with the typical space motion of nearby young (10--200\,Myr old) stars. Interestingly, 
the derived $U$, $V$, $W$ components of TYC\,5940 overlap with the Pleiades and the AB\,Dor 
moving groups. We note, however, that the direction and distance of this star are 
clearly inconsistent with those of both Pleiades and the currently known values 
of the AB\,Dor group, making its membership unlikely.

According to our analysis, TYC\,8105 has a wide-separation M5.5-type pre-main 
sequence companion. To determine its best-fit mass and isochrone age, we followed the Bayesian 
inference approach outlined in \citet{pascucci2016} for the comparison of the 
measured $L_*$ and $T_{\rm eff}$ parameters (from Table~\ref{tab:pairs}) 
with theoretical predictions from \citet{baraffe2015}.
We obtained an isochrone age of 75$^{+50}_{-30}$\,Myr. 
Although somewhat lower, this age estimate is 
consistent with the lithium and gyro ages within the uncertainties.
In fact, isochrones of \citet{baraffe2015} are available only for solar 
metallicity, while TYC\,8105 is a bit more metal rich at 
[Fe/H]$\sim$0.08, which might lead to a slight 
underestimation of age. Moreover, isochronal model tracks tend to 
underestimate the age of low-mass pre-main sequence stars \citep[e.g.,][]{pecaut2016}.

Taking into account the results of the different methods, we adopted ages of 
120$\pm$20\,Myr and 130$\pm$30\,Myr for TYC\,5940 and TYC\,8105, respectively.

\paragraph{TYC\,4209} falls between the Pleiades ($\sim$120\,Myr) and Hyades stars 
(600-800\,Myr old) in Fig.~\ref{fig:ewli}, and its measured EW$_{Li}$ value makes 
it similar to members of the 200--300\,Myr old open cluster NGC\,6475 
\citep{villanova2009,cummings2018,bossini2019}. The obtained $P_{\rm rot}$ of 5.07\,day also places the 
star between the sequences of Pleiades and Hyades/Praesepe in the period--$T_{\rm eff}$ 
plot (Fig.~\ref{fig:prot}). The average of the gyroage estimates is 295\,Myr with 
a range between 241 and 333\,Myr. That this object is somewhat older than the previous 
two stars is also supported by its location in Fig.~\ref{fig:uvw} outside the region 
of the young, 10--200\,Myr old groups. 
Based on these results, we adopted an age of 275$\pm$50\,Myr for TYC\,4209.

\paragraph{TYC\,4515 and TYC\,4946} are the hottest stars in our sample. 
Studying the Li abundances in different open clusters showed that similar mid F-type members with 
effective temperatures between 6300 and 6900\,K often have significant lithium depletion 
compared to co-eval but somewhat colder or hotter stars \citep[e.g.][]{boesgaard1986,balachandran1995}. 
This so-called Li dip has not been observed in the Pleiades, but according to \citet{steinhauser2004}
it already appears at an age of $\sim$150\,Myr. The measured EW$_{Li}$ values for TYC\,4946 and 
especially for TYC\,4515 are significantly lower than those of Pleiades stars, implying a serious Li 
depletion and suggesting a lower limit of 150\,Myr for their ages. Although the galactic velocity 
components of both stars are quite uncertain due to the uncertainty of radial velocities 
(Sect.~\ref{sec:companions}), in the case of TYC\,4946, the possible U,V,W range mostly deviates 
from the one typical for nearby young stars (Fig.~\ref{fig:uvw}) further supporting this age 
estimate. 
Considering all results, we adopted a lower limit of 150\,Myr for both systems.

\paragraph{TYC\,4479}
has a lithium EW lower than those of the similar temperature members of the NGC\,752 
(1.3--1.8\,Gyr) and roughly consistent with the upper limits measured for 
late G-type stars belonging to the $\sim$4\,Gyr old M67 open cluster (Fig.~\ref{fig:ewli}).  
In its TESS light curve, we found a variation with a period of $\sim$32\,days and an amplitude of 
$\sim$0.0012\,mag. The observed period is slightly longer than the rotational periods observed 
for late G stars of M67 (Fig.~\ref{fig:prot}). Assuming that the observed changes are also related to stellar rotation, 
 the above mentioned gyrochronologic relations yield age estimates 
between 5.4 and 6.5\,Gyr, with a mean of 6\,Gyr, for TYC\,4479. 
However, the calibrations of gyrochronology are based on well dated stars -- 
typically members of open clusters and moving groups -- whose metallicity differs only marginally from that of 
Sun, while TYC\,4479 has an [Fe/H] of $-$0.29. According to \citet{amard2020}, this could have a significant effect: 
the rotation of such a metal-poor star is expected to be faster than a similar mass star with a solar metallicity. 
This can lead to an underestimation of the stellar age using the available gyrochronology relations.
As Fig.~\ref{fig:uvw} demonstrates, the W velocity component of the star differs significantly 
from the typical W of nearby stellar groups younger than 800\,Myr. By applying a recent age-velocity 
relation proposed by \citet{almeida-fernandes2018}, we compiled  
a probability density function for the age of a star based on its UVW data. 
Based on this, the kinematic age is 4.8$^{+5.7}_{-1.6}$\,Gyr. 
All these results indicate that TYC\,4479 is definitely older than 2\,Gyr and its age is likely 
 at least comparable to that of the Sun. For this star we adopted an age of 5$\pm$2\,Gyr.   
  
\medskip

All six systems are found to be older than 100\,Myr, implying 
that the observed dust cannot be primordial material but rather  
is of secondary origin, i.e., these stars host 
extreme debris disks.

Using the obtained luminosity, effective temperature, and metallicity parameters (Tab.~\ref{tab:props}), 
we estimated the mass of the host stars based on the PARSEC evolutionary models 
\citep[ver. 1.2S,][]{bressan2012}. In this analysis, we took into account the estimated ages 
of the stars as a priori information. The resulting stellar masses are listed in Table~\ref{tab:props}. 
Using the same approach, we also estimated the masses of the companions. For
TYC\,4515B, TYC\,4209B, and TYC\,4479B, we used the PARSEC models, while for the late M-type TYC\,8105B 
the evolutionary models were taken from \citet{baraffe2015}. For these objects, we assumed 
metallicities and ages similar to those of the primary components. 
The obtained stellar masses are listed in Table~\ref{tab:pairs}.

\section{Characterization of the disks} \label{sec:characteizationofedds}

\subsection{WISE observations} \label{sec:wisedata}
The primary survey of {\sl WISE} is comprised of three phases. 
In the first 7 months of the survey operation (between January and 
August of 2010) the satellite performed observations in all four 
bands (4-Band Cryo Phase) scanning the sky 1.2 times before the 
exhaustion of frozen hydrogen from its outer cryogen tank. This phase 
was followed by a short ($\sim$2 months) operational period when 
the focal plane was cooled by the inner cryogen system, allowing the 
scanning of $\sim$30\% of the sky in the W1, W2, and W3 bands (3-Band 
Cryo Phase). Finally, after the end of the cryogenic mission, in the 
framework of the NEOWISE Post-Cryo survey phase approximately 70\% 
of the sky was observed in the W1 and W2 bands achieving thereby 
a complete second coverage of the sky at the shortest wavelengths. 
At the close of this phase the satellite was put into hibernation 
in February 2011. In October 2013, the spacecraft was reactivated 
and sky surveys were continued in the W1 and W2 bands. The recent 
NEOWISE 2020 Data Release provides photometric data obtained in 
the first six years of the Reactivation survey, in which period 
 the satellite scanned the sky approximately twelve times. 

By gathering single exposure observations obtained for our targets 
during the above-mentioned survey phases from the NASA/IPAC Infrared 
Science Archive (IRSA) database, we could extract W1 and W2 photometric 
data clustered in 14--15 observational windows. Apart from a larger gap 
due to the hibernation period these windows are typically spaced by 
$\sim$6\,months. 
In the case of TYC\,4479, these data are completed by W3 and W4 band 
photometry in the first two epochs. For the 
remaining sources, these long wavelength data are only available in the 
first epoch. Corresponding to the applied observing strategy of {\sl WISE} \citep{wright2010}, 
the individual observing windows of our targets 
are 0.7--34 days long with 10--300 single exposure measurements 
depending mainly on the ecliptic latitude of the source. 

In the following analysis we used only the best quality single exposure 
data points and discarded measurements where the frameset images had low quality \nobreak{(\texttt{qi\_fact} $=$ 0)}.
Following the Explanatory Supplement\footnote{\url{wise2.ipac.caltech.edu/docs/release/allwise/expsup/}}, 
in order to filter out framesets that could be contaminated by higher levels of 
charged particle hits or by scattered moon-light, we removed data points where 
the frameset was taken when the satellite was located within the nominal boundaries 
of the South Atlantic Anomaly (\texttt{saa\_sep}$<$5); or was performed within an area close 
to the Moon (\texttt{moon\_masked}=1). We also removed all measurements having 
contamination and confusion flags (\texttt{cc\_flags}) of either 'D', 'P', 'H', 'O', 'L', or 'R'.
Moreover, we found that data points where the reduced $\chi^2$ 
of the profile-fit photometry measurement (\texttt{w1rchi2} or \texttt{w2rchi2}) was higher than 5 
were frequently outlying from the bulk of the data, and therefore we rejected them 
as well.

\subsection{Additional infrared data} \label{sec:additionalirdata}
To expand our knowledge on the SED of the sources, 
we searched for additional infrared data in the literature. Using the {\sl IRAS} 
Point Source \citep{helou1988} and Faint Source \citep[FSC,][]{moshir89} catalogs,
we found possible counterparts for four of our targets (TYC\,4515, TYC\,4209, 
TYC\,8105, and TYC\,5940) in the FSC. For TYC\,4515, TYC\,4209, and TYC\,8105, 
the association with the nearby FSC source is unambiguous, since these targets 
are situated within the 1.5\,$\sigma$ {\sl IRAS} error ellipse, and their {\sl WISE} 
W3 band images show no other potentially contaminating objects in their environment 
(Fig.~\ref{fig:wiseimages}). Based on the {\sl IRAS} error ellipse plotted over on 
the W3 band image (Fig.~\ref{fig:wiseimages}), in the case of TYC\,5940, an additional fainter object, 
to the north-west of our 
target, can also contribute to the measured {\sl IRAS} flux. TYC\,8105 was detected 
both at 12 and 25\,\,{\micron}, for the other three sources, only 12\,{\micron} photometry 
is available. 

For TYC\,4209, an additional data point at 9\,{\micron} is available in the {\sl AKARI} 
IRC catalog \citep{ishihara2010}. TYC\,4515 was serendipitously covered 
by a 24\,{\micron} map obtained with the MIPS camera \citep[Multiband Imaging Photometer 
for Spitzer,][]{rieke2004} (AOR: 15184640, 
PI: David Jewitt).
In the latter case, MIPS images produced via the Enhanced Basic Calibrated Data (EBCD) pipeline 
(version S18.12.0)
were downloaded from the Spitzer Heritage Archive  and the MOPEX software package 
\citep[MOsaicking and Point source Extraction,][]{makovoz2005} 
was used to create a mosaic from the individual frames. Then point spread function (PSF) photometry was applied to 
obtain the flux density of TYC\,4515. The source has a positional offset of 0\farcs55 with respect to the 
proper motion corrected Gaia EDR3 position of the star.
The final photometric uncertainty was derived as 
the quadratic sum of the measurement noise and the conservative absolute calibrational uncertainty 
of the MIPS 24\,{\micron} array (4\%, MIPS Instrument Handbook\footnote{\url{https://irsa.ipac.caltech.edu/data/SPITZER/docs/mips/mipsinstrumenthandbook/}}).
As a result, we obtained a flux density of 25.2$\pm$1.4\,mJy at 24\,{\micron}.

TYC\,4479, which is located in the foreground of the Cep\,OB4, was observed 
at 3.6 and 4.5\,{\micron} in a mapping project that targeted this association with 
the IRAC intrument of {\sl Spitzer} (AOR: 68515840, PI: Joseph Hora).
We performed aperture photometry using those six 'corrected BCD' (cBCD) images that 
cover our target. 
The aperture radius was set to 3\,pixels (3\farcs66), the sky background was computed in an annulus 
between 12\,pixels and 20\,pixels. 
The background was estimated using an iterative sigma-clipping method,
where the threshold was set to 3\,$\sigma$. We applied aperture correction with a factor 
of 1.113 (IRAC Instrument Handbook\footnote{\url{https://irsa.ipac.caltech.edu/data/SPITZER/docs/irac/iracinstrumenthandbook/}})
in both filters.
Array location- and pixel phase-dependent photometric corrections were also done using 
the recipe from the IRAC Instrument Handbook. 
Finally, the six flux density values obtained in the different frames  
were averaged to get the final photometry. This approach yielded 
flux densities of 43.6$\pm$0.9\,mJy and 29.9$\pm$0.6\,mJy at 3.6\,{\micron} and 
4.5\,{\micron}, respectively.
The quoted uncertainties are quadratic sums of the measurement errors 
and the absolute calibrational uncertainty (2\%, IRAC Documentation\footnote{\url{https://irsa.ipac.caltech.edu/data/SPITZER/docs/irac/warmimgcharacteristics/}}).

%%%%%%%%%%%%%%%%%%%% TABLE 4 %%%%%%%%%%%%%%%%%%%%%%%%%%%%%%%%%%

\begin{deluxetable}{c|ccDDD}[h!]                                                
\tablecaption{Measured and predicted flux densities. \label{tab:sedtable}}      
\tablewidth{0pt}                                                                
\tabletypesize{\scriptsize}                                                     
\tablehead{                                                                     
\colhead{Name} &                                                                
\colhead{Instr.} &                                                              
\colhead{Epoch} &                                                               
\twocolhead{$\lambda$} &                                                        
\twocolhead{$P_\nu$\tablenotemark{a}} &                                                         
\twocolhead{$F_\nu$\tablenotemark{b}} \\                                        
\colhead{} &                                                                    
\colhead{} &                                                                    
\colhead{} &                                                                    
\twocolhead{({\micron})} &                                                      
\twocolhead{(mJy)} &                                                            
\twocolhead{(mJy)}                                                              
}                                                                               
\decimals                                                                       
\startdata                                                                      
      TYC\,4515 &      WISE &   2010 &    3.35 &    47.3 &    55.3$\pm$1.7 \\
                &      WISE &   2010 &    4.60 &    26.1 &    39.2$\pm$1.3 \\
                &      WISE &   2010 &   11.56 &     4.5 &    42.7$\pm$2.1 \\
                &      IRAS &   1983 &   12.00 &     4.1 &   98.9$\pm$20.8 \\
                &      WISE &   2010 &   22.09 &     1.2 &    30.5$\pm$3.7 \\
                &      MIPS &   2006 &   23.67 &     1.1 &    25.6$\pm$1.4 \\
\hline
      TYC\,5940 &      WISE &   2010 &    3.35 &    21.0 &    24.7$\pm$0.8 \\
                &      WISE &   2010 &    4.60 &    11.0 &    17.3$\pm$0.6 \\
                &      WISE &   2010 &   11.56 &     2.0 &    41.3$\pm$2.3 \\
                &      IRAS &   1983 &   12.00 &     1.9 &   98.6$\pm$22.7 \\
                &      WISE &   2010 &   22.09 &     0.6 &    26.5$\pm$2.4 \\
\hline
      TYC\,8105 &      WISE &   2010 &    3.35 &    33.3 &    34.5$\pm$1.2 \\
                &      WISE &   2010 &    4.60 &    16.8 &    21.9$\pm$0.8 \\
                &      WISE &   2010 &   11.56 &     3.2 &    80.5$\pm$3.9 \\
                &      IRAS &   1983 &   12.00 &     3.0 &  108.2$\pm$20.6 \\
                &      WISE &   2010 &   22.09 &     0.9 &   111.0$\pm$7.6 \\
                &      IRAS &   1983 &   25.00 &     0.7 &  125.9$\pm$18.9 \\
\hline
      TYC\,4946 &      WISE &   2010 &    3.35 &    48.8 &    58.2$\pm$1.9 \\
                &      WISE &   2010 &    4.60 &    26.8 &    39.9$\pm$1.6 \\
                &      WISE &   2010 &   11.56 &     4.6 &    20.6$\pm$1.1 \\
                &      WISE &   2010 &   22.09 &     1.3 &    11.5$\pm$1.6 \\
\hline
      TYC\,4209 &      WISE &   2010 &    3.35 &    25.6 &    42.7$\pm$1.4 \\
                &      WISE &   2010 &    4.60 &    13.6 &    40.9$\pm$1.6 \\
                &       IRC &   2006 &    9.00 &     4.0 &    89.6$\pm$5.5 \\
                &      WISE &   2010 &   11.56 &     2.4 &    87.7$\pm$4.2 \\
                &      IRAS &   1983 &   12.00 &     2.3 &   89.5$\pm$12.5 \\
                &      WISE &   2010 &   22.09 &     0.7 &    41.9$\pm$4.1 \\
\hline
      TYC\,4479 &      WISE &   2010 &    3.35 &    45.8 &    48.4$\pm$1.6 \\
                &      IRAC &   2019 &    3.55 &    41.3 &    43.1$\pm$0.9 \\
                &      IRAC &   2019 &    4.49 &    25.5 &    29.0$\pm$0.6 \\
                &      WISE &   2010 &    4.60 &    24.2 &    28.5$\pm$1.0 \\
                &      WISE &   2010 &   11.56 &     4.4 &    38.4$\pm$1.9 \\
                &      WISE &   2010 &   22.09 &     1.2 &    24.6$\pm$3.5 \\
\hline
\enddata                                                                                                      
\tablenotetext{a}{Predicted photospheric flux densities.} 
\tablenotetext{b}{Measured flux densities. The quoted values are color corrected.}                              
\end{deluxetable}                                                                                             

%%%%%%%%%%%%%%%%%%%% TABLE 4 %%%%%%%%%%%%%%%%%%%%%%%%%%%%%%%%%%

\subsection{Disk properties} \label{sec:diskprops}
To estimate the basic disk properties, we fitted a single temperature blackbody model 
to the observed IR excess calculated as the difference between the measured flux 
densities and the predicted photosphere. Mid-IR light curves showed evidence 
of variability at several EDDs \citep{melis2012,meng2012,meng2015,su2019}. To avoid
combining data points potentially related to different disk states,
for the fitting process, we used only the weighted averages of W1--W4 band {\sl WISE} single 
exposure photometry obtained in the first observing window of the satellite's primary 
mission. While the AllWISE catalog combines data from the {\sl WISE} cryogenic and NEOWISE 
post-cryogenic survey phases, our selected measurements form a homogeneous and 
simultaneously obtained data set in all four {\sl WISE} filters. 
The uncertainties of the measured flux densities were calculated as the quadratic sum 
of the calibration errors (WISE All-Sky Release Explanatory Supplement 
Sect.~4.4\footnote{ \url{http://wise2.ipac.caltech.edu/docs/release/allsky/expsup/sec4\_4h.html\#CalibrationU}}) 
and the errors of the weighted averages. In the computation of uncertainties of the 
 excesses, the uncertainties of the predicted photospheric fluxes were estimated to be 
3\% and added quadratically to the previously derived errors.  

\begin{figure*}
\begin{center}
\includegraphics[scale=.45,angle=0]{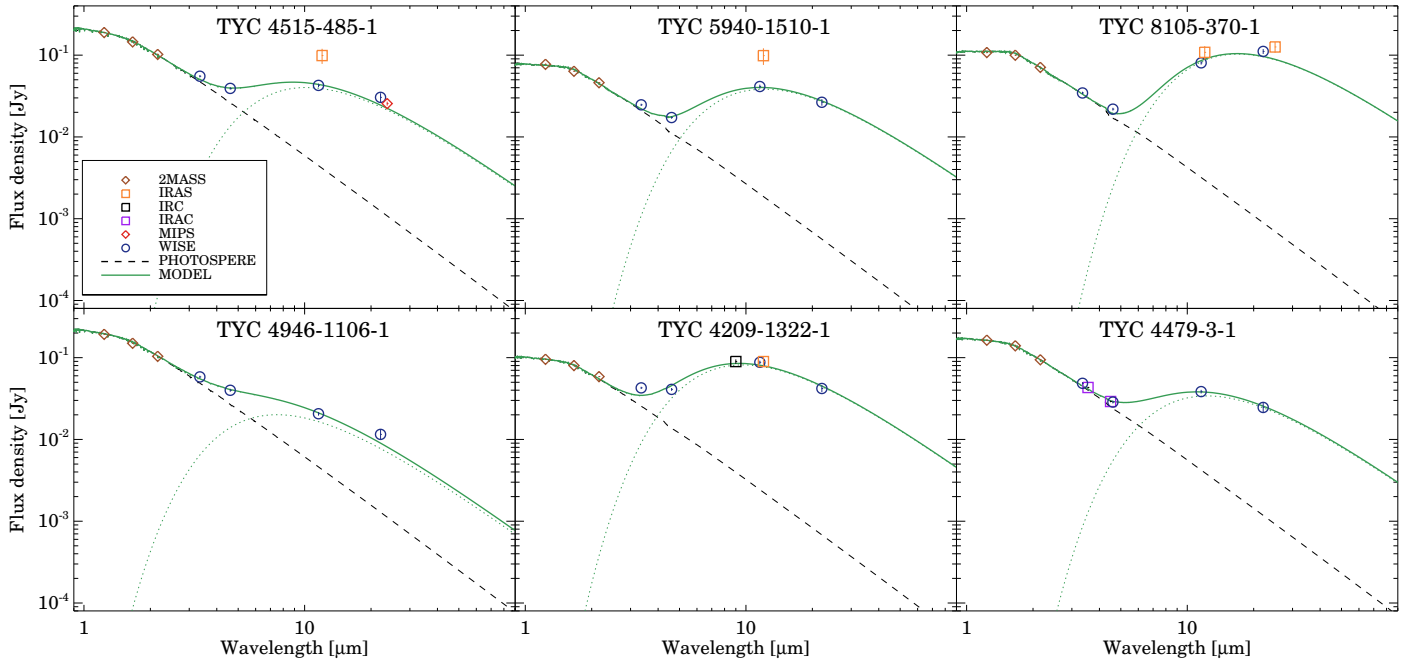}
\caption{{Color corrected SEDs overplotted with the photosphere and disk models. 
Model parameters are given in Table~\ref{tab:props}.}
\label{fig:sedplot}
}
\end{center}
\end{figure*}

Since the SED shapes differ from the $F_{\nu} \propto \nu^{-2}$ reference spectrum 
used in the {\sl WISE} catalog, the measured flux densities need to be color corrected. 
The color correction factors ($f_{\rm cc}$) were computed based on a simple SED
model (derived as the sum of the stellar photosphere and a blackbody fitted to the excess
computed from the original data). The $f_{\rm cc}$ values were less than 1\%, except the 
wide bandpass W3 measurements, where they could be as high as 10\%. The color corrected 
{\sl WISE} flux densities are plotted in Figure~\ref{fig:sedplot}.

The SED of the disk was computed as the excess above the photosphere (Fig.~\ref{fig:sedplot}).
In order to select the best-fitting blackbody model, we applied a Levenberg--Marquardt 
algorithm \citep{markwardt2009} using the characteristic dust temperature and the 
solid angle of the emitting region as free parameters. 
Color correction of other available IR data points (Sect.~\ref{sec:additionalirdata}) 
was also performed using the above models. Table~\ref{tab:sedtable} summarizes the color 
corrected infrared photometric data for all six objects. The obtained dust temperatures and 
fractional dust luminosities are listed in Table~\ref{tab:props}. 
Temperatures fall between 300 and 680\,K, while fractional luminosities range from 
0.01 to 0.07. Figure~\ref{fig:sedplot} shows, that the {\sl WISE} data points at 3.4\,{\micron} are 
systematically  
higher than the model fit. 
In the case of TYC\,4209, the measured flux density in the W1 band is 5.5\,$\sigma$ higher 
than the model fit, which likely indicates the presence of a hotter dust component in the disk.
For the rest of the sample, the deviations between the measured fluxes and the models 
are less than 3\,$\sigma$. Nevertheless, it cannot be excluded that some of these disks contain
additional even hotter dust as also suggested by 
color temperatures derived from W1, W2 data (see Fig.~\ref{fig:variability}).

Assuming an optically thin circumstellar dust ring with grains radiating like blackbodies, 
the ring radius can be calculated from the dust temperature and the stellar 
luminosity as ${R_{\rm bb}} [{\rm au}] = \left(\frac{L_{\rm star}}{L_{\sun}}\right)^{0.5} 
\left(\frac{278\,K}{T_{\rm dust}}\right)^2$.
The derived blackbody radii range from $\sim$0.3 to 0.7\,au (Table~\ref{tab:props}).

Several caveats are associated with our simple approach. Extreme debris disks 
tend to show strong solid state emission features \citep{olofsson2012}. 
Without any mid-IR spectral information, we do not know how the presence of 
such features modifies the shape of the SED. 
Furthermore, in the estimation of disk radii, we assumed large grains, 
behaving as blackbodies, and did not account for smaller particles that 
are ineffective emitters and thus become hotter compared to larger grains 
at the same location. The derived high fractional luminosity values ($\gtrsim$0.01)
indicate that the dust may be at least partly optically thick along the line 
of sight, which was also not considered in our simple model.

\section{Disk variability} \label{sec:diskvariability}
The available infrared data allow us to assess the mid-IR variability of the sources 
on several different timescales. Using single exposure {\sl WISE} data, we searched for 
flux changes on daily and annual timescales. For those sources where additional 
earlier IR photometry is available ({\sl IRAS}, {\sl AKARI}, {\sl Spitzer}) we could
also investigate variability on decadal timescales.

\subsection{Exploring the annual variability in W1/W2} \label{section:yearlyvariability}
To search for flux changes occuring on timescales of years, we used all single 
exposure W1 and W2 band data points not discarded in the quality check. 
We employed two different discriminants: the correlation-based Stetson 
$J$ index \citep[S$_{J}$,][]{stetson1996} and the scatter-based $\chi^2$ test 
\citep{sokolovsky2017}. By quantifying correlation of variability in two or more bands, 
the Stetson $J$ index provides a robust metric to search for brightness changes.
Involving the W1 and W2 bands, this index can be computed as:
\begin{equation}
S_J = \frac{\sum_{k=1}^n w_k sgn(P_k) \sqrt{|P_k|}}{\sum_{k=1}^n w_k},
\end{equation}   
where $n$ is the number of paired observations taken at the same time, $w_k$ is the 
weight of the $k$-th epoch (set to 1 uniformly in our case). $P_k$ is the product 
of the normalized residuals of two observations, 
$P_k = \left( \sqrt{\frac{n}{n-1}} \frac{W_{1,k} - \overline{W}_{1}} {\sigma_{W_{1,k}}} \right) 
\left(\sqrt{\frac{n}{n-1}} \frac{W_{2,k} - \overline{W}_{2}} {\sigma_{W_{2,k}}} \right),$
where $W_{1,k}$ and $W_{2,k}$ are the $k$-th measured magnitudes in the given bands, 
$\sigma_{W_{1,k}}$ and $\sigma_{W_{2,k}}$ are their uncertainties, while 
$\overline{W}_{1}$ and $\overline{W}_{2}$ are the means of the measured data points 
in the given bands.  
We also searched for variability in individual bands by computing the value of 
$\chi^2_{red}$ as
\begin{equation}
\chi^2_{\rm red} = \frac{1}{N-1} \sum_{k=1}^n \left( \frac{W_k - \overline{W}}{\sigma_{W_k}}  \right)^2, 
\end{equation}
where $W_k$ and $\sigma_{W_k}$ are the measured magnitudes and their uncertainties in 
the specific band. 

\begin{figure} 
\begin{center}
\includegraphics[scale=.50,angle=0]{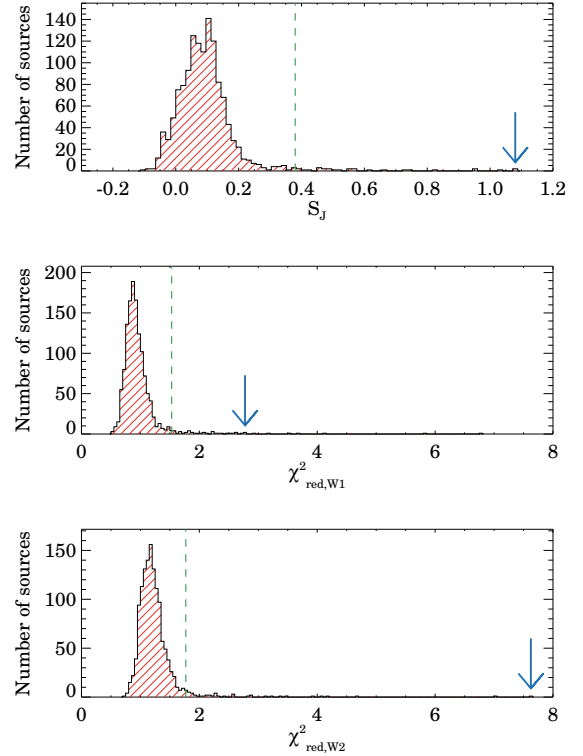}
\caption{ {Histogram of the Stetson, $\chi^2_{\rm red,W1}$, and $\chi^2_{\rm red,W2}$ variability 
indices computed for {\sl WISE} sources in the vicinity of TYC\,5940. Blue arrows show index values obtained for 
TYC\,5940, while vertical dashed green lines represent the derived cutoff values for 
the discrimination of non-variable/variable objects (see Sect.~\ref{sec:diskvariability}).
}
\label{fig:varindexexample}
}
\end{center}
\end{figure}

To define $S_J$ and $\chi^2_{\rm red}$ cutoff values for separating 
variable objects from non-variable ones, we used an empirical approach. We selected all objects 
from the AllWISE catalog that 1) are located within 2{\degr} of our targets (thus have similar 
time sampling), 2) have W2 band magnitudes within $\pm 1.0$\,mag to that of the specific EDD, 
and 3) are not marked by any contamination flag in the W1 and W2 bands ('cc\_flg'=0).
We gathered all single exposure photometric data for the selected objects and 
compiled their light curves by applying the same quality criteria as 
for our targets (see Sect.~\ref{sec:wisedata}). 
Finally we computed the $S_J$ and $\chi^2_{\rm red}$ values for the light curves and 
compiled the histograms of variability indices for all studied regions independently.
The obtained histograms typically show asymmetric behavior with a tail at positive values 
(see, e.g., Fig.~\ref{fig:varindexexample} for the case of TYC\,5940). 
Supposing that the majority of objects show no significant variations, this finding is consistent with 
our expectations. Variability in individual bands is expected to be accompanied with higher 
$\chi^2_{red}$ values. For non-variable objects with random noise, no correlation is expected between 
the different band observations thus the Stetson index should be close to zero, while in the case of 
the real correlated variability the index should be positive. 
To derive the mean and standard deviation of the obtained distributions ($\overline{D}, \sigma_D$) 
we performed an iterative 3$\sigma$ clipping process. Then the cutoff value was computed as 
$\overline{D} + 5\sigma_D$ (Fig.~\ref{fig:varindexexample}).

We found that the cutoff values differ from region to region, they range from 0.20 to 0.42 for the 
Stetson index, from 1.09 to 1.53 for the $\chi^2_{\rm red,W1}$, and 1.74 to 2.58 for the 
$\chi^2_{\rm red,W2}$ index. Interestingly, the means of the S$_J$ distributions were found to be 
larger than zero (0.08--0.1) in all cases. This is in line with the result of \citet{secrest2020}, 
who, examining the mid-IR variability of dwarf galaxies, found that the distribution of Pearson correlation 
coefficients between WISE W1 and W2 band photometry follows a normal distribution that is slightly 
shifted toward positive values (with a mean of 0.06 and a standard deviation of 0.09).
Table~\ref{tab:varindices} lists the obtained variability indices and the relevant critical 
cutoff values, the latter ones are in brackets. We identified four objects -- 
TYC\,8105, TYC\,5940,  
TYC\,4209, and TYC\,4479 -- that show significant variability. While the variation of
TYC\,8105 is limited to the W2 band, the other three systems exhibit flux changes in both {\sl WISE} 
bands in a correlated way. The light curves of TYC\,4515 and TYC\,4946 display no significant 
variability. By repeating the above procedure using only data points obtained in the NEOWISE Reactivation 
phase would not change our conclusions on the variability of the targets (see Table~\ref{tab:varindices}).

%%%%%%%%%%%%%%%%%%%% TABLE 5 %%%%%%%%%%%%%%%%%%%%%%%%%%%%%%%%%%

\begin{deluxetable}{l|CCCC}[h!!!!]                                               
\tablecaption{Variability indices \label{tab:varindices}} 
\tablecolumns{5}                      
\tablewidth{0pt}                                                                
\tablehead{                                                                     
\colhead{Name} & \colhead{$\chi_{\rm red,W1}$} & \colhead{$\chi_{\rm red,W2}$} & \colhead{$S_{\rm J}$} & \colhead{$\chi_{\rm red,W1-W2}$}}
\decimals                                                                       
\startdata                                                                      
\multicolumn{5}{c}{AllWISE + NEOWISE Reactivation data} \\
 TYC\,4515 &  0.99 (1.47)  & 1.76 (1.83)   & 0.21 (0.36) &  1.08 (1.48)\\
 TYC\,5940 &  2.78 (1.53)  & 7.61 (1.93)   & 1.08 (0.38) & 2.15 (1.55) \\
 TYC\,8105 &  1.25 (1.45)  & 2.51 (1.82)   & 0.20 (0.32) & 1.48 (1.49) \\  
 TYC\,4946 &  1.06 (1.48)  & 1.69 (2.58)   & 0.31 (0.42) & 0.98 (1.72) \\
 TYC\,4209 &  19.18 (1.09)  & 105.20 (1.74)& 5.72 (0.20) & 14.52 (1.19) \\
 TYC\,4479    &  2.37 (1.44)  &  8.96 (2.21)  & 1.08 (0.37) &  2.10 (1.60) \\
\multicolumn{5}{c}{NEOWISE Reactivation only} \\
 TYC\,4515  &  0.86 (1.44)  & 1.66 (1.98)   & 0.27 (0.40) & 0.88 (1.49) \\
 TYC\,5940 &  2.98 (1.49)  & 7.86 (2.01)   & 1.34 (0.40) & 1.82 (1.56) \\
 TYC\,8105  &  1.19 (1.30)  & 1.99 (1.77)   & 0.24 (0.32) & 1.19 (1.36) \\  
 TYC\,4946 &  1.01 (1.37)  & 1.50 (2.34)   & 0.25 (0.41) & 0.92 (1.61) \\
 TYC\,4209 &  20.46 (0.97)  & 113.04 (1.59)& 6.05 (0.19) & 15.56 (1.09) \\
 TYC\,4479 &   2.66 (1.35)  & 10.66 (1.93) & 1.34 (0.36) &  2.32 (1.48) \\ 
\enddata                                                                                                      
\end{deluxetable}  

%%%%%%%%%%%%%%%%%%%% TABLE 5 %%%%%%%%%%%%%%%%%%%%%%%%%%%%%%%%%%

We also examined whether the W1$-$W2 color index of our targets showed any significant variations
during the {\sl WISE} observations. By calculating the $\chi^2_{\rm red,W1-W2}$ values for our targets 
and for their comparison samples, and then using the above technique to set the thresholds, 
we found that the W1$-$W2 colors of TYC\,5940, TYC\,4209, and TYC\,4479 changed significantly 
between 2010 and 2019.

\begin{figure*} 
\begin{center}
\includegraphics[scale=.37,angle=0]{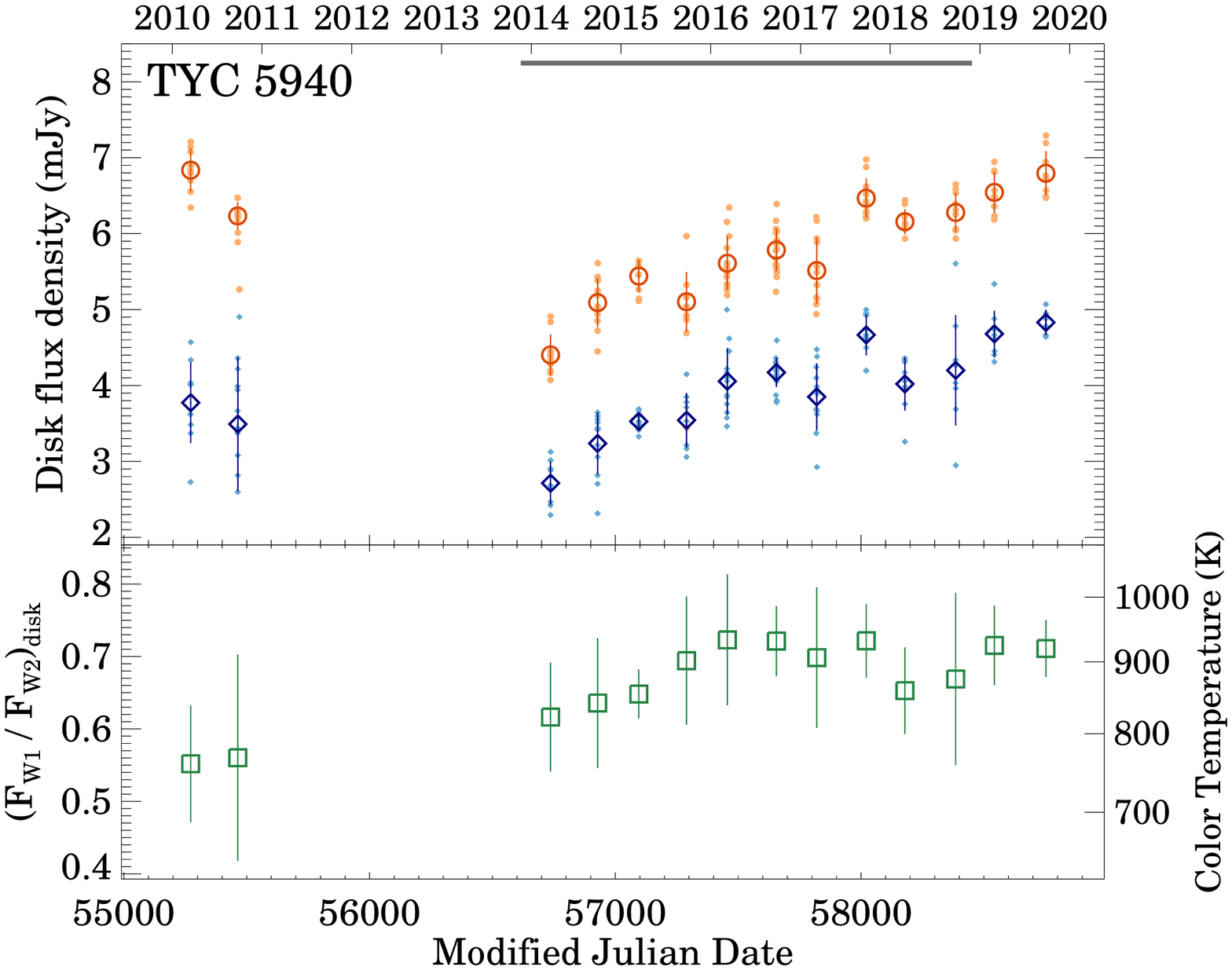}
\includegraphics[scale=.37,angle=0]{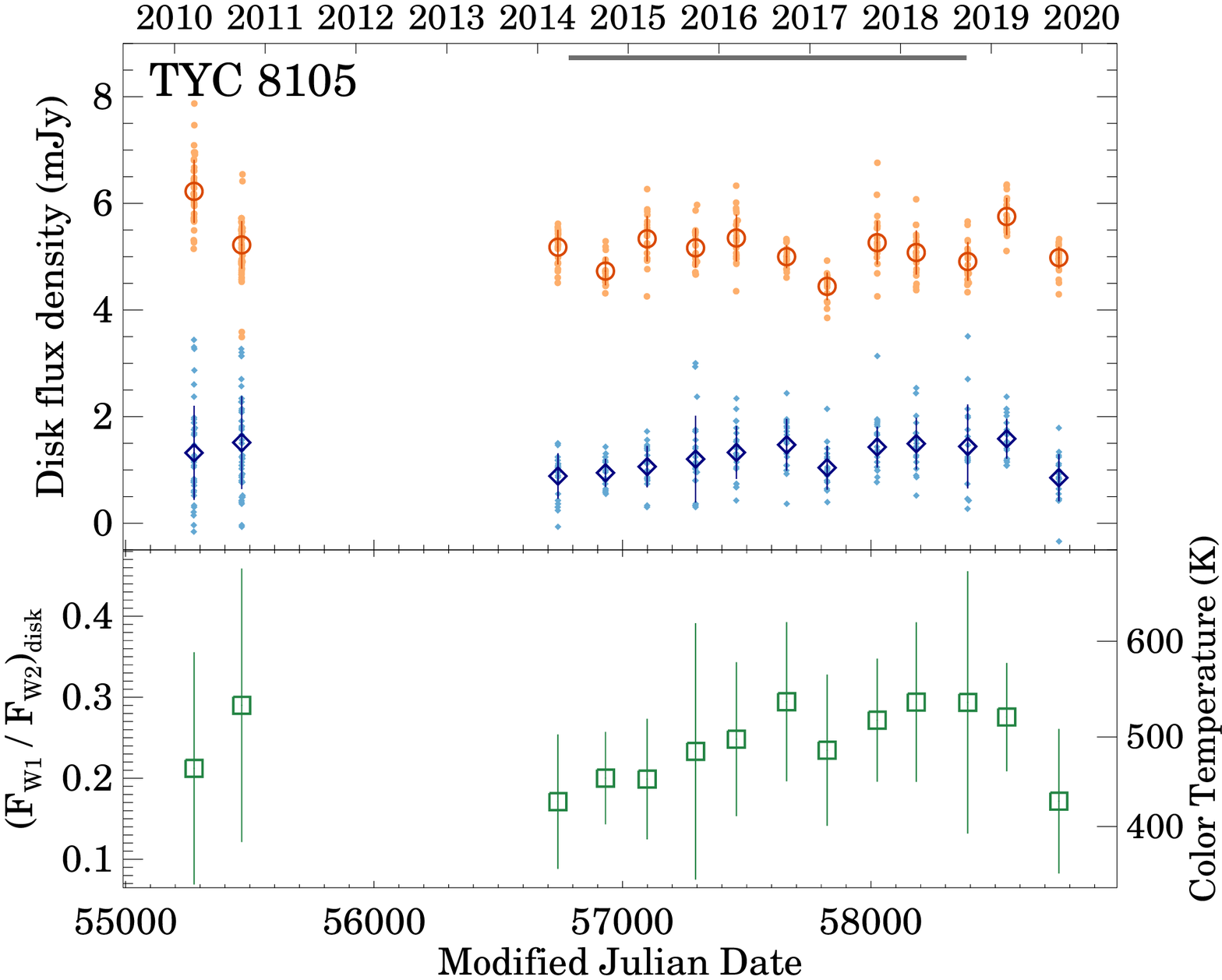} \\
\includegraphics[scale=.37,angle=0]{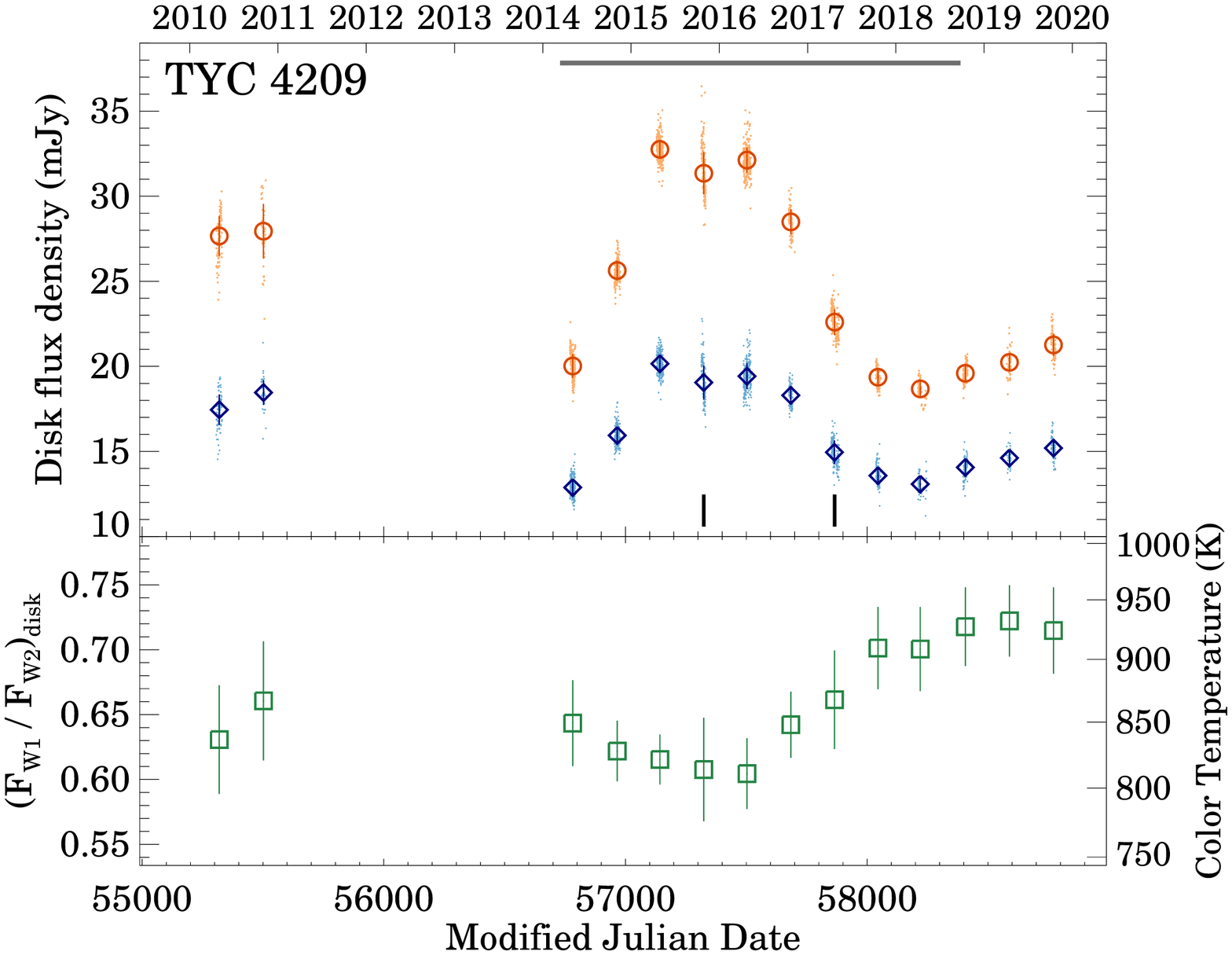}
\includegraphics[scale=.37,angle=0]{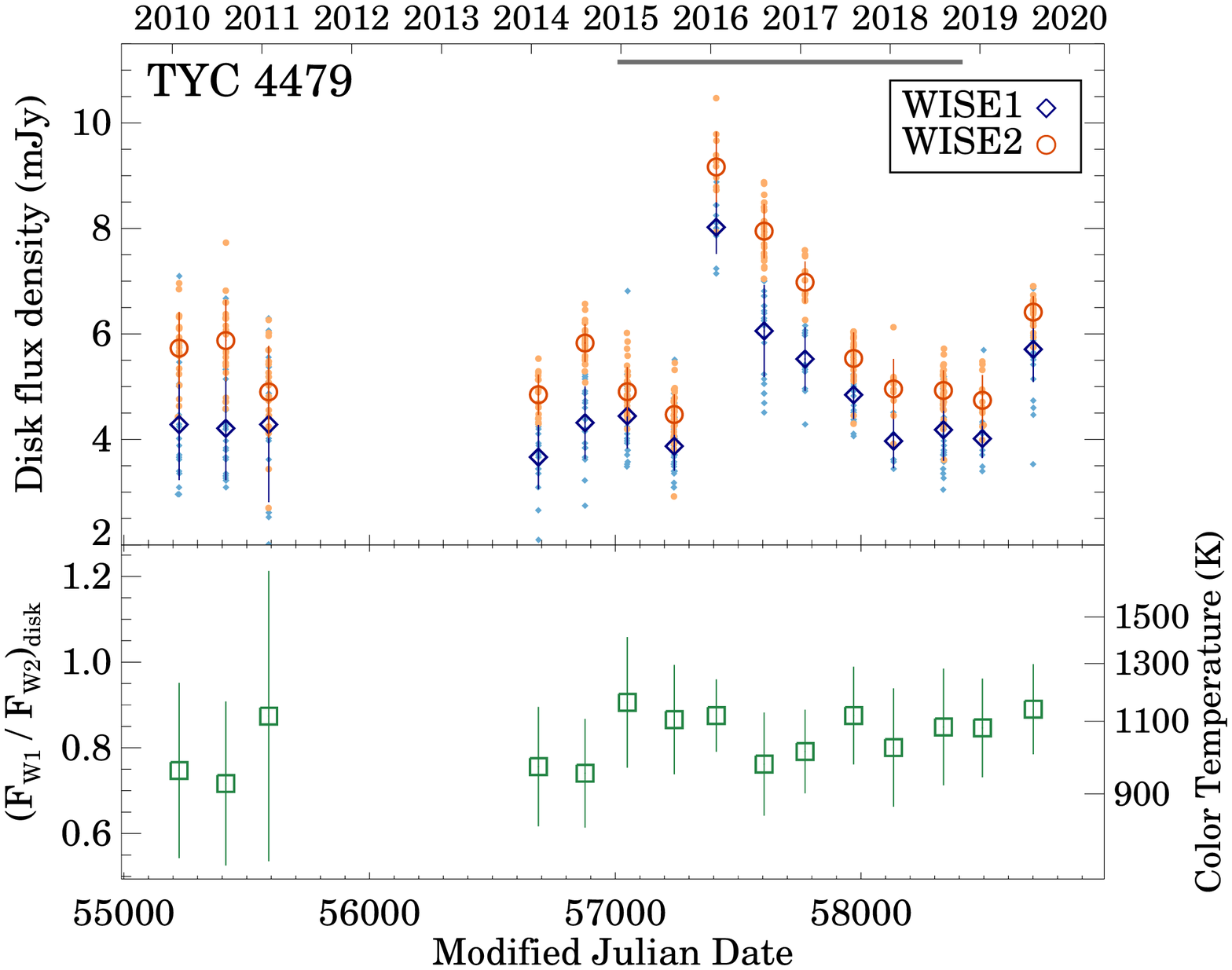} \\
\includegraphics[scale=.37,angle=0]{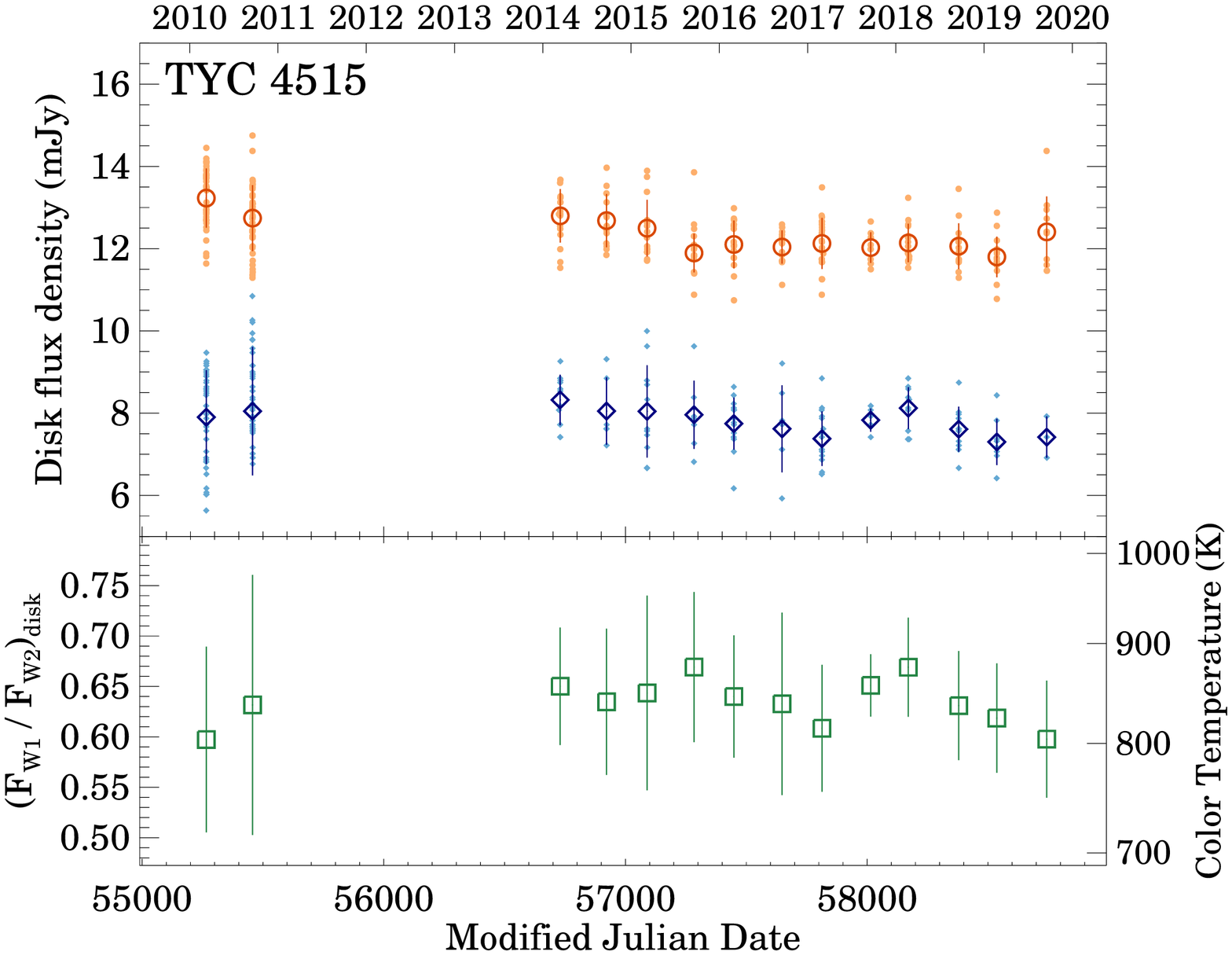}
\includegraphics[scale=.37,angle=0]{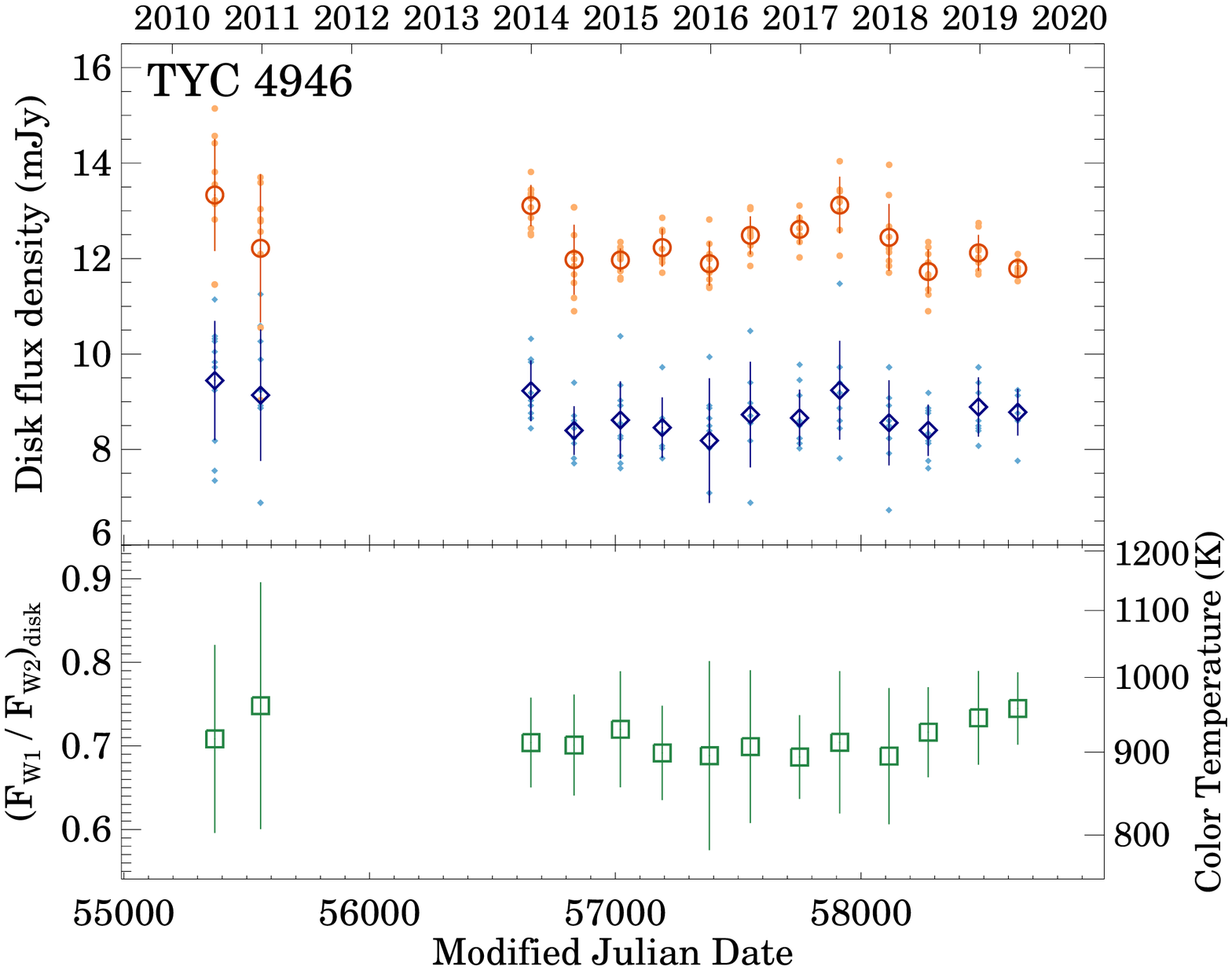}
\caption{ {WISE W1 (3.4\,{\micron}) and W2 (4.6\,{\micron}) band disk fluxes (the measured excess
emissions) for our targets between 2010 and 2019. Small dots show single exposure data points,
while larger symbols with error bars denote the seasonal averages. Horizontal gray lines show
the time ranges of available ASAS-SN photometric observations (Sect.~\ref{sec:opticalvariability}). 
Small vertical lines in the plot 
of TYC\,4209 mark those observational windows in which significant daily disk flux changes 
have been detected (Sect.~\ref{sect:dailyvar}).  
Ratios of W1 to W2 band seasonal averages are also shown (bottom panels) together 
with the corresponding color temperatures. 
}
\label{fig:variability}
}
\end{center}
\end{figure*}

Figure~\ref{fig:variability} shows how the disk fluxes in the W1 and W2 bands -- computed as the difference of 
the measured single exposure photometry and the photospheric flux densities -- of our targets   
changed between 2010 and 2019. In addition to the individual data points, their averages 
in each observing window (seasonal averages) were also plotted.  
In the course of averaging, an iterative sigma clipping algorithm was utilized where the 
clipping was set to 3$\times$ of the standard deviation. Formal uncertainties were computed 
from the dispersion of data points involved in the averaging. Ratios of the W1 and W2 band 
disk fluxes as well as the corresponding color temperatures were also derived and plotted 
in separate panels.

The four variable objects are displayed in the first four panels of
Fig.~\ref{fig:variability}. TYC\,5940 exhibited significant variability in both 
WISE bands. Following an initial drop, in 2014 the object started brightening, 
and the disk flux increased by 78\% in W1 and by 54\% in W2 over six years.  
TYC\,8105 displayed less pronounced variations, with no long term trend in the flux 
changes. According to our analysis, only the W2 band variability is formally significant, 
with a peak-to-peak amplitude of about 34\%. 
The disk of TYC\,4209 showed very prominent flux changes. After an intermediate flux level 
in 2010, its flux considerably dropped by 2014, which was followed by a dramatic brightening 
on a timescale of about one year. At 3.4 and 4.6\,{\micron} the disk became brighter 
by about 56\% and 64\%, respectively. Over the next one year the disk flux levels 
remained nearly constant, but after that a significant fading started. By late 2017,
the disk reached the flux level just preceding the brightening phase in 2014. The light 
curve between 2014 and 2017 was rather symmetric, the pace of rising and fading was roughly similar.
From 2018 the disk started brightening again.
TYC\,4479 was constant between 2010 and 2015. In early 2016, an abrupt flux rise by a 
factor of 2 occurred in less than 6 months. The peak was immediately followed 
by an exponential-like fading. By 2019, the source returned to its normal flux level, 
thus this asymmetric brightening event lasted for about 2.5 years. The last {\sl WISE} observation 
in 2019 August showed a higher flux level again. 
TYC\,4515 and TYC\,4946 exhibited no variability complying with the criteria of formal 
significance in our analysis. Nevertheless, the light curves of TYC\,4946 suggest a 
wave-like pattern between 2014 and 2019, repeated in both the W1 and W2 bands.

\subsection{Optical variability of the host stars} \label{sec:opticalvariability}
In the calculation of infrared excesses we assumed that the host stars 
did not exhibit variability trends on the timescales of our {\sl WISE} light curves 
in Fig.~\ref{fig:variability}. To evaluate this assumption and to exclude that 
the observed mid-IR flux changes could be driven by stellar variations, we examined 
the long term optical light curves of the host stars of the four variable objects.
The data were 
taken from the ASAS-SN survey \citep{shappee2014,kochanek2017} 
that provides V-band photometry for the whole sky with a cadence of $\sim$2--3\,days. 
These measurements allowed us to investigate the optical properties 
over only a part of the NEOWISE Reactivation mission (Fig.~\ref{fig:variability}). 
However, they are still relevant since the 
specific disks showed significant changes in mid-IR wavelengths during
this period as well.

Our analysis showed that all monitored stars were stable, their 
light curves were flat with root mean square (rms) noise 
$<$0.02\,magnitude. 
In the same time intervals, TYC\,5904, TYC\,4209, and TYC\,4479 displayed   
peak-to-peak variations of 0.14, 0.39, and 0.17\,mag in the W2 band averaged photometry, which thus 
 cannot be explained with variable stellar radiation. 
TYC\,8105 shows a more modest change in the W2 band with a peak-to-peak change of 
0.07\,mag during the NEOWISE Reactivation mission. Based on TESS data (Appendix~\ref{appendix:b}), 
within our sample, this star 
exhibits the largest rotational modulation. However, even in this case the 
amplitude is only 0.019\,mag. Moreover, the length of the {\sl WISE} observing windows are 
between 1.9 and 7.3\,days for this object, thus our averaging process within the windows 
mostly cancels the effect of the rotational variability, which has a period 
of $\sim$5\,days. A further indication that the observed W2 band change 
is associated to the disk is that the measured peak-to-peak change in the same time period 
in the W1 band -- where the disk's contribution is negligible -- was only 0.022\,mag.

\subsection{Exploring the hourly/daily variations} \label{sect:dailyvar}
To examine disk flux variations on shorter hourly/daily timescales, we turn to {\sl WISE} 
single exposures performed within a specific observational window. The length of individual 
data sets in these windows depends on the target's sky position. 
Due to its high ecliptic latitude of $\sim$87\degr, TYC\,4209 was observed 
for 17--34 days at a time. The other five targets have typical coverages 
shorter than a week. To identify objects showing significant variability 
on these shorter timescales, we used the same strategy as we described in 
Sect.~\ref{section:yearlyvariability}, i.e. the variability indices ($S_{\rm J}$ and 
$\chi^2_{\rm red}$ values) computed for our target were compared to those of nearby similarly 
bright sources that are supposed to be mostly non-variable. 

\begin{figure} 
\begin{center}
\includegraphics[scale=.34,angle=0,bb=20 16 370 481]{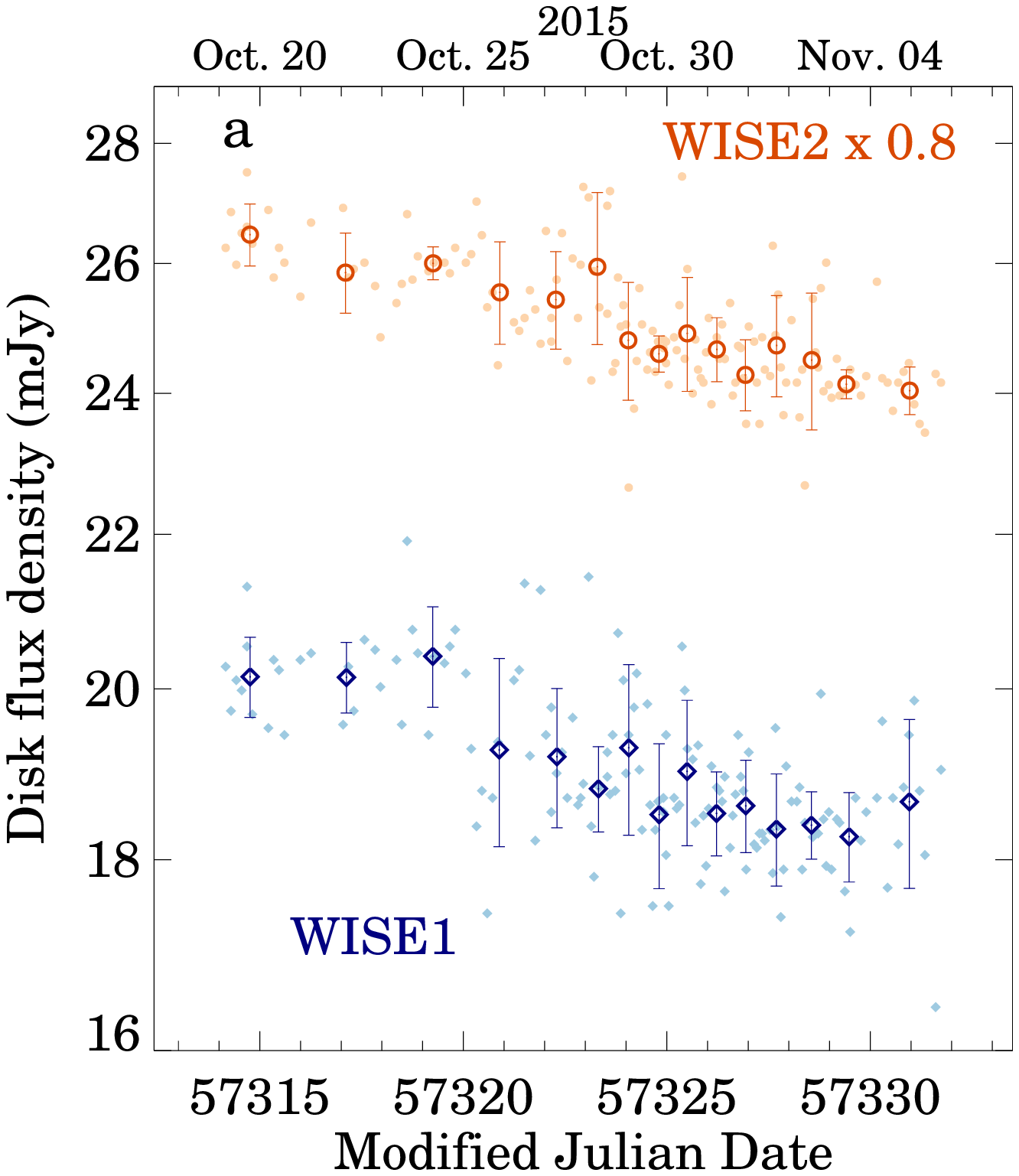}
\includegraphics[scale=.34,angle=0,bb=20 16 370 481]{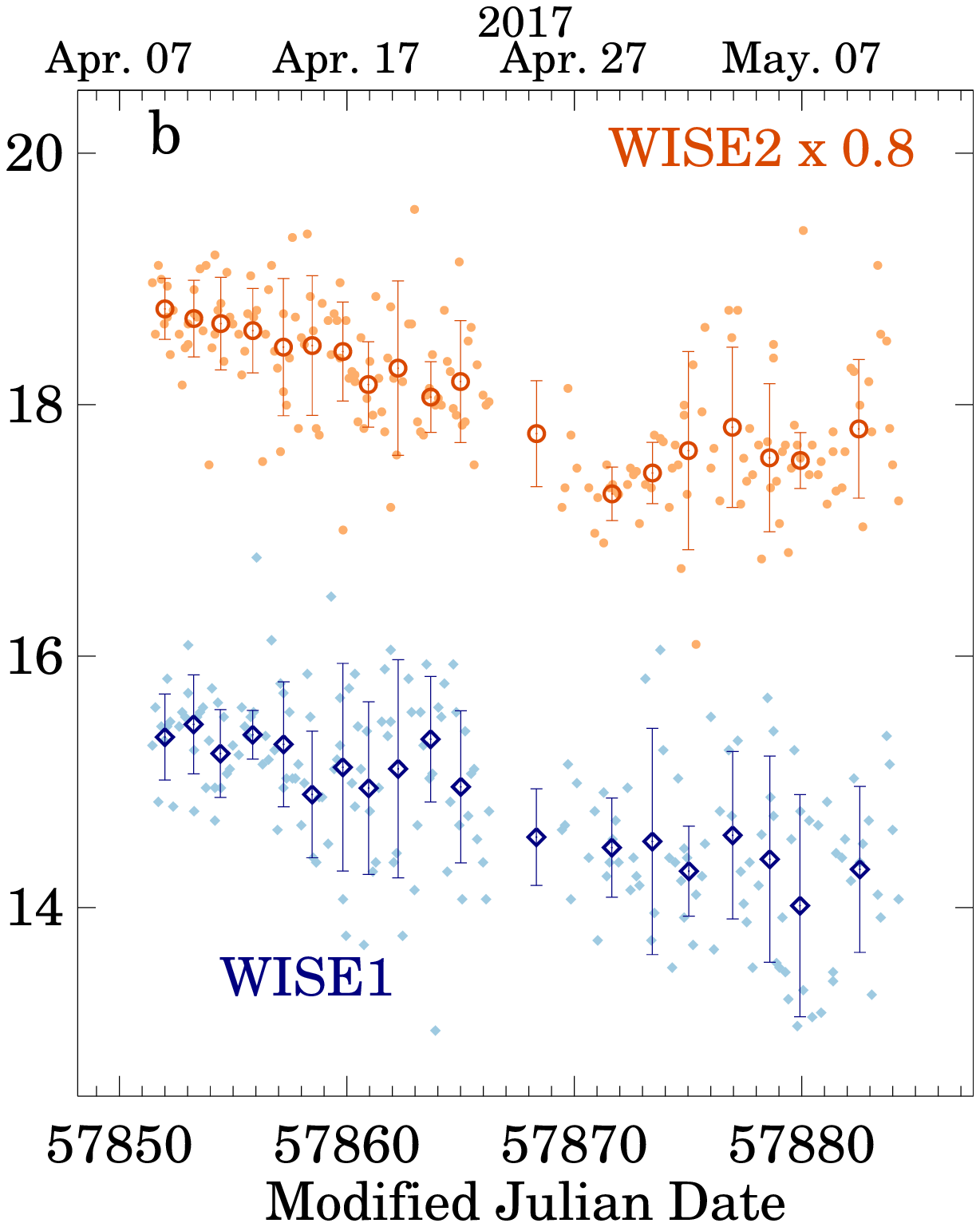}
\caption{ {WISE W1/W2 band disk fluxes of TYC\,4209 in the observational windows when significant 
flux changes were detected (Sect.~\ref{sect:dailyvar}). For better visualization, the W2 light 
curve was scaled by 0.8.
}
\label{fig:phasevar}
}
\end{center}
\end{figure}

In the W1 and W2 bands, where photometry was available for all epochs, we found 
significant variability only in one source, TYC\,4209, in two observational 
windows (2015 Oct 19 -- Nov 5; 2017 April 8 -- May 11; marked in Fig.~\ref{fig:variability}) 
in both bands. Note that this source has by far the broadest observing window in our sample. 
The {\sl WISE} light curves are displayed in Fig.~\ref{fig:phasevar}, showing both the individual 
single exposures and ten-point binned values. In both windows, a fading 
trend can be recognized. The observed trends were characterized by fitting a 
straight line to the binned data considering their error bars. These yielded 
slopes of $b_{W1}=-53 \pm 11$\,mJy~yr$^{-1}$ and $b_{W2}=-75 \pm 11$\,mJy~yr$^{-1}$ in the first 
window and $b_{W1}=-17 \pm 4$\,mJy~yr$^{-1}$ and $b_{W2}=-23 \pm 4$\,mJy~yr$^{-1}$ in the second 
window. All obtained slopes are significantly different from zero. We compared
these slopes with the long term flux changes in Fig.~\ref{fig:variability}. 
In the first window, our result is consistent with the fading since the  
previous window, but its pace is much higher. Extrapolating the rapid fading 
would result in the disappearance of the excess on a timescale of $\sim$0.5\,year. 
In contrast to this, the source became brighter by the next window 
$\sim$6\,months later. In the second window, the observed trend qualitatively matches 
the long term fading behavior, however the fading rate is faster. 
Similar few weeks long rapid variations are seen in {\sl Spitzer} light 
curves of other EDDs (ID\,8, P\,1121) of better sampling \citep{su2019}. In those objects, 
the quasi-periodic changes were found and
interpreted as orbital evolution of an optically thick dust cloud \citep{su2019}. 

In the W3/W4 bands, our data are limited to 1--2 observing windows in 2010--11.
Using the same strategy as above, we found no significant daily variability in any 
of the sources.

\subsection{Search for decadal changes} \label{sec:longterm}
For five objects, additional IR photometry is available from different 
space missions (Table~\ref{tab:sedtable}). In those four cases where IRAS 
measurements also exist, the data cover $\sim$30\,years. These additional data 
allow us to check for long-term variability. Fig.~\ref{fig:sedplot} shows the data points, 
together with the {\sl WISE} photometry and the models fitted to the {\sl WISE}
data (Sect.~\ref{sec:diskprops}). In most cases, the additional data match the 
model curves well. In TYC\,4515 and TYC\,5940, the IRAS 12\,{\micron} data points 
exceed the models, however, even in these cases the significance of the excess 
remains below 3$\sigma$. Moreover, in TYC\,5940 the 12\,{\micron} excess might 
be contaminated by a nearby source (Sect.~\ref{sec:additionalirdata}). Thus, the available observations do 
not support any long-term IR variability exceeding the formal uncertainties in 
our sample.

\section{Discussion} \label{sec:discussion}

\subsection{Steady-state or transient dust production?} \label{sec:originhotdust}
As a first step in analyzing the nature and origin of the observed abundant warm debris 
material around our targets, we examined whether it could be produced in a steady-state 
grind down of an in situ planetesimal belt, a massive exosolar analog of our Solar System's 
asteroid belt. If located close to the star, the evolution of such inner planetesimal 
belts is very rapid due to the short collisional timescale. Based on the analytical 
steady state evolutionary model of \citet{wyatt2007}, at any given age there is a maximum 
possible fractional luminosity ($f_{\rm d,max}$) of a debris ring. Following \citet{wyatt2007} 
we adopted a belt width of $dr = 0.5r$, a maximum asteroid size of 2000\,km, typical planetesimal 
strength of $Q_D^* = 200\,$J\,kg$^{-1}$, and planetesimal eccentricity of 0.05, which resulted in 
\begin{equation}
\begin{split}
  f_{\rm d,max} = 1.6\times10^{-4} 
     \left( \frac{R_{\rm disk}}{\rm 1\,AU} \right)^{7/3} 
     \left( \frac{M_*}{M_\odot} \right)^{-5/6} \\
     \left( \frac{L_*}{L_\odot}\right)^{-1/2} 
     \left( \frac{t}{\rm Myr} \right)^{-1}.
\end{split} 
\end{equation}      
\citet{wyatt2007} argued that, even by considering the uncertainties in their model, debris 
rings with $f_{\rm d}$ /$f_{\rm d,max}$ ratios higher than 1000 could not be produced 
through steady-state processes, but should be linked to some transient event instead. 
For computing 
$f_{\rm d,max}$, stellar and disk parameters were taken from Table~\ref{tab:props}. By 
assuming that the collisional cascade started early, for $t$ we adopted the age of the 
systems (Table~\ref{tab:props}).
For our targets, the ratios of the measured fractional luminosities to the theoretical maxima 
(Table~\ref{tab:props}) range between $\sim$8$\times$10$^4$ and 2.6$\times$10$^6$.   
We note, however, that $f_{\rm d,max}$ strongly depends on the disk radius 
which is not reliably known.
In Sect.~\ref{sec:diskprops} we used a simple blackbody model, that can lead to 
an underestimation of the disk radii. Based on debris disks spatially resolved by the 
{\sl Herschel Space Observatory}, in systems with Sun-like host stars, the ratios of the true 
disk radii to the blackbody radii are $<$5 \citep[fig.~4b in][]{pawellek2014}.
Entering 5$\times$ larger radius values into the equation results in $\sim$43$\times$ 
higher $f_{\rm d,max}$ values. However, even this way we still get $f_d/f_{\rm d,max}>1800$ 
for all sources, implying they are probably experiencing a transient phase.

\begin{figure*}
\begin{center}
\includegraphics[scale=1.05,angle=0]{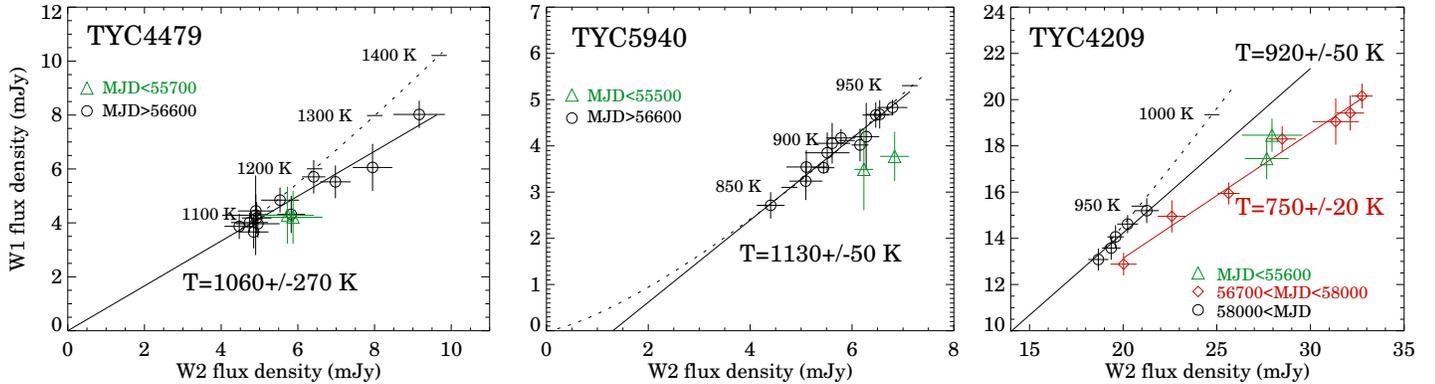}
\caption{Correlation of the WISE 3.4\,{\micron} and 4.6\,{\micron} flux 
densities using the seasonal averages of three EDD formally exhibiting 
variability (Sect.~\ref{sec:diskvariability}, Fig.~\ref{fig:variability}). Solid lines 
represent models of constant temperature with varying dust emitting surface, while dotted 
lines denote the emission of a dust population of increasing temperature with constant 
emitting surface. 
\label{fig:varorig}}
\end{center}
\end{figure*}

\subsection{The origin of disk variability} \label{sec:originvariability}

Figure~\ref{fig:variability} shows that in most objects, parallel to the detected mid-IR flux 
changes, the color temperature derived from the ratio of the 3.4\,{\micron} and 4.6\,{\micron} fluxes 
was also variable. In order to identify the physical processes behind these variations, we 
correlated the W1 and W2 multiepoch fluxes, by plotting the seasonal averages in Fig.~\ref{fig:varorig}. 
The photometric point distributions were then confronted with the predictions of two simple models: 
the first one assumes that the temperature of the circumstellar dust grains can change in time, 
while the second one keeps the temperatures constant but varies the total emitting dust surface area. 
Variability caused by the latter process would appear in Fig.~\ref{fig:varorig} as a straight line 
whose slope is related to the invariable temperature; while a temperature rise in the first model 
would outline a curve starting from the origin.

The distribution of the {\sl WISE} data points of TYC\,4479 in Fig.~\ref{fig:varorig} (left) is consistent 
with a line crossing the origin within the formal fit uncertainties. The derived 3.4/4.6\,{\micron} 
color temperature is 1060\,K. The points are inconsistent with a model where the temperature of the 
dust grains would increase above 1060\,K, because it would have made the resulting emission bluer 
which is not seen in the data (dotted curve). Thus, for the observed variability of TYC\,4479, the 
most plausible model is the change of the effective surface area. It may be caused by production of 
fresh dust, but it may also be due to geometrical rearrangement of the ring structure, which would 
temporarily decrease the dust column density and make the IR emission less optically thick. The 
minimum emitting area in the faintest state (assuming optically thin blackbody radiation) was 
0.00023\,au$^2$, while at 
the brightness peak it rose to 0.00046\,au$^2$, doubling the dust surface within a 
few months. Note that if the dust is partly optically thick then these estimates 
are lower limits. Taking into account the luminosity of the star (Tab.~\ref{tab:props}), 
the blackbody radius of the 1060\,K dust is $\approx$0.06~au.  

The infrared time variability of TYC\,5940 (Fig.~\ref{fig:varorig}, middle) can be explained by 
both of our simple models. The data points form a straight line, which, however, does not reach 
the origin. The non-zero intercept is significant at the 5$\sigma$ level. The data are also compatible 
within the measurement uncertainties with a model of increasing dust temperature (dotted curve). 
Thus, one possible interpretation is that we see a dust population, located at 0.06~au from the 
star at an equilibrium temperature of 1130\,K, which changed its emitting area between 0.00034~au$^2$ 
and 0.0006~au$^2$ over the observed period. The non-zero intercept may point to the existence of a 
colder dust population (probably at larger stellocentric radius), whose thermal emission at 
3.4\,{\micron} is still negligible, but at 4.6~$\mu$m becomes detectable. An equally plausible model 
is, however, that the area of the dust ring remained constant while its temperature increased from 
$\approx$850~K to $\approx$950~K. 
The origin of the energy needed for the extra heating is unlikely 
to arrive from the star (in Sect.~\ref{sec:opticalvariability} we argued about a low level of stellar variability), 
raising a question about the heat source.

The most complex case of variability is TYC\,4209. This source exhibited a broadly symmetric brightness 
peak in 2014--17 (Sect.~\ref{sec:diskvariability}), followed by smaller scale variations. The two periods show different 
color temperature behaviors (Fig.~\ref{fig:variability}). In order to examine the physical reasons 
of variability separately, we plotted in Fig.~\ref{fig:varorig}  (right) the data points of the 
two periods with different symbols. The flux changes after the brightness peak (MJD$>$58\,000, 
black circle) can be modeled with a straight line reaching the origin within the uncertainties of the 
fit parameters. It would imply an increasing emitting surface of a T=920\,K dust population, located 
at 0.11~au from the star, from 0.0041~au$^2$ to 0.0047~au$^2$. The observations are also marginally 
consistent with the warming up of this dust ring from $\approx$920~K to $\approx$950~K. The 2014--17 
brightness maximum, however, seems to be intrinsically different from the previous process. It is 
clearly inconsistent with any variable dust temperature model. Since its emission appears as an 
addition to the flux of the 920~K dust ring, we suggest that we witness the appearance and subsequent 
disappearance of a cloud of fresh dust at a temperature of T$\approx$750~K, with an emitting 
area of 0.006\,au$^2$. This temperature corresponds 
to a radius of 0.17~au, significantly farther than the first ring. Outside the most opaque core of the 
T=920~K ring the optical depth of the circumstellar dust is probably below unity, thus, a geometrical 
rearrangement of the dust at r=0.17~au would be an unlikely cause of the emission peak, leaving 
the formation of new dust via collisions as the most likely physical picture.

The last, formally variable object, TYC\,8105 could not be analysed with these techniques, because 
of the large error bars and the limited dynamic range of the data points.

The temperatures yielded above are higher than the characteristic 
dust temperatures derived from the blackbody fit to the 3.4--22\,{\micron} 
band IR excess (Sect.~\ref{sec:diskprops}). It may indicate that these 
disks present a range of dust temperatures and
their dust material is distributed over a broad radial range.

\subsection{The nature of the EDD phenomenon}
In addition to the 6 systems studied in this paper, 11 additional EDDs
are claimed in the literature (Appendix~\ref{appendix:c}).
Apart from HD\,113766 and HD\,145263, where the host stars are F2-type 
dwarfs, all these disks surround solar-type (F5-K3 type) stars.
The nature of two of these additional disks is still debated. RZ\,Psc was 
found to exhibit a weak accretion signature, 
suggesting that the disk may be of primordial nature rather than EDD \citep{potravnov2017,potravnov2019}.
Concerning HD\,166191, \citet{kennedy2014} raised the possibility that it is not 
a secondary disk but a primordial transition disk. However, based on recent millimeter continuum and 
line observations of this system, both the measured low CO gas mass and the low gas-to-dust mass ratio 
are more consistent with a debris disk nature \citep{garcia2019}.
Despite their debated classification, both objects are included in our analysis.
The case of TYC\,8241-2652-1 is quite special: before 2010 this young K-type star hosted 
a warm ($\sim$450\,K),  
unusually dust-rich disk, whose mid-IR luminosity then decayed significantly over less than 
two years, leaving behind a colder ($<$200\,K), tenuous disk \citep{melis2012}.  
Recent observations indicated that the disk has remained depleted \citep{guenther2017}.
Table~\ref{tab:knownedds} shows the parameters valid before 2009, 
which were consistent with an EDD.
In the following, we will consider the whole sample of EDDs, including all 
sources, proposed in the literature and the ones in this paper, to examine their 
 mid-infrared variability, age distribution, and the multiplicity of the host
 stars. We will also assess what these characteristics indicate about the 
 nature of EDDs.

\subsubsection{Mid-infrared variability of EDDs} \label{sec:midirvar}
Many of the previously known 11 EDDs were targets of multi-year photometric monitoring campaigns 
performed by {\sl Spitzer} at 3.6\,{\micron} and 4.5\,{\micron}. These studies 
revealed significant variability on monthly to yearly timescales -- in some cases with 
very complex variability patterns -- in seven of them: HD\,15407, HD\,23514, 
BD+20\,307, ID\,8, P\,1121, HD\,113766, and HD\,166191 \citep{meng2014,meng2015,su2019,su2019b,su2020}.
HD\,145263 was found to show no significant disk variations based 
on its {\sl Spitzer} observations in 2013 \citep{meng2015}. 

In addition to {\sl Spitzer} data, {\sl WISE} measurements 
are available for all 11 sources allowing us to study their mid-IR flux changes 
over the AllWISE and NEOWISE Reactivation mission phases between 2010 and 2019. 
The steps and results of this analysis are presented in Appendix~\ref{appendix:c}.
The three brightest sources, HD\,15407, HD\,113766, and HD\,166191, are saturated in both 
bands, preventing us from drawing reliable conclusions regarding 
their variability (we note that their {\sl Spitzer} data point to substantial variability). 
A similar issue affects the W1 photometry of HD\,145263 and BD+20\,307. The W2 light 
curves of the latter two objects displayed no significant variability over the given 
period.

As Figure~\ref{fig:knowneddsvar} demonstrates, for HD\,23514, ID\,8, and P\,1121, {\sl WISE} measurements 
confirm the previous Spitzer-based results, proving that 
these disks exhibit well detected flux changes between 3 and 5\,{\micron}. While during the 
{\sl Spitzer} 
observations in 2013, the disk of HD\,23514 showed a fading 
trend, its NEOWISE Reactivation measurements indicate a slow brightening since 
2014. 
For ID\,8, {\sl WISE} measurements from 2018--2019 complement the previous {\sl Spitzer} data,
showing that the disk continues to display significant changes. 
The {\sl WISE} light curves of RZ\,Psc \citep[see also in][]{kennedy2017} and 
V488\,Per show evidences of strong variability in both the W1 and W2 bands. It is noteworthy that 
between January and August 2019, the disk of V488\,Per became $\sim$7 and $\sim$4 times brighter 
at 3.6 and 4.5\,{\micron}, respectively, and it is now brighter at these wavelengths 
than ever observed before.
This is the largest increase we have seen in any EDDs at any given 
six month interval.
Although the {\sl WISE} fluxes of TYC\,8241-2652-1 did not show significant
changes between 2010 and 2019, before 
this period, its disk underwent a dramatic fading \citep{melis2012}.

Thus, in the case of the previously known EDDs there were significant flux changes in 10 out of 
the 11 sources, while in our sample 4 disks out of the 6 proved to be variable.  
In summary, 14 out of the 17 EDDs have shown significant variability in the wavelength range of 3 to 5\,{\micron} 
over the past roughly ten years. Flux changes can be observed even in the oldest EDD systems, 
both TYC\,4209 and TYC\,4479 exhibit substantial variations.
These results indicate that the variability phenomenon is an inherent characteristic of EDDs.

Although the light curves of EDDs show a large variety in their shape, some of them exhibit similar patterns. 
Between 2015 and 2019, the disks of HD\,23514 (Fig.~\ref{fig:knowneddsvar}) and TYC\,5940 
(Fig.~\ref{fig:variability}) showed analogous slowly rising trends both at 3.4 and 4.6\,{\micron}. 
The light curves of ID\,8 display a wavy pattern (Fig.~\ref{fig:knowneddsvar}), the
successive brightening and fading periods occur on a yearly timescale and probably represent 
the aftermath of violent impact events \citep{su2019}. As Fig.~\ref{fig:variability} 
demonstrates, the {\sl WISE} data of TYC\,4209 and TYC\,4479 show similar ripples, although 
in these cases, 
in contrast to ID\,8, the rising phase is steeper than the declining one. These bumps
might also be attributed to significant dust releasing events. In the case of ID\,8, 
\citet{su2019} found that the observed 4.5\,{\micron} flux change associated with the decay 
in 2013 corresponds to a total cross section change of 0.0021\,au$^2$. 
Based on {\sl WISE} data, the inferred cross section change at TYC\,4209 is $\sim$3$\times$ 
higher, while at TYC\,4479, it is $\sim$9$\times$ lower (Sect.~\ref{sec:originvariability}).

\subsubsection{Wide separation companions} \label{sec:widecompanions}
Examining a smaller EDD sample available at that time (five disks 
around 30--100\,Myr old solar-type stars), \citet{zuckerman2015} 
suggested that these disks are situated preferentially in wide binaries.
Considering that five of our disks are also found in 
similar systems, it is worth looking again at this possible relationship 
using the now much larger sample. 
As it is mentioned above, out of the previously known 11 EDDs, 9 have solar-type 
(F5-K3 type) host stars, from which so far five are proven multiples. 
BD+20\,307 is composed of two late F-type main-sequence stars in a close orbit 
\citep[P$\sim$3.4\,day,][]{weinberger2008,zuckerman2008}. 
Using Gaia\,DR2 astrometric data, \citet{hartman2020} identified a new common 
proper motion companion of this star at a projected separation of 8\farcs4 
(980\,au). This wide companion 
is classified as a white dwarf candidate with a mass of 0.48--0.58\,M$_\odot$ 
\citep[depending on whether a pure-He or pure-H atmosphere is supposed,][]{fusillo2019}.
Recently, \citet{kennedy2020} reported the discovery of a low mass companion of RZ\,Psc 
at a projected separation of 23\,au.
The other 
three stars, HD\,15407, HD\,23514, and V488\,Per, are all reported to have wide-orbit companions, 
the projected separations are $\rho_{\rm p} =$ 1050, 365, and 12300\,au, respectively 
\citep{melis2010,rodriguez2012,zuckerman2015}. V488~Per possibly hosts a second low-mass 
companion at $\rho_{\rm p}\sim12\,000$\,au \citep{zuckerman2015}.

Using the method and requirements we described in Sect.~\ref{sec:companions}, we searched 
the Gaia EDR3 catalog for previously unknown common proper motion pairs of these nine 
stars. This confirmed the findings of \citet{hartman2020} for BD+20\,307 and 
yielded new faint candidate pairs for three more stars, V488\,Per 
(five candidates, all with $\rho_{\rm p} > 0.39$\,pc), HD\,23514 (one candidate, $\rho_{\rm p} = 0.41$\,pc), and 
P\,1121 (one candidate, $\rho_{\rm p} = 0.74$\,pc). 
However, all of these stars are members of young open clusters (V488\,Per: $\alpha$\,Per open cluster, 
 HD\,23514: Pleiades, P\,1121: NGC\,2422). 
By studying wide binary systems in the $\alpha$\,Per, Pleiades, and Praesepe open clusters, 
\citet{deacon2020} demonstrated that at projected separations $>$3000\,au 
it is difficult to separate true binaries and unrelated cluster members. 
Although it cannot be completely ruled out that some of the revealed candidates 
are true binaries, the separations suggest that they instead fall into the 
latter category. Therefore, these were not taken into account in our further 
analysis. Based on the same arguments, \citet{deacon2020} do not list 
V488\,Per as a probable wide binary, because, due to their large separations, 
the previously reported companions (at $\rho_{\rm p}\sim$12000\,au, see above) 
could also be unrelated cluster members. 
Therefore, in the following, we do not consider this system as a justified binary.

Putting it all together with our results, 
out of the 15 EDDs with solar-type hosts, there are 8 systems (53.3$^{+11.6}_{-12.4}$\%)
with wide separation (365--6010\,au) companions. Remarkably, all of them are older 
than 100\,Myr. Thus, at ages $<$100\,Myr (5 EDDs out of the 15) no wide binaries were found, 
while at ages $>$100\,Myr (10 EDDs), the multiplicity fraction is 80$^{+7.1}_{-17.2}$\%. 
Considering the small sample, the uncertainties were derived by adopting a binomial 
distribution \citep{burgasser2003}.
For the majority of the EDD systems the information about wide companions 
is based on data from the Gaia EDR3 catalog. There are 
possible limitations of this approach. By testing the small-scale 
completeness of this data release based on source-pair distances in a 
small dense field near the Galactic plane, \citet{fabricius2020} claimed 
that the completeness falls rapidly at separations less than 0\farcs7. 
This value corresponds to $\sim$300\,au at P\,1121, 
which is the farthest object in the sample. Of course the detection sensitivity of possible 
companions depends not only on the angular separation but on the brightness ratio 
of the pair as well \citep{brandeker2019}. Even in pairs with large angular 
separations, it is possible that the companion is too faint to have a 
proper astrometric solution in the current Gaia data release. 
Future observations with ground-based high-contrast facilities and with Gaia, 
have the potential to reveal additional fainter and less wide 
companions in these systems.

In order to put these results into context, we need to know 
the incidence of similar binaries among solar-type stars. 
Based on the log-normal distribution derived by \citet{raghavan2010} for separations 
of companions around F6--K3 type main-sequence stars, about 13\% of such dwarfs 
have at least one companion at separations between 300\,au and 1\,pc. 
This fraction is $\sim$4$\times$ lower than what we observe among solar-type EDD stars.
By applying a binomial test (using the R statistical programming language's 
\texttt{binom.test} function) 
we obtain a probability of 2.3$\times$10$^{-4}$ for the null hypothesis 
that the fraction of wide binaries in the EDD sample is equal to or lower than in
the comparison sample. If we consider only objects older than 100\,Myr from the EDD sample, 
the contrast is even stronger, the observed wide companion fraction in that subsample 
is $\sim$6$\times$ higher than among normal solar-type stars. In this case, a binomial test 
yields a probability of 3$\times$10$^{-6}$ for the abovementioned null hypothesis, 
i.e. it could be rejected with a high statistical reliability.

However, while most EDD host stars are likely younger than 300\,Myr,
the comparison sample predominantly contains older stars. This must be taken into account 
in the comparison, since the frequency of weakly bound wide binaries can decline 
with time as a result of gravitational influence due to encounters with other stars 
and giant molecular clouds \citep{weinberg1987}. 
In a recent survey of the 10\,Myr old Upper Scorpius association, \citet{tokovinin2020} 
found that the fraction of 100--10$^4$\,au pairs with solar type primaries 
resembles that of stars in the field. By scrutinizing members of nearby young moving groups 
(5--100\,Myr old), \citet{elliott2015} also claimed that the multiplicity frequency of solar 
type stars in the separation range of 10--1000\,au is similar to that in the field star 
sample compiled by \citet{raghavan2010}. \citet{deacon2020} confirmed these results by reaching 
the conclusion that the fraction of pairs with separations of $>$300\,au among
FGK-type stars belonging to nearby 10--200\,Myr old young moving groups 
and to the $\sim$125\,Myr old Pisces-Eridanus stream shows no significant 
excess over that of similar 
type older field stars. In the same study, they showed that compared to these 
low density formation environments and field stars, there appears to be a deficit of wide (300--3000\,au) 
binaries in open clusters. Actually, the $\alpha$\,Per, Pleiades, and Praesepe clusters
 have an average binary fraction of 3\% (with FGK-type primaries) in 
 this projected separation range. We note that in the 'old' ($>$100\,Myr) EDD sample
the corresponding fraction is 6/10.
Based on these results, there is no indication that the incidence of wide 
binaries would be significantly higher in younger samples confirming that our 
previous comparisons with the field star sample do not require corrections.
Our results thus imply that very wide-orbit pairs are more common in EDD systems 
than in the normal stellar population.

\subsubsection{Age distribution and disk evolution} \label{sec:age}
\begin{figure} 
\begin{center}
\includegraphics[scale=.45,angle=0]{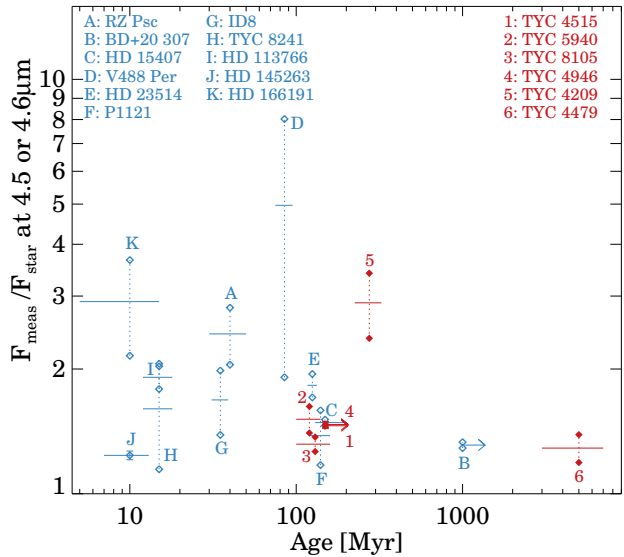}
\caption{Ratios of the measured WISE2 band (4.6$\mu$m) flux densities to stellar 
photospheric fluxes as a function of age for extreme debris disks. Blue empty symbols 
show the eleven previously known objects, while red filled symbols denote our targets from 
Table~\ref{tab:props}. 
Variability ranges are
shown by vertical dotted lines.  
\label{fig:fluxratio}
}
\end{center}
\end{figure}

EDDs are thought to be produced in violent collisions occuring between 
planetary embryos during the final accumulation of terrestrial planets
\citep{jackson2012,genda2015,su2019}. If so, then by studying 
their age distribution, we can put observational constraints on the timeline 
of rocky planet formation around Sun-like main-sequence stars. Figure~\ref{fig:fluxratio}
shows the ratios of the measured 4.5 or 4.6\,{\micron} flux densities to the predicted 
stellar photospheric fluxes as a function of ages for all 17 EDDs. 
Variability flux ranges, when applicable, are shown by vertical dotted lines.
For BD+20\,307, HD\,15407, and HD\,113766, we used {\sl Spitzer} 4.5\,{\micron} photometry  
taken from the literature \citep{meng2015,su2019}. HD\,166191 exhibited a strong brightening in 
2019 \citep{su2019b}, the corresponding individual {\sl Spitzer} data points 
are from Su et al. (2021, in preparation).
For the rest of the sample {\sl WISE}
4.6\,{\micron} data are plotted, where the minimum and maximum fluxes 
were taken from the light curves plotted in Figs.~\ref{fig:variability} and \ref{fig:knowneddsvar}.
In the case of TYC\,8241-2652-1, only the minimum flux was obtained in 
this way, the maximum flux at 4.6\,{\micron} was estimated from 
its disk model constructed by \citet{melis2012} considering the IRAS 
measurements of the source.  
For HD\,145263, which is constant based on both {\sl Spitzer} and {\sl WISE} 
time-domain data, AllWISE W2 
band photometry was utilized, after correcting for saturation by applying 
the method described in \citet{cotten2016}

In our solar system, the era of giant impacts probably ended by a collision 
between the proto-Earth and another body that led to the formation of the 
Earth-Moon system \citep{canup2001}. Analyses of radioactive isotopes yield age estimates between 
30 and 110\,Myr for this event \citep{jacobsen2005,touboul2007,halliday2008,kleine2009}. 
\citet{quintana2016} used $N$-body simulations to examine the formation of planets 
in the terrestrial zone of a Sun-like star that harbors
giant planets analogous to Jupiter 
and Saturn. Their initial disk included 26 Mars-sized planetary embryos  and 260 
smaller, approximately Moon-sized bodies. They found that most of the collisions 
took place in less than 100\,Myr: in models where fragmentations were enabled, 
90\% of collisions occurred within the first 82\,Myr, and 50\% of them in the first 20\,Myr. 
Nevertheless, based on the simulations, large collisions can happen 
even after hundreds of millions of years (up to 1\,Gyr), although these are quite rare. 
\citet{genda2015} reached a similar conclusion in their simulations without giant planets: 
the final giant impacts were found to occur at 73$\pm$74\,Myr. The age distribution of 
EDDs differs significantly from these results.  
Although most of the systems are probably 
younger than 500\,Myr, and thus can even be the result of a similar major 
collision, the majority of them ($\sim$60\% of the sample) are older than 100\,Myr, and at least 3 of them ($\sim$20\%)
have an age of $>$200\,Myr. Thus if EDDs are associated with rocky planet formation, then 
our results displayed in Fig.~\ref{fig:fluxratio} 
hint that the intensity of these processes -- contrary to what solar system 
experienced and what is predicted 
by the simulations -- 
does not decay significantly even after 100\,Myr. Obviously, the simulations examined only a 
fraction of the possible parameter space.  The presence of giant planets and their architecture,
as well as the mass and location of embryos/planetesimals in the terrestrial zone at the beginning of the 
giant impact era may have a serious influence on the timeline of planet formation 
\citep{quintana2016,barclay2017}. However, it is questionable whether all these can 
cause such prolongation of rocky planet formation processes as suggested by our results.

\subsubsection{Alternative scenarios for the formation of EDDs} \label{sec:alternativescenarios}

Alternatively, it is possible that some other mechanism(s) can also produce EDDs and
this process is more typical in slightly older systems. When speculating on such 
mechanisms, we must consider that, based on the available observational 
characteristics, the older EDDs are quite similar to the younger ones. Though the average 
fractional luminosity of disks older than 100\,Myr is somewhat lower than that of the 
younger objects, their characteristic dust temperatures are similar and they display 
mid-IR variations with the same frequency and comparable amplitudes. 
A possible alternative mechanism should be able to explain these properties.

\paragraph{Single giant collision in an inner planetesimal belt}
Although the formation of EDDs is certainly not explicable by the long term steady state 
collisional evolution of an in situ planetesimal belt (Sect.~\ref{sec:originhotdust}), 
individual giant impacts may happen in a collisionally depleted belt as well. However, considering 
that during the collisional evolution the number of large bodies, whose destruction can 
reproduce the observed dust quantity, is also decreasing, the probability of witnessing such 
an event falls off as $\propto t_{\rm age}^{-2}$ \citep{wyatt2007}. This is inconsistent with 
the observed age distribution. Actually, according to \citet{wyatt2007}, disks with high 
$f_{\rm d}$ /$f_{\rm d,max}$ (Sect.~\ref{sec:originhotdust}), which cannot be the result of 
steady state evolution of a 
planetesimal belt, cannot be explained with such single giant collisions either.

\paragraph{Sublimation and distruption cometary bodies}
Comets transported from an outer reservoir into the inner regions are also widely 
considered as possible sources of warm debris material 
\citep[e.g.,][]{wyatt2007,morales2011,bonsor2012,ballering2017,marino2017}. 
Actually, in our solar system the zodiacal dust particles dominantly 
stem from comets \citep{nesvorny2010,rowanrobinson2013}. 
Detections of non-photospheric, 
sometimes variable, absorption features toward several stars imply cometary activity 
in other systems as well \citep[e.g.,][]{beust1991,kiefer2014,welsh2015,iglesias2018,rebollido2020}. 
Due to the subtantially longer collisional times, the depletion 
of cold outer planetesimal belts is much slower than that of inner Asteroid Belt analogs.  
Therefore, minor bodies originating from such reservoirs can supply dust in the inner zones even
long after the local belt has largely been exhausted. These icy bodies can contribute to the 
production of warm dust in several ways. Dust grains can arise from the sublimation and  
disruption of comets, as well as from collisions of the inwardly scattered objects with each 
other and/or with inner rocky bodies. 

Planetesimals can evolve into more eccentric orbits under the gravitational influence of 
planets or stellar companions, or because of a close stellar encounter. Investigation of 
scattering processes revealed that efficient, long term inward transport of planetesimals 
 from an outer belt requires a special architecture: the presence of a chain of closely 
 spaced low mass planets \citep{bonsor2012,bonsor2014,marino2018}. As \citet{faramaz2017} 
demonstrated, interactions between planetesimals and an outer eccentric planet can also 
drive a vast number of icy bodies to highly eccentric orbits, which then can produce 
warm dust grains in the inner system on gigayear time scales. Transient events with a shorter 
time scale may also be of interest. A rearrangement of the planetary 
system following a dynamical instability -- akin to the hypothesized Late Heavy Bombardment scenario 
in our solar system \citep{tera1974,booth2009,bottke2017} -- can cause shorter lived, 
significantly enhanced cometary activity.
A companion or a star passing close-by might also result in a transient spike in the 
warm dust production by initiating a comet shower from an outer planetesimal belt or from a
 reservoir analogous to the solar system's Oort cloud.

In the following, we examine whether the cometary scenario can explain the peculiarly high dust content of
EDDs. 
As a minimal EDD, we consider a 0.25\,au wide debris belt that is located at 
0.5 au around a Sun analog star ($L_* = 1~L_\odot, M_* = 1~M_\odot$) and has a fractional
luminosity of 0.01. To estimate its mass loss rate, we
used eq.~29 from \citet{wyatt2007} and obtained $\sim$0.17\,M$_\oplus$/Myr.
Supposing a dust to ice mass ratio of 1 in the comets and that their material is converted 
to dust grains with a 100\% efficiency, the sustenance of the above model EDD over 100\,Myr
would require the destruction of 34\,M$_\oplus$ of icy planetesimals in the inner zones. 
By simulating various multiplanet architectures, \citet{marino2018} found that at most $\sim$7\% 
of planetesimals scattered from an outer belt via the planets reached the inner regions. Taking this 
maximum fraction, the long term supply of the disk in a steady state manner would require an unrealistically 
high total planetesimal mass of $\gtrsim$500\,M$_\oplus$ in the outer reservoir. 
This suggests that even in the cometary scenario, it is more likely that the observed phenomenon 
is related to a short-lived transient event rather than a continuous long term comet flux. 

Based on the cometary model developed by \citet{marboeuf2016}, we can make an 
estimate of how many active comets would be needed to provide the above derived dust 
production rate. Using their eq.~17-19, we find that a comet with a radius of 1\,km 
located 0.5\,au from a star with $L_* = 1~L_\odot$ releases dust with a rate of 
$\sim$5300~kg~s$^{-1}$. Roughly 6 million such active comets are needed to supply our 
minimum EDD. Supposing larger icy planetesimals with a radius of 100\,km, i.e., with a size 
roughly corresponding to the largest Centaurs in the solar system, the required number would
 be $\sim$600. Since a large fraction of scattered icy bodies never reaches 
 the snow line and even those that do, spend only a part of their 
 orbital time within the line, the total number of comets involved in the process 
 has to be much higher than this.
However, this calculation only considers the dust production driven by the sublimation of icy bodies. 
According to \citet{nesvorny2010} in the Solar system the spontaneous disruption of 
Jupiter family comets (JFCs) is the main source of the zodiacal dust. 
Moreover, the presence of a large number of comets in the terrestrial zone 
increases the chance of collisions with inner planets and other comets.
Thus, it is entirely possible that in EDDs the latter two processes dominate the supply of 
dust and, therefore, much fewer active comets are sufficient to explain the observed excess.

Disruption of large comets and giant impacts is accompanied by rapid, 
vigorous dust release that can also explain the high amplitude mid-IR changes seen in EDDs.
For TYC\,4479 and TYC\,4209,
the observed mid-IR brightening requires a minimum dust cross section change of 
 2.3$\times$10$^{-4}$\,au$^2$ and 6.0$\times$10$^{-3}$\,au$^2$, respectively (Sect.~\ref{sec:originvariability}). 
How large should a disintegrating comet be to produce the necessary amount of dust?
In the solar system, disruption of JFCs produces mm to cm-sized particles
that later collisionally grind down to smaller fragments \citep{nesvorny2010}. 
Assuming that the dust material of the comet (half of its total mass) turns into 
0.5\,{\micron}--1\,mm grains in a power-law size distribution with an index 
of $-$3.5 and taking an dust/ice density ratio of 3, the reproduction of the 
derived cross sections needs the disruption of a planetesimal with a radius 
of $\sim$50\,km and $\sim$160\,km for TYC\,4479 and TYC\,4209, respectively. 
These sizes should be considered as lower limits since 1) the disruption may
not be so efficient; 2) the average size of the emerging
particles is probably underestimated; and 3) the estimated cross section values 
are also lower limits (Sect.~\ref{sec:originvariability}).
We note that collisions between objects of the sizes computed above 
 can also provide a good explanation for 
the observed flux changes \citep[e.g.,][]{su2019}.

\paragraph{The possible role of dynamical instabilities}
As mentioned earlier, the temporarily enhanced comet activity 
could be due to a dynamical instability of the planetary system. On their changed orbit,
planets can perturb neighboring planetesimal populations, leading to their 
excited eccentricities, causing an increased flux of infalling icy 
planetesimals in the terrestrial regions. Although the aftermath of such instabilities 
can last for up to a few 10\,Myr, the enhanced level of exozodiacal dust 
typically decays on a time scale of a few million years \citep{bonsor2013}. 
Figure~2 of \citet{bonsor2013} suggests that even within this period, there 
could be a shorter peak in the comet flux. By the end of such an instability 
event, the outer planetesimal belt can be substantially depleted \citep{booth2009}.
The mass inflow of comets might be sufficiently high to 
produce an extreme debris disk. However, although the NICE model of the solar system 
demonstrates that a dynamical instability can be delayed to hundred million years 
\citep{gomes2005}, \citet{bonsor2013} claims 
that such instabilities are more frequent in younger systems (typically $<$10\,Myr).
Thus, this mechanism, similarly to the rocky planet formation scenario, is unable 
to offer a solution for the observed age distribution, although it may explain 
the origin of some individual systems.    

\paragraph{The possible influence of wide companions}
Intriguingly, we found that very wide-orbit pairs are more common in EDD systems 
than in the normal stellar population (Sect.~\ref{sec:widecompanions}). This is particularly true for  
EDDs older than 100\,Myr: out of the 10 objects, 8 are located in wide binaries.
This raises the question whether the two phenomena may be physically connected.
One possibility is that, as a consequence of impulses from passing stars and torques 
from the Galactic tide, orbits of very wide ($a\gtrsim$1000\,au) binary stars can become very 
eccentric after a time \citep{jiang2010,kaib2013,bonsor2015}. Due to the decreasing 
pericenter radius, the companion can launch a comet shower when approaching the outer 
population of planetesimals.
Indeed, the 365--6000\,au projected separations found in the sample 
are comparable to the size
 of the scattered disk or the Oort cloud in the solar system. 
This mechanism is able to produce an excess population of long period comets. A favorable
 aspect of this model is that the orbital change of the companion requires time, thus this 
 process could occur much later than 100\,Myr, too. 
 However, this hypothesis has several drawbacks. Long period comets are on very eccentric orbits and 
spend only a small fraction of their revolution within the snow line thus the dust production 
is limited. Moreover, simulations 
of perturbations caused by a companion or a stellar flyby predict a comet flux increase of at most 
a few orders of magnitude, which means a few tens of such comets per year \citep{davis1984,berski2016}, 
far below the required level. Therefore, this scenario alone probably 
cannot explain the observed dust contents. 
 
A companion with an initial inclination ($i_{\rm c}$) between $\sim$39{\degr} and 
$\sim$141{\degr} with respect to the orbital plane of a third body (e.g. a planetesimal or a planet)
 can cause oscillations in the eccentricity and inclination of this body via the Kozai-Lidov 
 (KL) mechanism \citep{naoz2016}.  The timescale of
this effect depends on the mass ($m_{\rm c}$), orbital period ($P_{\rm c}$), and 
eccentricity ($e_{\rm c}$) of the companion, the total mass of the primary, 
the perturber, and the disk 
($m_{\rm tot} = m_{\rm pr} + m_{\rm c} + m_{\rm disk}$), and the orbital period 
of the disk's particles ($P$) \citep{nesvold2016}: 
\begin{equation}
t_{\rm KL} \sim \frac{m_{\rm tot}}{m_{\rm c}} \frac{P_{\rm c}^2}{P} (1 - e_{\rm c}^2)^{3/2}.
\end{equation}
The first eccentricity peak is achieved at $t_{\rm KL}/2$, the maximum of the eccentricity 
can be approximated as $\sqrt{1 - (5/3) \cos^2{i_{\rm c}}}$ \citep{naoz2016,nesvold2016}. 
Assuming that the detected wide companions fulfill the inclination criteria above 
($39{\degr} \leq i_c \leq 141{\degr}$) and are on a circular orbit, and adopting 
the projected separations as an estimate of the semi-major axes 
 we can calculate 
the innermost radius where a third body can undergo the maximum eccentricity phase during the 
lifetime of system. 
In the cases of TYC\,4209 and TYC\,8105, we deduced radii of $\sim$140\,au and 
$\sim$240\,au, respectively.
Even the largest known 
exo-Kuiper belts around solar-type stars have radii $<$200\,au \citep{loehne2012,faramaz2019,sepulveda2019}.
Considering this, and that the eccentricities already started increasing inside 
the above derived radii, it is conceivable that 
in TYC\,4209 the companion can affect the orbit of planetesimals in an 
outer belt. A direct entry of a perturbed planetesimal into the sublimation region 
inside the snow line is possible but would require 
a high eccentricity (e.g. $e>0.99$ for a comet originating from a belt at $>$100\,au), and 
thus would require a high inclination of the companion ($i_{\rm c}\gtrsim84{\degr}$ for 
the previous example).      
However, even with smaller eccentricities, a planetesimal can cross a chaotic zone of a planet  
from where it can be transported inward, potentially by multiple scattering in a chain of planets.
In BD+20\,307, HD\,23514, HD\,15407, TYC\,4479, TYC\,4946, and TYC\,4515, the eccentricity of 
a tertiary body can peak within the available time (the age of the system) already at 
$<$1.5\,au, $\sim$2\,au, 4\,au, 11\,au, $<$8\,au, and $<$15\,au, respectively,  due to the K-L mechanism.
It means that in these cases, not only the outer planetesimal belt 
could be affected, but also the possible planetary system within the belt. Changing eccentricities 
of the planets can lead to a dynamical instability that can enhance the inward transport of 
comets further (see above). It is also a delayed effect, which can be consistent with the 
existence of EDDs at $>$100\,Myr. 
%\edit1{For ages of BD+20\,307, TYC\,4946, and TYC\,4515 
%only lower limits are available. Therefore, especially in the case of BD+20\,307, where the inner dust ring 
%is located at 0.7--0.9 au (Appendix~\ref{appendix:c}), it is conceivable that the K-L mechanism 
%could have a direct influence on a planetesimal belt co-located with this dust. 
%Investigation of the white dwarf companion in this system can provide a more accurate age 
%determination \citep[e.g.][]{fouesneau2019} and  
%allows studying the dynamical evolution of this interesting triple system.} 

In this calculation, for the probable white dwarf companion of 
BD+20\,307 we adopted a mass of 0.48\,M$_\odot$ 
\citep{fusillo2019} and a semi-major axis of 980\,au. However, the progenitor
 of this companion was more massive and the mass loss during 
the stellar evolution was accompanied by a change in its orbit, making 
our result on the K-L mechanism in this system uncertain.
Further investigation of the white dwarf companion would be  
important to provide a more accurate age 
determination \citep[e.g.][]{fouesneau2019} and to
study the dynamical evolution of this interesting triple 
system. This could lead to a better understanding of the 
formation of the dusty warm disk as well.

For the semi-major axes of the companions, we adopted the observed projected 
separations, which are likely smaller than the true semi-major axes.
Considering a random distribution of orbital inclinations and the observed eccentricity 
distribution of binary systems \citet{fischer1992} found that the 
true semi-major axes are 1.26$\times$ larger on average than the projected separations. 
Taking into account this multiplicative factor in a statistical manner 
results in $\sim$1.6$\times$ larger inner radii for the area affected by the 
K-L mechanism. This would not change our main conclusions.

Most cometary models assume the presence of an outer, cold planetesimal belt in 
the system. Based on currently available observations, 
none of the EDDs with age $>$100\,Myr show evidence for cold debris belts. 
Three systems have deep far-infrared photometry obtained with the {\sl Herschel 
Space Observatory}. For HD\,15407 and HD\,23514, these observations implied the presence 
of two temperature components, however, even the colder dust has 
relatively high temperatures; 334\,K and 168\,K, respectively
\citep{vican2016}. These temperatures are higher than what is required for sublimation of 
water ice, and thus they do not indicate regions in which icy planetesimals 
can persist for a long time. Considering their fractional luminosity, these components, 
even if they indicated separate rings, could still be of transient origin 
\citep{vican2016}. In the case of BD+20~307, the 70 and 100\,{\micron} 
{\sl Herschel} photometries are consistent with the emission of 
a single warm component. P\,1121 harbors no cold dust either \citep{meng2015,su2019}.
As for our six objects, the {\sl IRAS} FSC catalog 
provides upper limits at 60\,{\micron} for four of them (TYC\,4515, TYC\,5940, TYC\,8105, 
and TYC\,4209). For the other two targets, we used the IRAS SCANPI 
tool\footnote{\url{https://irsa.ipac.caltech.edu/applications/Scanpi/}} to derive 
 upper limits in the same band. On the basis of these IRAS data, even in the best cases, 
 it could only be established that there are no cold (30--80\,K) disks with a fractional 
 luminosity greater than 0.01 in our systems. Though there is no direct evidence 
for colder belts, in most cases the current data cannot exclude the existence 
of massive outer debris disks.

It is worth noting that in our previous analyses, we tacitly assumed that only 
grains for which the ratio of radiation pressure to gravitational force, $\beta$, exceeds 
0.5 are blown out from the system ($\beta_{\rm bl}=0.5$), which is valid if 
the parent bodies are on a circular orbit.
Considering parent bodies on eccentric orbits ($e_{\rm pb}>0$), such as comets, 
the $\beta_{\rm bl}$ can be computed as 
$\beta_{\rm bl} = 0.5 \left( \frac{1-e_{\rm pb}^2}{1+e_{\rm pb}\cos{\Phi}} \right),$ where $\Phi$ 
is the longitude of the release position on the orbit \citep{murray1999,sezestre2019}.
This means that depending on the eccentricity, particles with much smaller $\beta$ parameters 
(with much larger sizes) can become unbound and thus the released dust is removed 
more rapidly than in the circular case. Given typical comet orbits, this obviously requires 
higher replenishment rates 
than what obtained from our calculations.

\section{Summary} \label{sec:summary}
In this study, we conducted a survey to search for EDDs around Sun-like stars. 
Using the AllWISE infrared photometric and the Gaia TGAS astrometric catalogs, and applying 
careful selection criteria, we found six new EDDs: TYC\,4515, TYC\,5940, TYC\,8105, 
TYC\,4946, TYC\,4209, and TYC\,4479. Previously, only one of them, TYC\,4479, was 
identified as a debris disk -- but not as an EDD -- in the literature \citep{cotten2016}. 
Host stars of these disks are main-sequence F5--G9 stars located at distances 
between 164 and 279\,pc from the Sun.  For their age estimates, we combined different empirical 
diagnostic methods based on the stars' lithium content, rotation, and kinematic properties. 
While the youngest objects have ages similar to that of the Pleiades ($\sim$120--130\,Myr), the 
oldest one, TYC\,4479, has an age of $\sim$5\,Gyr. 
Actually, the latter system is by far the oldest known EDD.  
Using astrometric data from Gaia\,EDR3, we found that five among the six
host stars have co-moving pairs with projected separations ranging from 1820 
to 6000\,au. All these wide separation 
companions are low-mass, M-type stars. 
  
To estimate the basic disk properties, we fitted a simple blackbody model to the observed 
IR excesses. 
The dust temperatures are higher than 300\,K in all cases, and the 
derived fractional luminosities range between 0.01 and 0.07.
Considering the ages, the observed high amount of warm dust indicates 
that these systems likely underwent a recent transient event of dust 
production. Using time-domain photometric data at 3.4 and 4.6\,{\micron} from 
the {\sl WISE} all sky surveys between 2010 and 2019, we concluded 
that the light curves of four systems (TYC\,5940, TYC\,8105, TYC\,4209, and 
TYC\,4479) show evidence of variable mid-IR 
emission on yearly timescales. We deduced that these variations
are associated with the disks and not to the stars. 
By analyzing the observed 3.4 and 4.6\,{\micron} variations, we found that 
at TYC\,4209 and TYC\,4479, the observed brightening events seem to be 
inconsistent with a model where the emitting area of the dust remains 
constant while its temperature increases. It is more likely that 
new dust was created temporarily.
At TYC\,4209, we discovered mid-IR variations at daily timescales as well.

With the six disks studied in this paper, the number of known EDDs increased to 
17. Fifteen of them surround Sun-like (F5-K2) stars, while two 
have earlier F-type hosts. Using this new, unified sample we assessed
what they indicate about the nature of these interesting objects. 
From the 15 Sun-like stars at least 9 reside in multiple systems, 
8 of which are wide separation binaries with projected separations $>$365\,au.
By comparing this incidence with the corresponding statistics for 
solar-type field stars, we found that very
wide-orbit pairs are significantly more common in EDD systems than
in the normal stellar population. 
The contrast is even stronger if we consider only EDDs older than 
100\,Myr (10 objects), since all 8 wide binary systems belong to this subsample.
Based on the literature and on our analysis,
14 of the 17 known 
EDDs showed changes at 3--5\,{\micron} over the past decade. This suggests that 
the mid-IR variability is an inherent characteristic of EDDs.

The formation of EDDs is generally thought to be linked to giant impacts occuring during
the final, chaotic growth phase of terrestrial planets. 
In our solar system, this phase may have lasted up to $\sim$100 million years.
General numerical simulations of rocky planet formation predict that this era could extend up 
to a few hundred million years, but the vast majority of giant collisions happen in the 
first 100\,Myr. The observed age distribution of the currently known EDDs is inconsistent 
with this picture. Only seven of them are younger than 100\,Myr and we know of at least 
two EDDs that are certainly older than 1\,Gyr. If EDDs are indeed linked to 
rocky planet formation, then these results suggest that the intensity of this process 
does not decline significantly even after 100\,Myr. 
Alternatively, some other mechanism(s) also produce EDDs. 
One possible explanation is that the observed dust comes from the disruption and/or 
collisions of comets delivered from an outer reservoir 
into the inner regions. 

A high fraction of the oldest EDDs -- whose existence 
is most difficult to explain by the classical model -- are situated in wide binaries.   
This raises the possibility that these wide companions may play a role in 
initiating/maintaining the inward comet transport, e.g., by the Kozai-Lidov mechanism, 
or by launching a comet shower when an eccentric companion 
approaches an outer population of icy planetesimals.
Both invoked mechanisms require time to be activated, and thus could explain 
the existence of older EDDs. However, further detailed analysis is needed to assess
whether these scenarios are able to explain all the observed features of EDDs.

\acknowledgments
The authors are grateful to the anonymous referee for the comments that improved 
the quality of this manuscript.
This publication makes use of data products from the Wide-field Infrared Survey 
Explorer, which is a joint project of the University of California, Los Angeles, 
and the Jet Propulsion Laboratory/California Institute of Technology, and NEOWISE, 
which is a project of the Jet Propulsion Laboratory/California Institute of Technology. 
WISE and NEOWISE are funded by the National Aeronautics and Space Administration (NASA).
This publication makes use of data products from the Two Micron All Sky Survey, 
which is a joint project of the University of Massachusetts and the Infrared
Processing and Analysis Center/California Institute of Technology, funded by the 
National Aeronautics and Space Administration and the National Science Foundation.
 This work has made use of data from the European Space Agency (ESA) mission
{\it Gaia} (\url{https://www.cosmos.esa.int/gaia}), processed by the {\it Gaia}
Data Processing and Analysis Consortium (DPAC,
\url{https://www.cosmos.esa.int/web/gaia/dpac/consortium}). Funding for the DPAC
has been provided by national institutions, in particular the institutions
participating in the {\it Gaia} Multilateral Agreement.
This research has made use of the NASA/ IPAC Infrared Science Archive, which is 
operated by the Jet Propulsion Laboratory, California Institute of Technology, 
under contract with the National Aeronautics and Space Administration.
This research has also made use of the WEBDA database, operated at the Department of 
Theoretical Physics and Astrophysics of the Masaryk University
We used the VizieR catalogue access tool and the Simbad object data base at CDS to 
gather data. 
Our work is partly based on observations obtained with the Apache Point Observatory
3.5-meter telescope, which is owned and operated by the Astrophysical
Research Consortium. This project has been supported by the KH130526, K-125015, K-131508, 
K-119517, and GINOP-2.3.2-15-2016-00003 grants of the National Research, Development 
and Innovation Office (NKFIH, Hungary) as well as the Lend\"ulet Program  of the 
Hungarian Academy of Sciences, project No. LP2018-7/2019. 
Zs.M.Sz is supported by the \'UNKP-20-2 New National Excellence Program of the Ministry for 
Innovation and Technology from the source of the National Research, Development and Innovation Fund.
AD was supported by the \'UNKP-20-5 New National Excellence Program of the Ministry for Innovation and 
Technology from the source of the National Research, Development amid Innovation Fund and the 
J\'anos Bolyai Research Scholarship of the Hungarian Academy of 
Sciences. A.D and Gy.M.Sz would like to thank the City of Szombathely for support under 
Agreement No. 67.177-21/2016. G.C is supported by NAOJ ALMA Scientific Research Grant Number 2019-13B.


\vspace{5mm}
\facilities{Akari, IRAS, NEOWISE, Spitzer, TESS, WISE}


\software{mpfit \citep{markwardt2009}, FITSH \citep{pal2012}, MUFRAN \citep{kollath1990}}




\clearpage

\appendix
\restartappendixnumbering

\section{Checking source confusion} \label{appendix:a}

While the emission is dominated by the stellar photosphere for all our sources 
in the W1 band, 
at longer wavelengths the contribution of excess increases (Table~\ref{tab:sedtable}). 
In the W3 and W4 bands, the measured flux densities exceed the predicted photospheric 
contributions by at least $\sim$4.5 and $\sim$8.7 times, respectively.
In the case of TYC\,4209, the observed excess is dominant even in the W2 band.
If the excess is caused by a neighboring source(s), not resolved by {\sl WISE}, 
then the position of the emitting source measured in W1 and in the 
other bands would be different. Another indication would be
if the shape of the targets' profile does not match the typical profile of point sources.
Examining the shape can also help to detect the effects of possible nearby nebulosity.

Since the AllWISE catalog does not include separate positional information for 
the different bands, we performed our analysis using the original {\sl WISE} measurements.  
We utilized the ``unWISE'' coadds \citep{lang2014} of the {\sl WISE} all sky 
survey, which are designed to preserve the intrinsic resolution of the {\sl WISE} images 
($\sim$6\farcs1, $\sim$6\farcs4, $\sim$6\farcs5, and $\sim$12\farcs0 in the W1, W2, 
W3, and W4 bands, respectively). This provides a factor of 
$\sim$1.4 
narrower PSFs 
than those of the intentionally blurred 
"Atlas Image" products of the AllWISE Release. Figure~\ref{fig:wiseimages} shows 
the vicinity of the six selected systems in all four {\sl WISE} bands based on the
unWISE data. We fitted two-dimensional Gaussians to the measured brightness profiles 
of our targets, determining their positions, the mean of the
full width at half-maxima 
(FWHM) of the major and minor axes ($m_A$), and the ratio of the FWHM of the minor
to the major axis ($r_A$). The same parameters were derived using the unWISE 
data of a comparison sample.
For the comparison sample, we kept only those objects from the previously compiled AllWISE 
list (Sect.~\ref{sec:sampleselection}) for which the S/N of the flux measurement in each band is greater than the 
minimum S/N for our six sources. This shortened list was also cross-matched with 
the TGAS catalog, but this time we did not remove giant stars and 
we made no cuts based on the parallax data, resulting in a database with 
3732 stars. 

For the six EDD candidates, the largest W2 and W3 band offsets compared to 
the W1 position were 0\farcs062 and 0\farcs415.  
These do not count as outliers because 15\% and 6\% of the objects in the comparison 
sample display similar 
or larger offsets in the W2 and W3 bands, respectively. The W4 images of  
TYC\,4946 and TYC\,4479 show some extended nebulosity around 
the sources (in the case TYC\,4479 this can already be noticed to some extent in 
the W3 image). This may explain that while for the other four objects the W4 offsets 
remain below 0\farcs7, for these two targets we measured offsets of 1\farcs1. 
Nevertheless, even these latter values are consistent with those found for 
 comparison objects with similar W4 band S/N (32\% and 27\% of them 
 have larger offsets than that of TYC\,4946 and TYC\,4479, respectively).   
Neither the size nor the shape of the EDD candidates indicate significant 
contamination: we found that their $m_A$ and $r_A$ parameters fall within 
2.3$\sigma$ of the comparison sample means in all four bands. 

For three targets, additional higher spatial resolution mid-IR data are 
available. TYC\,4479 was serendipitiously observed with the IRAC camera onboard 
{\sl Spitzer} at 3.6\,{\micron} and 4.5\,{\micron} 
as part of a larger mapping project (Sect.~\ref{sec:additionalirdata}), while TYC\,4209 was the target 
of an extensive IRAC monitoring program at the same wavelengths (Mo\'or et al. 2021,
in preparation). Our targets appear as single sources in the obtained images that have
three times better spatial resolution (FWHM$\sim$2{\arcsec}) than that of {\sl WISE} 
at the corresponding wavelengths. The recently released unWISE Catalog 
\citep{schlafly2019} -- which considers all W1 and W2 band images obtained 
in the {\sl WISE} and the first four year of the NEOWISE Reactivation missions 
-- lists an object which is located 6\farcs4 away from TYC\,4515.
This faint object ($\sim$20$\times$ fainter than TYC\,4515 in both bands) is actually an 
M-type companion of TYC\,4515 (Sect.~\ref{sec:companions}). However, the observed excess emission 
surely originates from the primary component. This is justified not only by the 
$WISE$ observations, but by a 24\,{\micron} $Spitzer$ image where the position of the 
detected source coincides well with the optical position of TYC\,4515 (Sect.~\ref{sec:additionalirdata}). 
Thus, based on currently available IR data, we found no sign of confusion or 
significant contamination at our targets.

\section{Analysis of the TESS data} \label{appendix:b}

\subsection{Processing of TESS data}
Five of our stars were covered by the {\sl TESS} spacecraft providing high quality 
30 minute cadence photometric data. 
Table~\ref{tab:tess} shows the TESS Input Catalog (TIC) identifiers 
of our observed targets, as well as the log of their TESS measurements.
We started the data reduction with the calibrated 
full-frame images which were downloaded from the MAST 
archive\footnote{\url{https://mast.stsci.edu}}. As a first step,
following the outline given in \citet{pal2020}, we derived the plate solution 
using matched sources from the Gaia DR2 catalog. Considering the resemblance 
of the {\sl TESS} throughput to that of Gaia $G_{\rm RP}$ \citep{ricker2015,jordi2010}, 
these matched objects were also utilized to compute a zero-point flux reference based on 
the $G_{\rm RP}$ magnitudes. By analyzing various {\sl TESS} full-frame images, we 
found that the RMS of this flux calibration is $\sim$0.015\,mag. To extract the photometry 
of our sources, we performed differential image analysis utilizing the \texttt{ficonv} 
and \texttt{fiphot} tasks of the FITSH package \citep{pal2012}. The photometry was performed 
on 128$\times$128\,pixel subframes centered on the targets. The reference frame was constructed 
by computing a stray light-free median of 11 individual images measured close to the middle 
of the whole observing sequence. Lastly we identified and removed data points which 
were affected by momentum wheel desaturation or significant stray light \citep{pal2020}. 
If the target was measured in more than one sector, then the above described procedure was carried out 
separately in each sector.

\begin{deluxetable}{c|ccccc}[h!]
\tablecaption{TESS data \label{tab:tess}}
\tablecolumns{6}
\tablewidth{0pt}
\tabletypesize{\scriptsize}
\tablehead{\colhead{Name} &
\colhead{TIC ID} &
\colhead{Sectors} &
\colhead{Time range} &
\colhead{$P_{\rm rot}$ (d)} & 
\colhead{Amplitude (mag)}
}
\startdata
TYC\,4515  & TIC\,142013492  & S19     & 11/27/2019--12/24/2019 &  2.455\tablenotemark{a} & 0.00026\\
TYC\,5940  & TIC\,123977701  & S6      & 12/11/2018--01/07/2019 &  3.756\tablenotemark{b} & 0.0083 \\
TYC\,8105  & TIC\,231921033  & S5--7   & 11/15/2018--02/02/2019 &  5.042\tablenotemark{c} & 0.0187 \\           
TYC\,4209  & TIC\,233128866  & S15--17 & 08/15/2019--11/02/2019 &  5.065\tablenotemark{d} & 0.0032 \\
           &                 & S19--20 & 11/27/2019--01/21/2020 &   &  \\
	   &                 & S22--26 & 02/18/2020--07/04/2020 &   &  \\              
TYC\,4479  & TIC\,368374360  & S17--18 & 10/07/2019--11/27/2019 &  32.08\tablenotemark{e} & 0.0012 \\
           &                 & S24--25 & 04/16/2020--06/08/2020 &   &  \\
\enddata
\tablenotetext{a}{There are two additional weaker peaks at $P=0.585$ and $P=2.638$ days. 
ASAS light curve shows no convincing signals.}
\tablenotetext{b}{There are weaker Fourier-peaks at double/half frequencies. ASAS 
data show two close peaks ($P=3.637 $ and $P=3.648$) confirming the 
findings from TESS observations.}
\tablenotetext{c}{Two weaker peaks are found at $P=5.636$ and $P=2.509$ days. Some 
further irregular variations are still present after pre-whitening with these signals. ASAS 
light curve indicates one signal at $P=5.003$ days confirming the TESS analysis.}
\tablenotetext{d}{There are additional nearby weak peaks, possibly due to differential rotation, 
and a signal at double frequency. ASAS light curve confirms the five-day signal ($P=5.126d$).}
\tablenotetext{e}{Some weaker Fourier peaks with periods of $\approx 20$\,days and 
amplitudes of $<$50\% that of the main signal
are also present.}
\end{deluxetable}

\subsection{Period analysis }

To identify periodic signals in the obtained TESS light curves, we performed a
Fourier analysis with the MUlti FRequency ANalysis (MUFRAN) 
tool\footnote{http://www.konkoly.hu/staff/kollath/mufran.html}
developed by \citet{kollath1990}. Using MUFRAN, we prepared the discrete Fourier-transformation 
of the light curve for frequencies 0--10 1/days. We selected the most significant peaks, and 
after fitting the frequencies we pre-whitened the light curve, then iterated these steps until 
no more convincing Fourier peaks were found in the data. In many cases, longer trends (20-30 days) 
were found in the data. These might have  astrophysical background or instrumental trends, but 
these cannot be confirmed with TESS data only, as their lengths are comparable to the observations 
themselves. 

The TESS data showed periodic or quasi-periodic signals in all cases. In some targets, close to 
the main Fourier-peak additional smaller peaks were found: these are generally attributed to 
differential rotation of the stellar surface, i.e., regions of different latitudes showing 
slightly different rotational periods. Where available, we also checked light curves from the 
ASAS database to validate our findings from the TESS data. The derived main periods and amplitudes 
are summarized in Table~\ref{tab:tess}.

\section{Properties and long term mid-infrared variability of previously 
known EDDs} \label{appendix:c}
\restartappendixnumbering

%%%%%%%%%%%%%%%%%%%% TABLE C1 %%%%%%%%%%%%%%%%%%%%%%%%%%%%%%%%%%

\begin{deluxetable}{lcccccccccccc}
\tabletypesize{\scriptsize} 
\tablecaption{Stellar and disk properties of extreme debris systems 
              identified prior to our study \label{tab:knownedds}} 
\tablecolumns{12} 
%\tablewidth{260pt} 
\tablehead{
\colhead{Name} &
\colhead{SpT}  &
\colhead{Dist.}  &
\colhead{Lum.}  &
\colhead{$T_{\rm eff}$} &
\colhead{Group} &
\colhead{Age}  &
\colhead{Mult.}  &
\colhead{$T_{\rm BB}$} &
\colhead{$R_{\rm BB}$} &
\colhead{$f_{\rm d}$} &
\colhead{Disk var.}  &  
\colhead{Refs.} \\
\colhead{} &
\colhead{}  &
\colhead{(pc)}  &
\colhead{($L_\odot$)}  &
\colhead{(K)} &
\colhead{}  &
\colhead{(Myr)}  &
\colhead{}  &
\colhead{(K)} &
\colhead{(au)} &
\colhead{} &
\colhead{}  &  
\colhead{}
}
\colnumbers
\startdata
RZ\,Psc   &  K0IV & 184.7 &  1.0 &  5350 &   -       &  30$^{+10}_{-10}$ &  Y  & 340--500 & 0.3--0.7 & 0.049--0.08 & Y (W) & 9,17,21,35,37,38  \\
BD+20\,307&  F9V  & 116.9 &  2.8 &  6000 &   -          & $\gtrsim$1000 &  Y  & 358--440 & 0.7--0.9 & 0.01--0.04 & Y (S) & 5,11,16,19,29,43,48,49,50,51,52,53  \\
HD\,15407 &  F5V  &  49.3       &  3.25&  6500 & ABDor        & 149$^{+51}_{-19}$ &  Y  & 500--1020 & 0.13-0.6 & 0.006--0.0097 & Y (S) &  3,26,29,32,49 \\
V488\,Per &  K3V  & 172.9 &  0.3 &  4900 & $\alpha$ Per & 85$^{+10}_{-10}$ &  Y?  & 820 & 0.06 & 0.16 & Y (W) & 2,8,24,54,55  \\
HD\,23514 &  F5V  & 139.1 &  3.0 &  6450 & Pleiades	& 125$^{+8}_{-8}$ &  Y  & 600--1080 & 0.1--0.4 & 0.02 & Y (S,W) & 7,29,39,40,44,49 \\
P\,1121   &  F9V  & 440.4 &  1.6 &  6050 & NGC\,2422     & 140$^{+20}_{-20}$ &  N  & 460 & 0.5 & 0.014--0.020 & Y (S,W) & 1,5,6,13,29,45  \\
ID\,8     &  G6V  & 355.7 &  0.7 & 5500  & NGC\,2547	& 35$^{+4}_{-4}$ &  N  & 400 & 0.4 & 0.025--0.032 & Y (S,W) & 14,18,28,30,31,45  \\
TYC\,8241-2652-1\tablenotemark{a} & K2V  & 121.3  &  0.6 &  4950 & LCC & 15$^{+3}_{-3}$ &  N  & 450 & 0.3 & 0.11 & Y (O,W) &  15,27,34 \\
HD\,113766&  F2V  & 108.6 & 3.9 &  6800 & LCC & 15$^{+3}_{-3}$ &  Y  & 490 & 0.64 & 0.017 & Y (S) & 4,10,22,32,33,47  \\
HD\,145263&  F2V  & 141.2 &  4.5 & 6800 & US & 10$^{+3}_{-3}$ &  N  & 240--290 & 2.0--2.9 & 0.01--0.02 & N (S,W) & 10,12,19,23,25,30,33,42  \\
HD\,166191&  G0  &  100.8 & 4.0 & 6000 &  HD\,166191\tablenotemark{b} & 10$^{+5}_{-5}$ &  N  & 760 & 0.3 & 0.06 & Y (S) & 20,32,36,41,46  \\
\enddata
\tablecomments{
Column (1): Name of the star. 
Column (2): Spectral type. 
Column (3): Distance \citep{cbj2020}.
Column (4): Luminosity.  Literature data are scaled corresponding to the new distances from Col.~3 when needed.
Column (5): Effective temperature.
Column (6): Group membership. ABDor: AB Doradus moving group; 
            LCC: Lower Centaurus Crux association; US: Upper Scorpius association.
Column (7): Stellar age. In the case of group members, age of the corresponding group is quoted.
Column (8): Multiplicity. See Sect.~\ref{sec:widecompanions} for more details.
Column (9): Dust temperature or temperature range when several different estimates are available in the literature. 
In cases where the SED was fitted with two temperature components, 
the warmer one is considered.
Column (10): Disk radius. For comparability with our results presented in Table~\ref{tab:props} we quote 
the blackbody radii that were computed using luminosities and 
dust temperatures from Col. (4) and Col. (9). For possible caveats related to this approach, see Sect.~\ref{sec:diskprops}.
Column (11): Fractional luminosity. If this is a range, then it shows the scatter of the 
literature data and is not because of variability.
Column (12): Disk variability in the wavelength regime between 3 and 5{\micron}. 
Entries in parentheses show what type of measurements were used to examine the variability -- 
S: {\sl Spitzer} photometry; W: {\sl WISE} photometry; G: ground based instruments.  
Column (13): References: (1) \citet{babusiaux2018}; (2) \citet{barrado2004}; (3) \citet{bell2015}; (4) \citet{chen2006}; (5) \citet{cotten2016}; (6) \citet{cummings2018}; (7) \citet{david2015}; (8) \citet{deacon2020}; (9) \citet{dewit2013}; (10) \citet{dezeeuw1999}; (11) \citet{fekel2012}; (12) \citet{fujiwara2013}; (13) \citet{gorlova2004}; (14) \citet{gorlova2007}; (15) \citet{guenther2017}; (16) \citet{hartman2020}; (17) \citet{herbig1960}; (18) \citet{jeffries2005}; (19) \citet{kennedy2013}; (20) \citet{kennedy2014}; (21) \citet{kennedy2020}; (22) \citet{lisse2017}; (23) \citet{lisse2020}; (24) \citet{luo2019}; (25) \citet{mannings1998}; (26) \citet{melis2010}; (27) \citet{melis2012}; (28) \citet{meng2014}; (29) \citet{meng2015}; (30) \citet{mittal2015}; (31) \citet{olofsson2012}; (32) \citet{oudmaijer1992}; (33) \citet{pecaut2012}; (34) \citet{pecaut2016}; (35) \citet{potravnov2014}; (36) \citet{potravnov2018}; (37) \citet{potravnov2019}; (38) \citet{punzi2018}; (39) \citet{rhee2008}; (40) \citet{rodriguez2012}; (41) \citet{schneider2013}; (42) \citet{smith2008}; (43) \citet{song2005}; (44) \citet{stauffer1998}; (45) \citet{su2019}; (46) \citet{su2019b}; (47) \citet{su2020}; (48) \citet{thomson2019}; (49) \citet{vican2016}; (50) \citet{weinberger2008}; (51) \citet{weinberger2011}; (52) \citet{whitelock1991}; (53) \citet{zuckerman2008}; (54) \citet{zuckerman2012}; (55) \citet{zuckerman2015}
}
\tablenotetext{a}{TYC\,8241-2652-1 hosted an unusually dust-rich disk before 2009, whose mid-IR 
luminosity then decayed significantly over less than two years, leaving behind a colder ($<$200K), 
more tenuous disk \citep{melis2012}. The quoted data show the disk parameters measured before 2009.}
\tablenotetext{b}{\citet{kennedy2014} found that HD\,166191 is comoving with the well known Herbig Ae star, HD\,163296. 
Later, \citet{potravnov2018} identified some additional young stars showing similar kinematics and located close to 
HD\,166191 and they argued that all these objects belong to a $\sim$10\,Myr old kinematic group. 
}
\end{deluxetable}

%%%%%%%%%%%%%%%%%%%% TABLE C1 %%%%%%%%%%%%%%%%%%%%%%%%%%%%%%%%%%

Prior to our study, eleven F--K type stars were reported to host extreme debris disks. 
In Table~\ref{tab:knownedds} we summarize the fundamental properties of 
these systems.

Using the available single-exposure photometry acquired in the W1 and W2 bands 
over the different {\sl WISE} mission phases, we explored the long term 
infrared variability of the 11 previously known EDDs listed in Table~\ref{tab:knownedds}. 
Some of these 
targets are brighter than 8 and 7\,mags in the W1 and W2 bands, respectively, and 
thus are potentially affected by non-linearity and saturation effects. As \citet{mainzer2014} showed 
for sources brighter than the saturation limits, fluxes measured in the reactivated 
NEOWISE mission could be significantly overestimated relative to the AllWISE 
mission data (see their fig.~6). Therefore, our previously used quality criteria (Sect.~\ref{sec:wisedata}) 
were supplemented with an additional one: we ignored those 
data points where the fraction 
of the saturated pixels in the profile-fitting area exceeds 0.05 (\texttt{w1sat}$>$0.05, 
\texttt{w2sat}$>$0.05). For the three brightest sources, HD\,15407, HD\,113766, and HD\,166191, 
this criterion removes essentially all data points. In the cases of BD+20\,307 and 
HD\,145263, the W1 band data turned out to be heavily saturated, but we could extract 
the light curve in the W2 band. From the rest of the sample, 
only the W1 photometry of HD\,23514 was somewhat affected by this issue, but it did not 
prevent the compilation of the light curve. The photometry of ID8 has a different 
quality issue: the reduced $\chi^2$ of all of its W1 band profile-fit photometry measurements (\texttt{w1rchi2})
obtained during the AllWISE mission phase are higher than the critical value we defined 
($>$5, Sect.~\ref{sec:wisedata}).  However, looking at the specific observation windows,
the standard deviation of the points did not appear to be substantially higher than 
in the other windows and therefore we have waived the application of this 
quality criterion. 
  
\begin{figure*} 
\begin{center}
\includegraphics[scale=.37,angle=0]{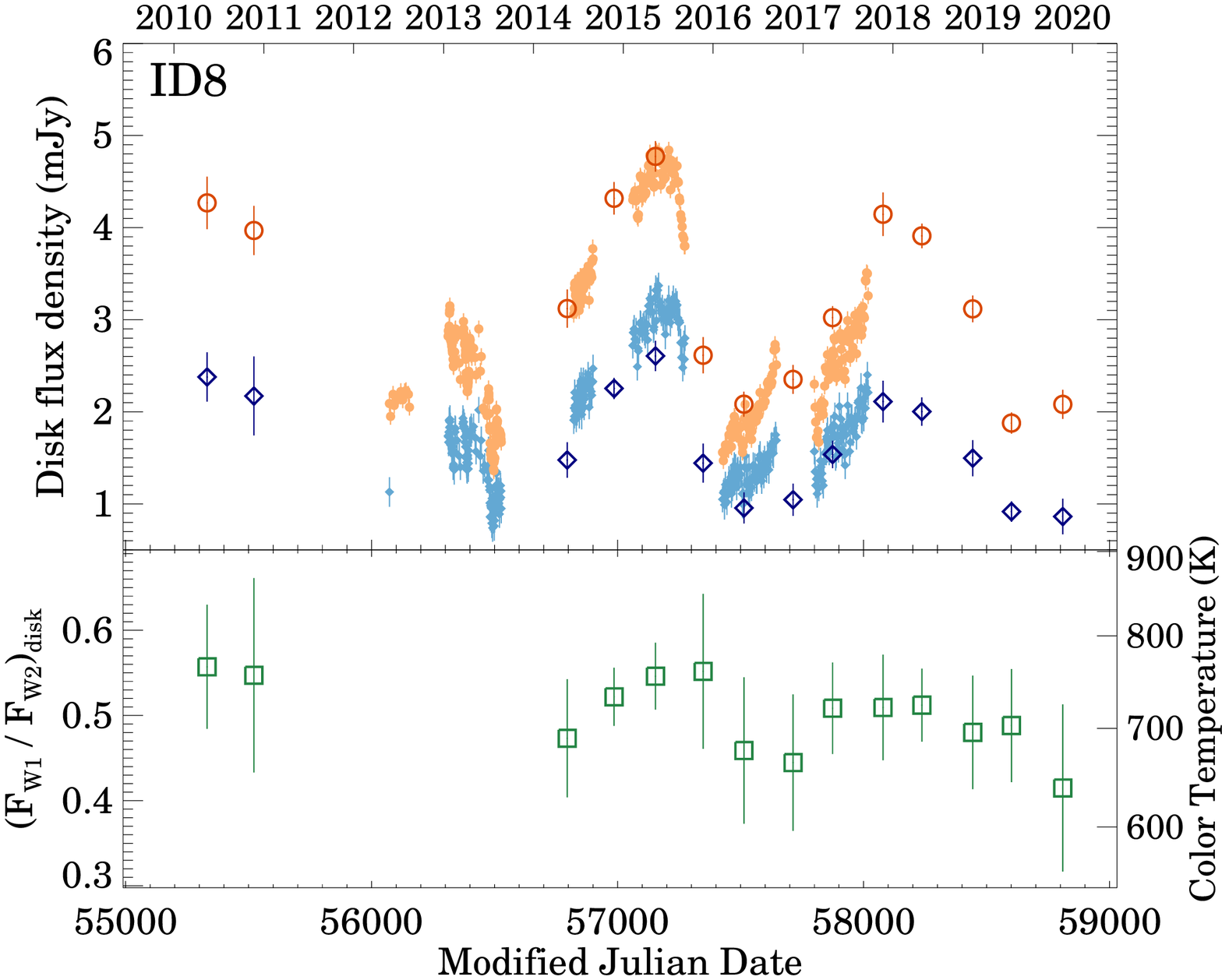}
\includegraphics[scale=.37,angle=0]{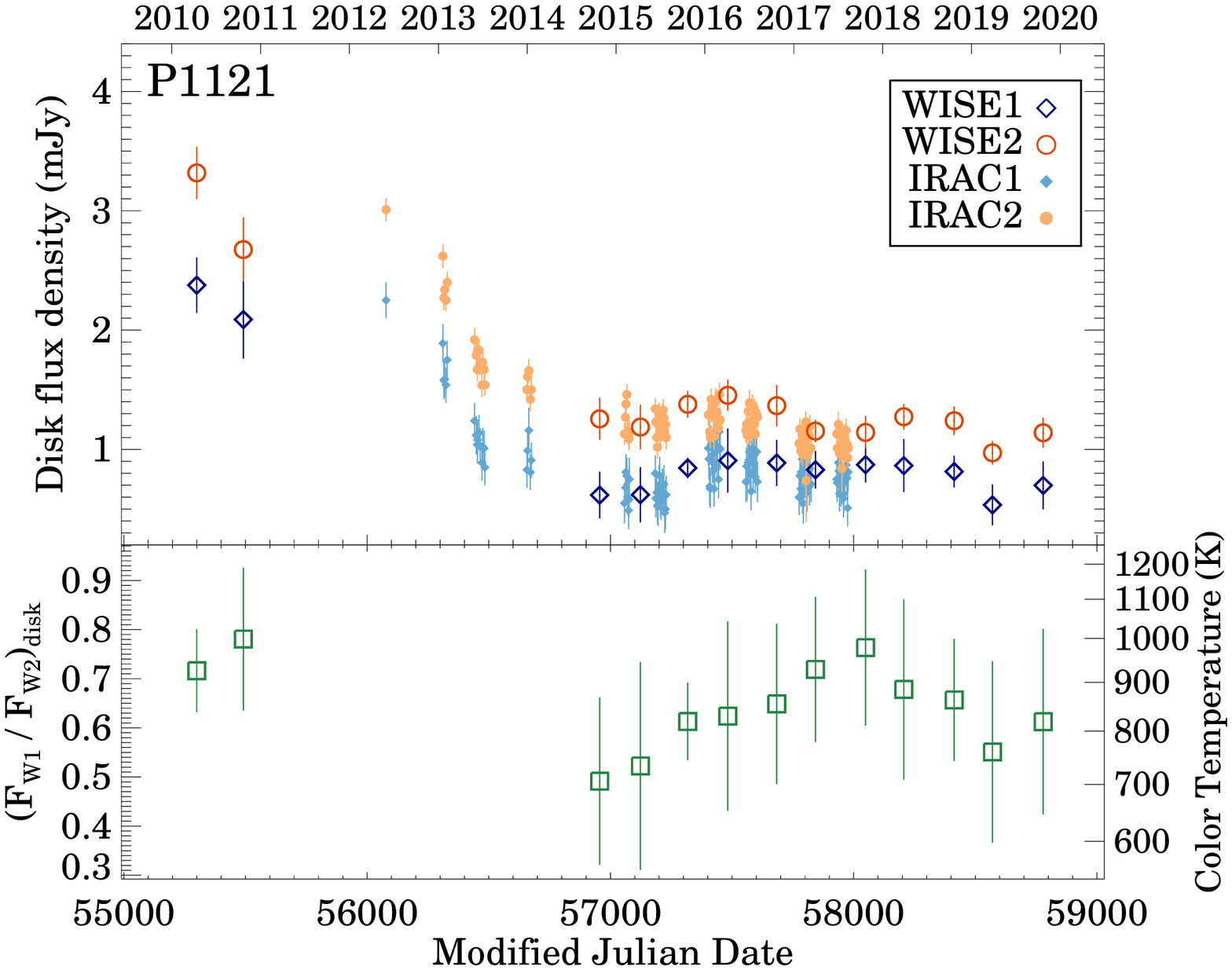} \\
\includegraphics[scale=.37,angle=0]{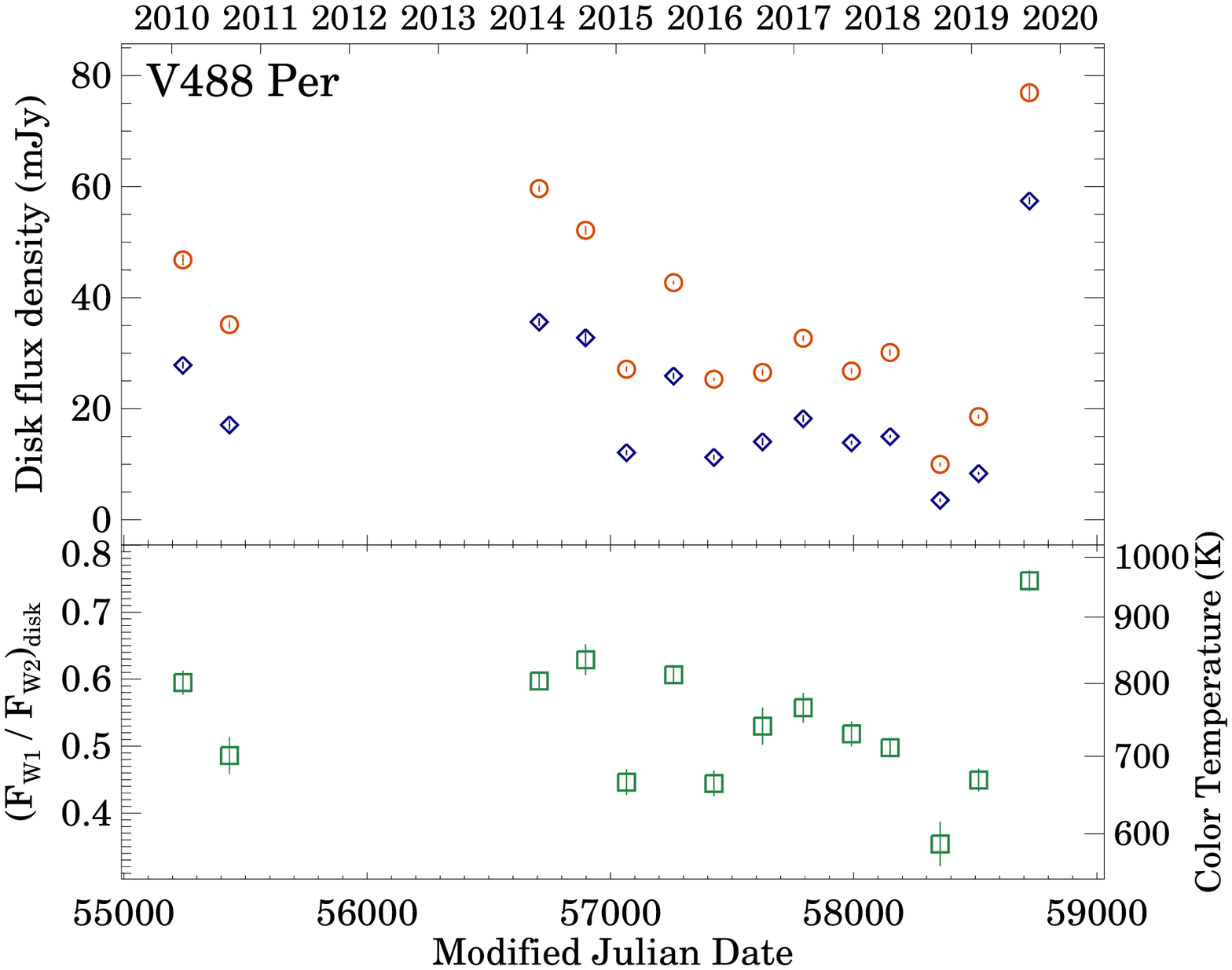}
\includegraphics[scale=.37,angle=0]{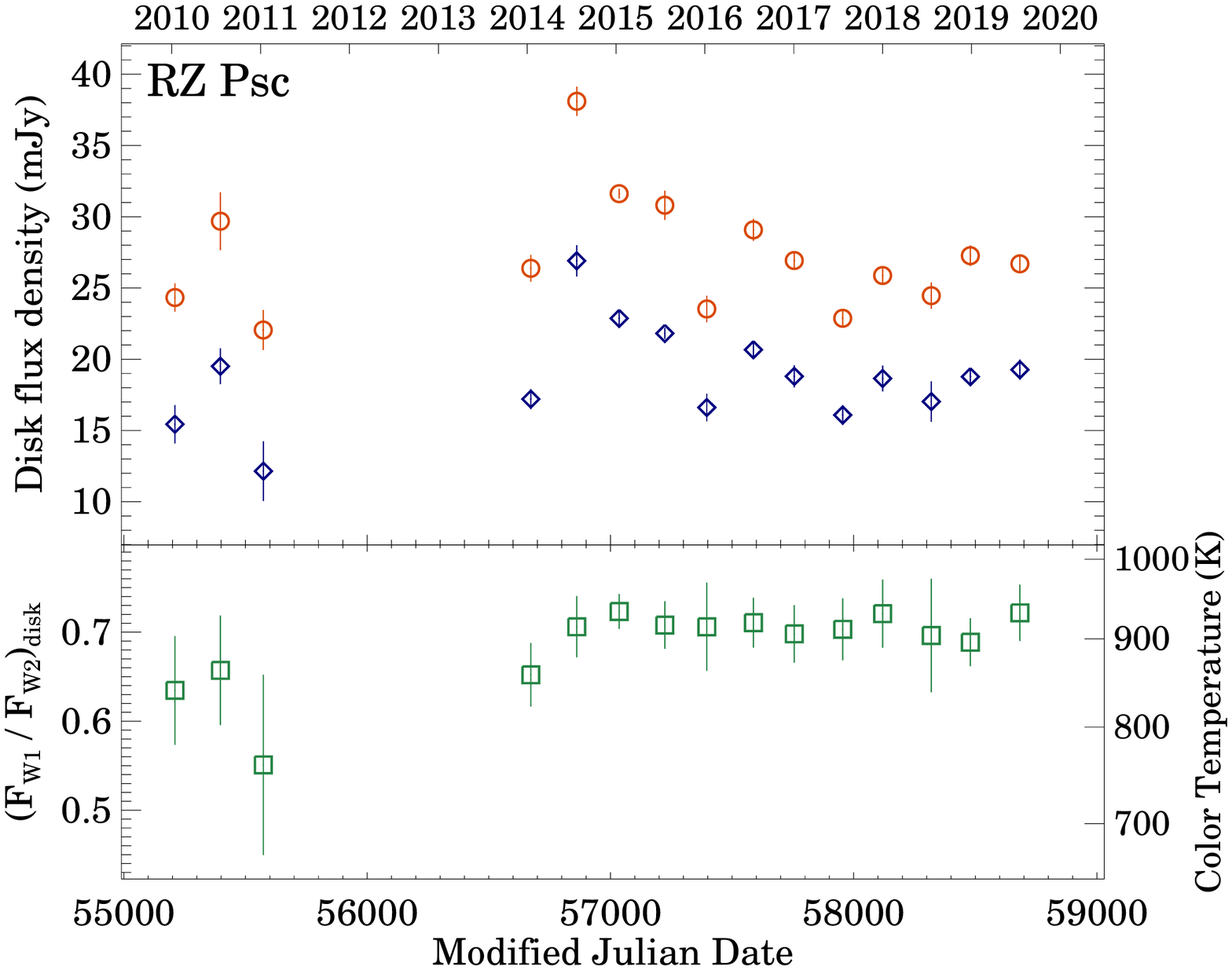} 
\includegraphics[scale=.37,angle=0]{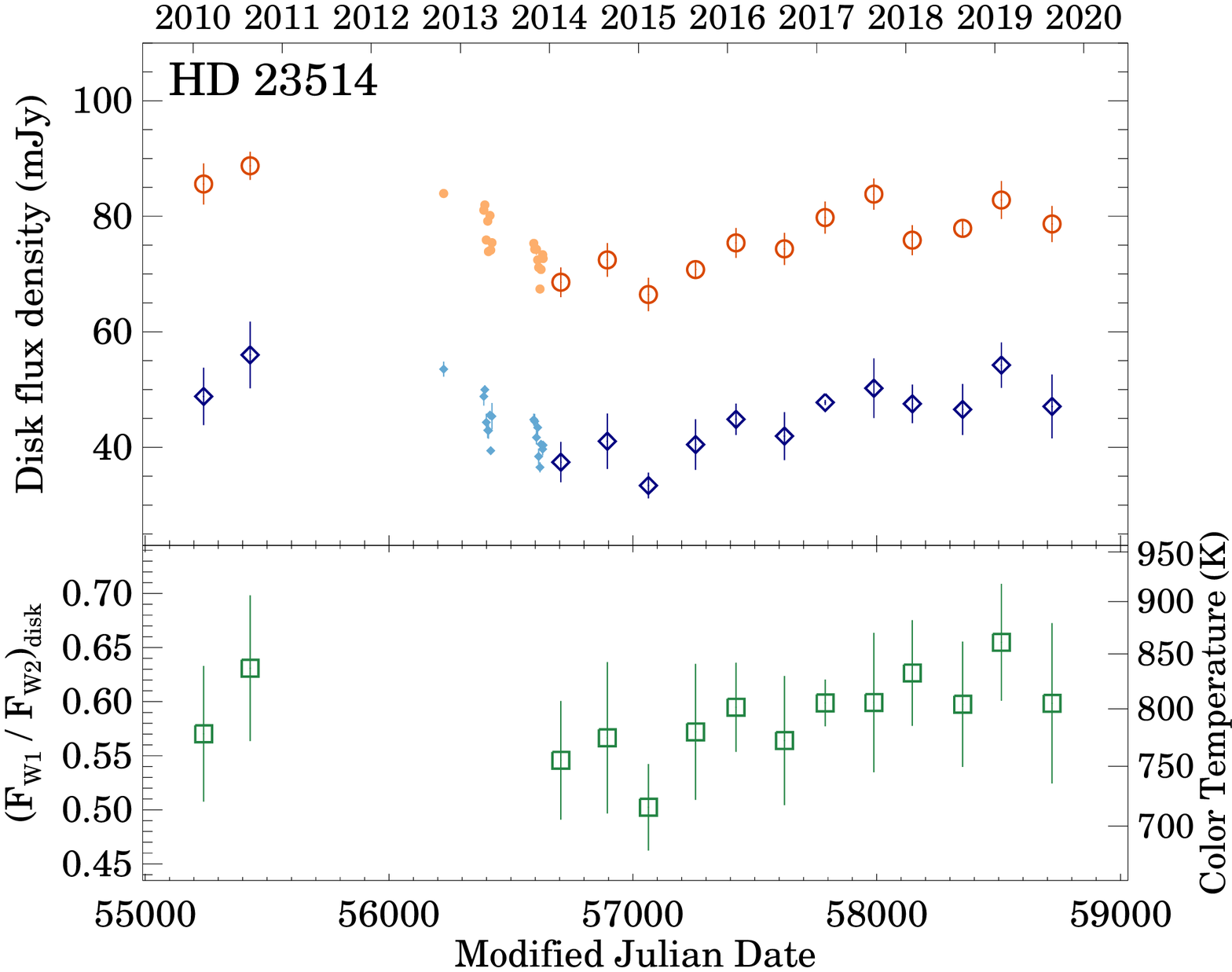}
\caption{ {Seasonal averages of W1 and W2 band disk fluxes (top panels) and their ratios 
(bottom panels) between 2010 and 2019 for five previously known EDDs. Disk fluxes 
derived from multi-year Spitzer photometric monitoring of HD\,23514, ID\,8, and P\,1121 \citep{meng2015,su2019} 
are also plotted with smaller filled symbols. 
}
\label{fig:knowneddsvar}
}
\end{center}
\end{figure*}

In order to assess the long term variability of the targets, we used the same strategy 
as described in Sect.~\ref{section:yearlyvariability}. For BD+20\,307 and 
HD\,145263,
we applied only the $\chi^2$ test and found both objects to be constant within the 
errors of the 4.6\,{\micron} observations. The mid-IR flux of the disk around TYC\,8241-2652-1 
dropped significantly before the start of the {\sl WISE} mission \citep{melis2012} and
recent observations indicated that the disk has remained depleted \citep{guenther2017}. 
In the W1/W2 bands, the source does not exhibit significant excess, thus it is not surprising 
that we found no variability in either filter. 
The other five objects (ID\,8, P\,1121, V488\,Per, RZ\,Psc, and HD\,23514), however, 
proved to be variable at both 3.4 and 4.6\,{\micron}.
Apart from HD\,23514, significant color (W1$-$W2) variations were also detected.
Figure~\ref{fig:knowneddsvar} displays their average disk flux densities in each 
observational window, typically separated by $\sim$6 months, between 2010 and 2019. 
To compute these 
data, we subtracted the photospheric contributions from the measured 
flux densities. For photosphere models, the stellar parameters were taken from the 
literature and the corresponding ATLAS9 model was fitted to the available optical and 
near-infrared photometric data of the given source. 
The bottom panels of each plot show the ratio of W1/W2 disk fluxes and the corresponding color temperatures.

%%%%%%%%%%%%%%%%%%%%%%%%%%%%%%%%%%%%%%%%%%%%%%%%%%%%%%%%%5

\clearpage



\bibliographystyle{apj}
%\bibliography{adssample}

%\begin{thebibliography}{}
%\expandafter\ifx\csname natexlab\endcsname\relax\def\natexlab#1{#1}\fi

%\end{thebibliography}



\end{document}